\documentclass[prd, aps, nofootinbib, preprintnumbers, showpacs, superscriptaddress, twocolumn, floatfix]{revtex4}

\usepackage{aas_macros}

\usepackage{amsmath,amssymb,wasysym}
\usepackage[english]{babel}
\usepackage{ifpdf}
\usepackage[latin9]{inputenc}
\usepackage{mathpazo,mathrsfs,float,psfrag}
\usepackage[usenames,dvipsnames]{color}

\ifpdf
\usepackage{hyperref}
\fi

\usepackage[pdftex]{graphicx}
\usepackage{epstopdf}

\graphicspath{{pdf/}}

\def\text#1{{\rm #1}}

\def\pd{\partial}
\def\ji{\varphi}
\def\re{\textrm{Re}}
\def\im{\textrm{Im}}

\def\cm{\textrm{cm}}

\def\gr{\textrm{g}}
\def\sec{\textrm{s}}
\def\ms{\textrm{ms}}

\def\Hz{\textrm{Hz}}
\def\kHz{\textrm{kHz}}
\def\MSun{\mathcal M_{\astrosun}}

\def\eqref#1{(\ref{#1})}

\newcommand{\thickhline}{\noalign{\hrule height 0.8pt}}

\def\Quilt{\textsc{Quilt}}
\def\Cactus{\textsc{Cactus}}
\def\Carpet{\textsc{Carpet}}
\def\Thor{\textsc{Thor}}
\def\rnsid{\textsc{rns}}

\begin{document}
\title
[Self-Gravitating Disks]
{Stability of general-relativistic accretion disks}

\author{Oleg Korobkin}
\affiliation{Department of Physics \& Astronomy, Louisiana State
  University, USA}
\affiliation{Center for Computation \& Technology, Louisiana State
  University, USA}

\author{Ernazar B. Abdikamalov}
\affiliation{Center for Computation \& Technology, Louisiana State
  University, USA}

\author{Erik Schnetter}
\affiliation{Center for Computation \& Technology, Louisiana State
  University, USA}
\affiliation{Department of Physics \& Astronomy, Louisiana State
  University, USA}

\author{Nikolaos Stergioulas}
\affiliation{Department of Physics, Aristotle University of
  Thessaloniki, Greece}

\author{Burkhard Zink}
\affiliation{Theoretical Astrophysics, University of T\"ubingen, Germany}

\date{November 12, 2010}

\begin{abstract}
   Self-gravitating relativistic disks around black holes can
   form as transient structures in a number of astrophysical scenarios
   such as binary neutron star and black hole-neutron star coalescences, as well
   as the core-collapse of massive stars. We explore the stability of such
   disks against runaway and non-axisymmetric instabilities using
   three-dimensional hydrodynamics simulations in full
   general relativity using the \Thor{} code.
   We model the disk matter using the ideal fluid approximation
   with a $\Gamma$-law equation of state with $\Gamma=4/3$. We explore
   three disk models around non-rotating black holes with
   disk-to-black hole mass
   ratios of $0.24$, $0.17$ and $0.11$. Due to metric blending in
   our initial data, all of our initial models contain an initial
   axisymmetric perturbation which induces radial disk
   oscillations. Despite these oscillations, our models do not
   develop the runaway instability during the first several orbital
   periods. Instead, all of the models develop unstable
   non-axisymmetric modes on a dynamical timescale. We observe two
   distinct types of instabilities: the Papaloizou-Pringle and the
   so-called intermediate type instabilities. The development of the
   non-axisymmetric mode with azimuthal number $m=1$ is accompanied by an
   outspiraling motion of the black hole, which significantly amplifies the growth
   rate of the $m=1$ mode in some cases. Overall, our
   simulations show that the properties of the unstable
   non-axisymmetric modes in our disk models are qualitatively similar
   to those in Newtonian theory.
\end{abstract}

\maketitle

\section{Introduction}
\label{S:Intro}

Thick relativistic accretion disks and tori around black holes (BHs) can
form as transient structures in several astrophysical scenarios,
including core-collapse of massive stars~\cite{Woosley93b, Woosley06,
SekiguchiShibata10, Proga03a, Fujimoto06a, Dessart08a, OConnor10a,
Nagataki09a, Harikae09a, Lopez-Camara09a, MacFadyen99a, MacFadyen01a}
and the merger of neutron star (NS)~\cite{ShibataTaniguchi06, Baiotti08,
  Baiotti09, Liu08a, Giacomazzo09a, Kiuchi09a, Duez10a, Rezzolla10b,
  Ruffert96a, Ruffert96b, Ruffert97, Ruffert98a, OechslinJanka06a,
  Anderson08a, Anderson08b} and NS-BH binaries~\cite{Chawla10,
  Duez10a, RuffertJanka10a, ShibataUryu06a, ShibataUryu07a, Foucart10a,
  Shibata09a, Yamamoto08a, Etienne09a, Duez10b, Rantsiou08a}. 
Many models of gamma-ray bursts (GRBs) rely on the existence of
massive dense accretion disks around BHs~\cite{Popham99a, Woosley93b,
Piran04a, Lee07, SekiguchiShibata10}. The observed $ \sim 10^{51}
$ erg energy powering GRBs \cite{Piran04a, Meszaros92, Gehrels09} is believed
to be coming either from the accretion disk and/or rotation of the central object. 
If this energy comes from the disk~\footnote{
  In alternative models, the GRB is powered by the rotation energy of the 
  central object. If the latter is a BH, then the rotation can be converted 
  through Blandford-Znajek mechanism~\cite{Blandford77}. If the central object 
  is a NS, then its energy can be transformed through magnetic 
  fields~\cite{Usov92, Metzger10}.}, 
then -- assuming that the efficiency of converting disk energy into that of 
GRB can at most be $ \sim 10 \% $ as in many other astrophysical
scenarios~\cite{Frank02} -- the accretion disk should have a mass of $
\gtrsim 0.01 \ \MSun $. 
Recent numerical simulations have demonstrated that the mass of the disk 
resulting from binary NS-NS (BH-NS) mergers, which are thought to be 
candidates for central engine of short GRBs (see, e.g.~\cite{Rosswog02a,
Lee07}, but also~\cite{Lyutikov09a}), 
can be in the range of $\sim 0.01 - 0.2\;\MSun$~\cite{ShibataTaniguchi06,
Shibata09a, Rezzolla10b, Duez10a}.
Due to their larger initial mass, in core-collapse of massive stars,
which are believed to be progenitors of long GRBs\footnote{Note that
  although all of the long GRBs are believed to be produced by
  core-collapse of massive stars, not all of the latter can produce
  long GRBs: In order to produce a GRB, precollapse star is probably
  required to be rapidly rotating \cite{Woosley93b, MacFadyen01a,
    Woosley02}.} \cite{Woosley93b, Woosley06, Dessart08a}, significantly more mass
is available for forming
disks~\cite{SekiguchiShibata07,SekiguchiShibata10}. 
   
The neutrino-annihilation mechanism for triggering GRBs \cite{Lee07, Dessart09a, Rosswog03a},
in which neutrinos emitted by the disk annihilate predominantly at the
rotation axis to produce $e^+-e^-$ pairs and deposit energy behind the
jet, rely on super-Eddington accretion rates, which can take place
only in high-density ($\rho \sim 10^{12} \ \mathrm{g \ cm^{-3}}$)
disks. Moreover, the efficiency of neutrino annihilation at the
rotation axis and the requirement of a small baryon load of the
relativistic ejecta~\cite{Meszaros92, Rees92, Piran04a} was shown to
strongly favor a toroidal structure for accreting
matter~\cite{Goodman87b, Jaroszynski93}.          

The duration of prompt $\gamma$-ray emission is $> 2 \ \sec$ 
($ \lesssim 2 \ \sec $) for long (short) GRBs, while that of the later-time 
non-$\gamma$ ray
emission can be as long as $ \sim
10^5 \ \sec$ (e.g., \cite{Gehrels09}). 
Recent observations of GRB afterglows have revealed a variety of
late-time emission processes, including X-ray plateaus, flares, and
chromatic breaks~\cite{Gehrels09}, some of which can persist up to
$\sim 10^5 \ \sec$ following the initial GRB prompt emission (see,
e.g., \cite{Gehrels09} for a recent review). 
The amount of energy released in the late-time emission phase can be
comparable or even larger than that produced during the prompt emission phase. 
Both the prompt and the late-time emissions can be explained as the result of
the activity of the central engine (see, e.g., \cite{Meszaros06, Gehrels09}),
although alternatives models exist (see, e.g., \cite{Kobayashi07, NakarPiran03,
ZhangEA06}).
If the emission energy comes from the disk, such long durations of observed GRB
emission either require the disks to accrete in a quasi-stable manner for a
sufficiently long period of time, or require the engine to be
restarted in some way.

Early studies of the stability of accretion disks have revealed
that they can be subject to several types of
axisymmetric and/or non-axisymmetric instabilities in a number of
circumstances~\cite{Abramowicz83, PPI, PPII,
  Kojima86a, WTH94, Font02a, Zanotti03a}. Instabilities can lead
to highly variable and unstable accretion rates, posing a serious
challenge to the viability of ``accretion-powered'' GRB models.  
The so-called dynamical runaway instability of thick accretion disks around
BHs was first discovered by Abramowicz, Calvani \& Nobili
\cite{Abramowicz83}. 
This instability is similar to the dynamical instability in close binary
systems, when the more massive companion overflows its Roche lobe.
In such a case, the radius of the Roche lobe shrinks faster than the radius of
the companion, leading to a catastrophic disruption of the latter.
In disk+BH systems, a toroidal surface analogous to the Roche lobe can be found.
A meridional cut of this surface has a cusp located at the $L_1$
Lagrange point.
If the disk is overflowing this toroidal ``Roche lobe'', then the mass-transfer
through the cusp will advance the cusp outwards inside the disk.
This can result in a catastrophic growth of the mass-transfer and disruption
of the disk in just a few dynamical timescales.

Abramowicz et al.~\cite{Abramowicz83} studied the properties of mass
transfer using many simplifying assumptions: a pseudo-Newtonian
potential for BH gravity~\cite{Paczynsky80}, constant specific angular
momentum of the disk, and approximate treatment of the disk
self-gravity. They found that the runaway instability occurs for a
large range of parameters, such as the disk radius and the disk-to-BH
mass ratio $ M_\mathrm{D} / M_\mathrm{BH} $. Subsequent and somewhat
more refined studies found that the rotation of the BH has a
stabilizing effect~\cite{Wilson84, Abramowicz98}, and a non-constant
radial profile angular momentum was found to strongly disfavor the
instability~\cite{Abramowicz98, Daigne97a, Font02b,
  DaigneFont04}. On the other hand, studies using a Newtonian pseudopotential for
the BH~\cite{Khanna92, Masuda98a} and relativistic calculations with
fixed spacetime background~\cite{Nishida96, Font02a} found indications
of the self-gravity of the disk to favor the instability. However,
Montero et al.~\cite{Montero10} recently performed the first
self-consistent and fully general relativistic simulations of thick 
accretion disks around BHs in axisymmetry for a few dynamical
timescales. They found no signatures of a runaway instability during the
simulated time, perhaps implying that the self-gravity of the disk
does not play a critical role in favoring the instability, at least
during the first few dynamical timescales. 

The problem of the existence and development of non-axisymmetric instabilities
has a long history. For thin Keplerian self-gravitating disks it was found
that the Toomre
parameter~\cite{Safronov60b, Toomre64, Gammie01} can be used to
determine stability against both local clumping or fragmentation, and
formation of global non-axisymmetric modes. 
For thick pressure-supported disks, Papaloizou and Pringle
discovered~\cite{PPI} the existence of a global hydrodynamical instability that
develops on a dynamical timescale in disks with negligible self-gravity and
constant specific angular momentum.
A follow-up publication~\cite{PPII} found this instability also in the disks
with power law distribution of specific angular momentum $\ell(r) =
\ell_0(r/r_c)^{2-q}$ for all $q>\sqrt{3}$.
Kojima~\cite{Kojima86a, Kojima86b} found the Papaloizou-Pringle (PP)
instability in equilibrium tori on a fixed Schwarzschild
background~\cite{AJS78:AccretingDisks} using a linearized perturbative approach.

Subsequent works clarified the nature of the PP instability~\cite{GGN86,
Blaes86, NGG87, FrankRobertson88}, established how it redistributes specific
angular momentum~\cite{Zurek86a} and discovered that accretion has a stabilizing
effect on the disk~\cite{Blaes87a, BlaesHawley88}.
In particular, Narayan et al.~\cite{GGN86} showed that the PP modes are formed
by two boundary wave-like perturbations with energy and angular momentum of
opposite signs that are coupled across a forbidden region near the mode corotation
radius.
For wide disks around BHs, the accretion suppresses the development of the inner
boundary wave and therefore has a stabilizing effect on PP
modes~\cite{Blaes87a, BlaesHawley88}. 
For slender disks, the development of PP modes is mostly unaffected by
accretion~\cite{BlaesHawley88}.
The PP instability itself amplifies accretion by exerting torques on the
disk and redistributing specific angular momentum~\cite{Zurek86a}.

When the self-gravity of the disk has been taken into account, it was
found~\cite{GoodmanNarayan88a, ChristodoulouNarayan92, Christodoulou93} that
two new types of non-axisymmetric instabilities appear, while the PP instability
disappears for most of the models except slender ones with weak self-gravity.
The first type of unstable modes (J-modes) appears in strong self-gravity
regime, and it is an analog of the classical Jeans instability.
The second type of unstable modes was found in the strong and medium self-gravity
regimes~\cite{GoodmanNarayan88a}.
The modes of this type are referred to as intermediate modes (I-modes) and
represent elliptic deformations of the disk (or triangular, square, etc.\
deformations for higher azimuthal numbers -- see e.g.~\cite{Christodoulou93}).

Yet another type of instability, the so-called ``eccentric instability'', was
discovered in~\cite{ARS89} for thin nearly Keplerian self-gravitating disks
when the central mass was allowed to move.
An elaborate mechanism called ``SLING amplification''~\cite{Shu90b} was proposed
to describe this instability.
Subsequent investigation~\cite{Heemskerk92} of this instability in thin disks
suggested a different mechanism and predicted that the system is dynamically
unstable only when the mass of the disk exceeds the mass of the central object.

Finally, Woodward, Tohline, and Hachisu \cite{WTH94} presented an
extensive parameter study of thick self-gravitating disks 
in Newtonian gravity to determine the types, growth rates and pattern speeds
of non-axisymmetric modes.
The central mass in their simulations was allowed to move, and they used 3D
time evolutions of the disk models with wide range of parameters, including
disk-to-central object mass ratios $M_D/M_c=0, \, 1/5, \, 1$ and $\infty$.

Several recent publications address accretion disks and instabilities in these
disks in context of GRB central engines.
In~\cite{Rezzolla10b}, Rezzolla et al.\ studied the properties of accretion
disks resulting from binary NS mergers.
They obtained thin accretion disks with masses $\sim0.01-0.2\ \MSun$ and no
evidence of growing non-axisymmetric modes or runaway instability.
In~\cite{TMP10}, Taylor, Miller, and Podsiadlowski used 3D SPH simulations with
detailed microphysics and neutrino transport to follow a collapse of an
iron core up to the formation of a thin massive accretion disk and development
of global non-axisymmetric modes.
They found that torques created by the non-axisymmetric modes provide the
main mechanism for angular momentum transport, leading to high accretion rates
of $\sim0.1-1\ \MSun/\sec$, which may create a favorable conditions for
powering GRBs.

The stability of disks to runaway and non-axisymmetric instabilities
is a three-dimensional problem that has to be addressed in the
framework of full GR. 
One issue of importance is to understand if the runaway instability is
affected by non-axisymmetric instabilities, and vice versa.  
Despite significant theoretical and computational effort, previous studies of
the runaway instability do not give a definite answer to this
question. To our knowledge, only the work by Rezzolla et
al.~\cite{Rezzolla10b} explored the stability of the disks in 3D and full GR
(for the disks that form in their binary NS merger simulations).
However, it is important to create a comprehensive overall picture of the
stability of accretion disks for a richer variety of disk models,
which would require investigating a wider range of parameters. In 
our work, we study in detail the stability of slender and moderately slender
disks with constant distribution of specific angular momentum, which are
expected to be more unstable both to runaway~\cite{Daigne97a, Font02b}
and to non-axisymmetric instabilities~\cite{Blaes87a, BlaesHawley88}
compared to models of Rezzolla et al.~\cite{Rezzolla10b}. Our study is
based on three-dimensional hydrodynamics simulations in full GR.

We model our disks using the ideal fluid approximation (i.e.\ without
viscosity) with a $\Gamma$-law equation of state (EOS) and $\Gamma=4/3$.
We do not include additional physics such as magnetic fields or
neutrino and radiation transport due to the complexity and
computational cost of the resulting problem. 
Nevertheless, the adopted approach will allow us to identify GR effects which
can operate in more complex setting that include more realistic microphysics,
neutrino/radiation transport and magnetic fields.
Also, we limit ourselves to the case of non-rotating BHs, while the
case of rotating BHs will be studied in future publications.

Another aim of this paper is to estimate the detectability of the
gravitational waves (GWs) by the accretion disk dynamics. The radial
and/or non-axisymmetric oscillations of accretion disks can be a source
of strong GWs. If non-axisymmetric deformations persist for long
enough time, then the emitted GWs can be detectable by current and
future GW detectors \cite{vanPutten01f}, provided the source is not
too far away.
Further work on the stability properties
of accretion disk could shed more light on the number of cycles a
non-axisymmetric deformation can persist, and thus on the prospects of detecting
GWs from such systems. 

This paper is organized as follows: 
Section~\ref{S:NumericalMethod} describes the formulations and numerical
methods used in this paper, including multiblock grids
(subsection~\ref{SS:Multiblock}) and the formulations used to evolve the
general relativistic (subsection~\ref{S:GREvolution}) and hydrodynamic
(subsection~\ref{S:HydroEvolution}) equations.
Section~\ref{S:IDSetup} describes the grid setup, initial data construction
procedure and the analysis techniques for the non-axisymmetric instabilities.
Sections~\ref{S:TimeEv} and~\ref{S:Nonaxisym} present the
results of the time evolution of self-gravitating disks and the analysis of
non-axisymmetric instabilities.

Throughout the paper we use CGS and geometrized units $G = c = 1$.

\section{Numerical methods}
\label{S:NumericalMethod}

The numerical time-evolution scheme used in our study can be split into
two main parts: the spacetime evolution, and the fluid dynamics
equations.
These two parts are evolved in a coupled manner. 
The numerical code that we use has been developed within the \Cactus{}
computational infrastructure~\cite{Goodale02a, cactus_web}, and uses
the \Carpet{} mesh
refinement and multiblock driver~\cite{Schnetter04a, carpet_web}.
A separate module based on \Carpet{} provides a range of multiblock systems to
represent a variety of computational domains for 3D evolution
codes~\cite{Schnetter:2006pg}.
The module for evolving the GR hydrodynamics equations uses the \Thor{}
multiblock code~\cite{Zink08}, which has been coupled to the
multiblock-based module \Quilt{} for evolving the spacetime~\cite{Pazos07}.
The latter implements the Generalized Harmonic formulation of the Einstein
equations in first-order form~\cite{LSKOR06}.
Below we give a brief description of the methods implemented in each module.

\subsection{Multiblock approach}
\label{SS:Multiblock}

  \begin{figure}[!htp]
  \begin{center}
  \begin{tabular}{cc}
  \includegraphics[width=0.23\textwidth]{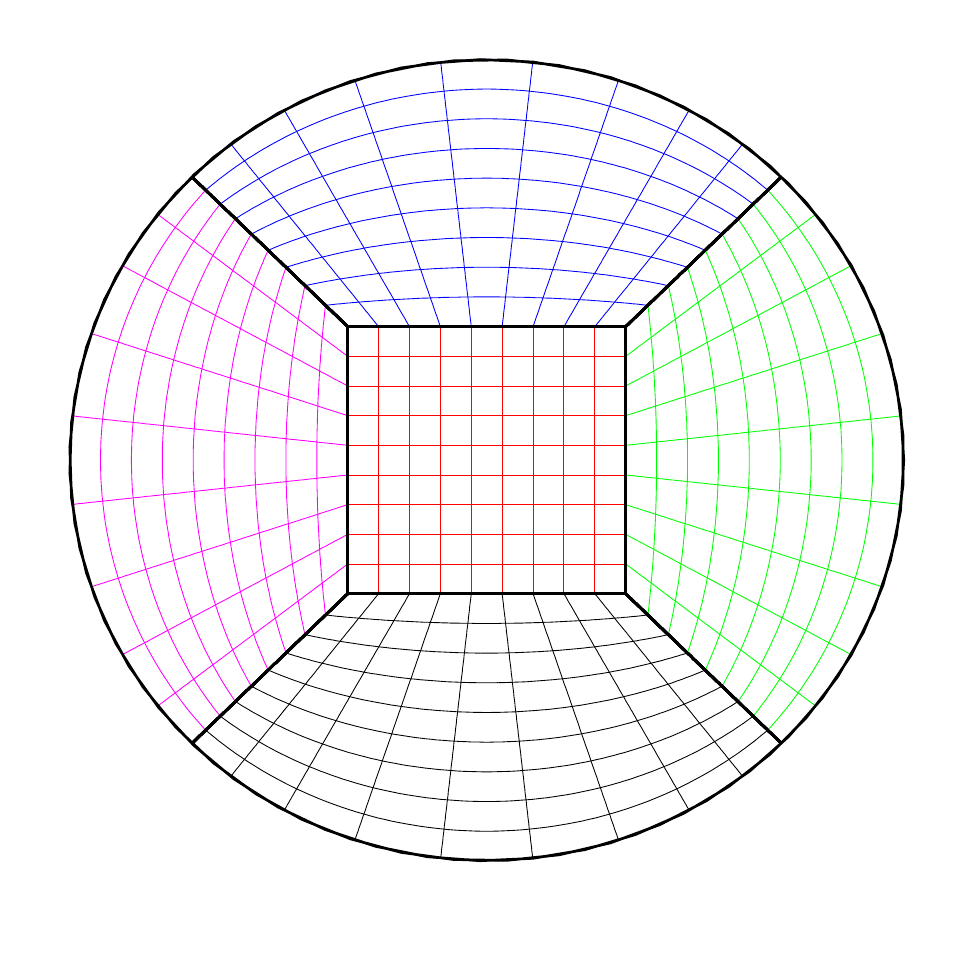} &
  \includegraphics[width=0.23\textwidth]{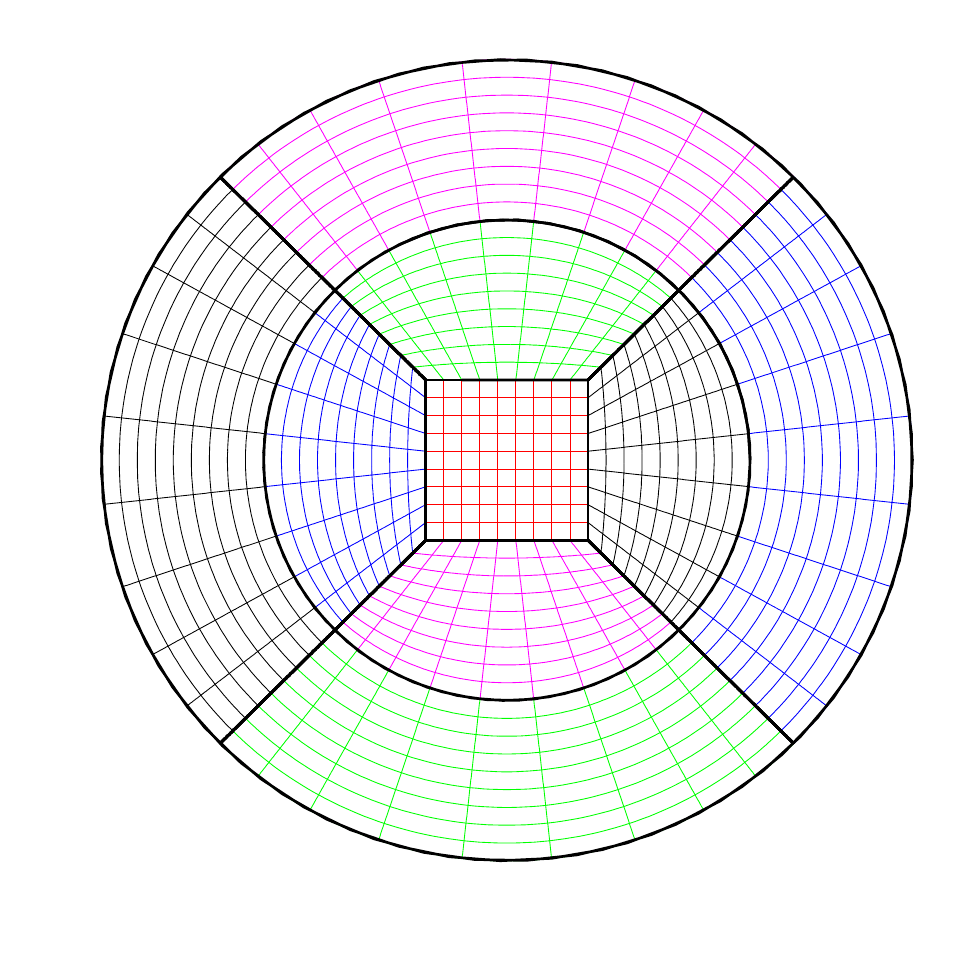} \\
  (a) & (b) \\
  \includegraphics[width=0.23\textwidth]{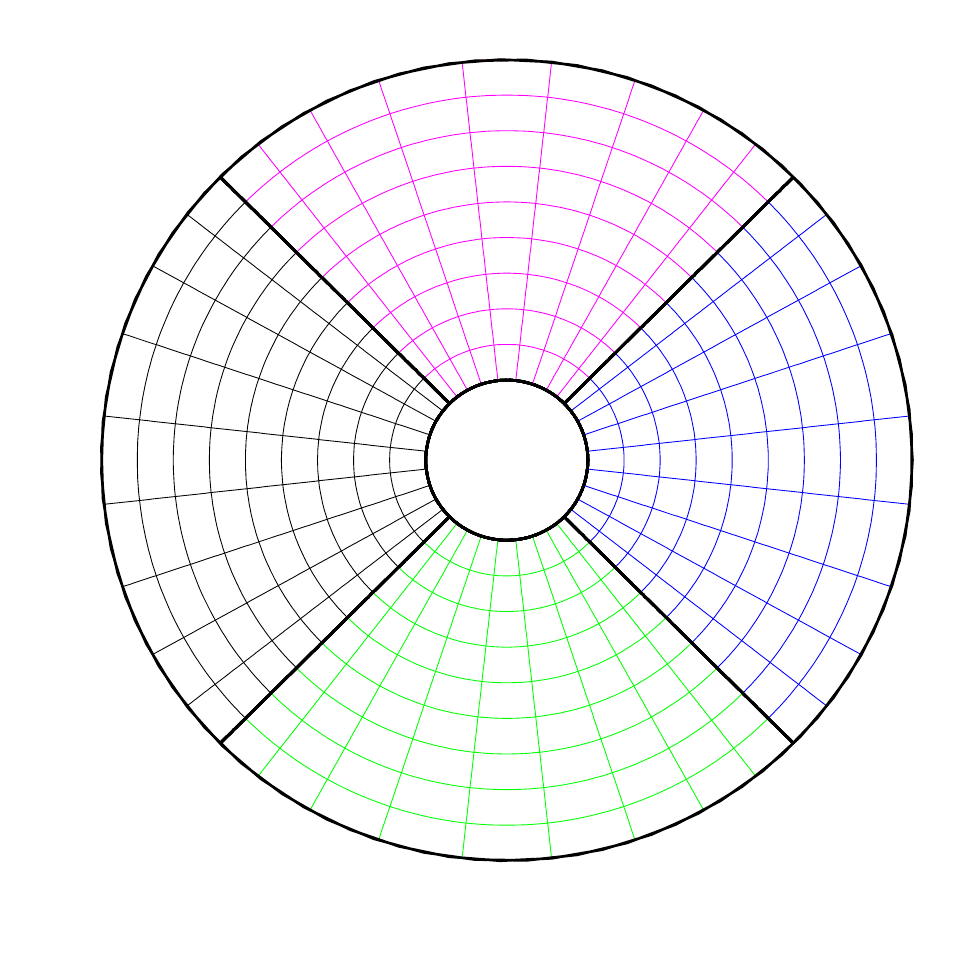} &
  \includegraphics[width=0.23\textwidth]{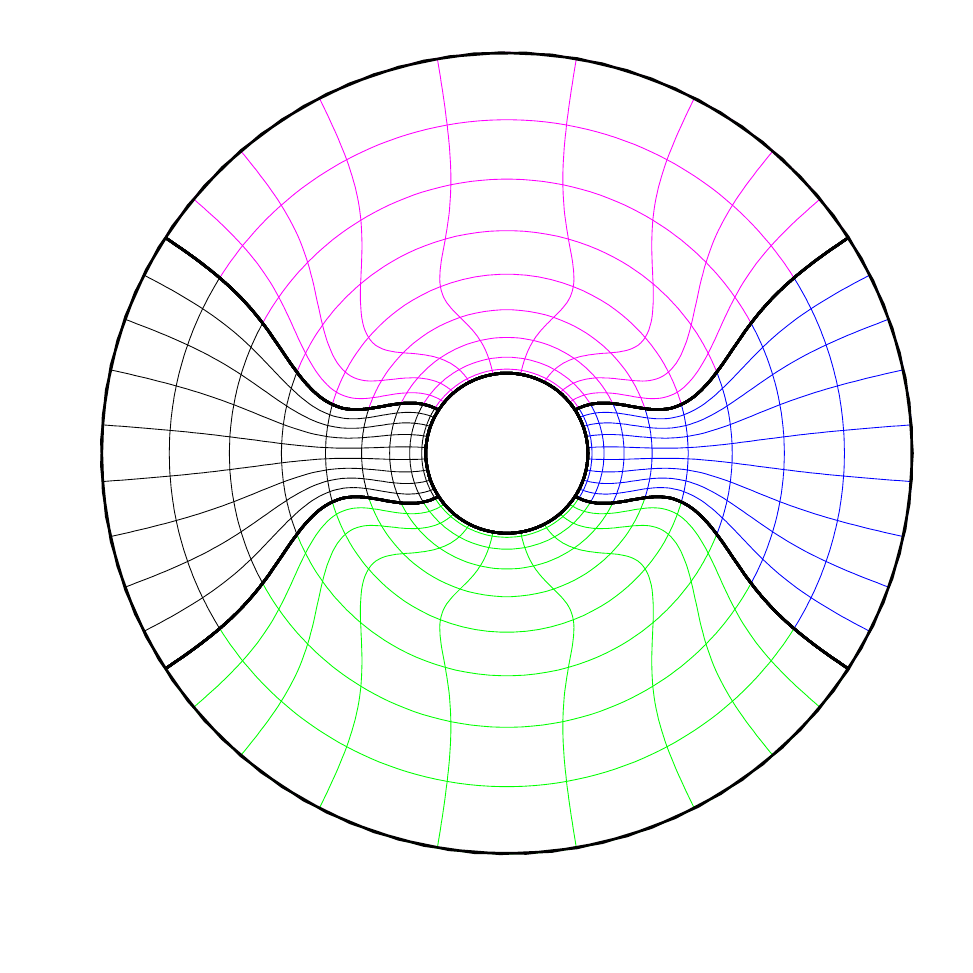} \\
  (c) & (d) \\
  \end{tabular}
  \caption{
    Cross-sections of the patch systems used in this work in the $xz$-plane:
    (a) seven-block system,
    (b) thirteen-block system,
    (c) six-block system with straight coordinate lines,
    (d) six-block system with distorted radial coordinate lines, described in
    Section~\ref{S:AdMesh} below.
  }
  \label{F:MyPatchsystems}
  \end{center}
  \end{figure}

For mesh generation and parallelization purposes, we employ the multiblock
infrastructure developed by~\cite{Schnetter:2006pg}.
The multiblock approach is widely used in astrophysical and numerical
relativity simulations (see~\cite{Thornburg87, Thornburg93, Gomez97a, 
Gomez97b, Gomez98a, Bonazzola99a, Kidder01a, Gourgoulhon01,
Grandclement-etal-2000:multi-domain-spectral-method,
Pfeiffer03, Thornburg2004:multipatch-BH-excision, Calabrese2004:boosted-bh,
Scheel-etal-2006:dual-frame, Pfeiffer:2007, Foucart08, Duez08, Pazos09,
Fragile09, Scheel09, Duez10b} and references therein).

Figure~\ref{F:MyPatchsystems} shows meridional cuts (cuts in the $xz$-plane)
of the block systems that we use.
The first two block systems on Fig.~\ref{F:MyPatchsystems} are the
seven-block (also known as a ``cubed sphere''~\cite{Ronchi96}) and the
thirteen-block systems.
They both have a spherical outer boundary and no coordinate singularities.
The thirteen-block system additionally has spherical grid coordinate surfaces
in the outer layer of blocks, which is very convenient for computing
spherical harmonics and extracting gravitational waves. 
The thirteen-block system is well adapted for simulating processes that
involve a small source of waves, surrounded by an extended spherical wave
propagation region. 
The central cubical block can have a high resolution and can be used to
model accurately the dynamics of the source, the intermediate layer
can represent a near zone, while the outer layer of blocks model the
radiation zone, which can carry an outgoing radiation using, e.g.,
constant angular and radial resolutions. These two systems are used in
the Appendix for test evolutions of rotating
and non-rotating polytropic stars.

The remaining block systems on Fig.~\ref{F:MyPatchsystems}c
and~\ref{F:MyPatchsystems}d are six-block systems, and these systems are more
suitable for modeling configurations with a central black hole. 
The block system displayed on Fig.~\ref{F:MyPatchsystems}d is similar
to the block systems that we use for simulating accretion disks around black
holes (see Section~\ref{S:AdMesh}).
These two block systems contain empty spherical region, which is used for
excising interior of the BH\@.

The multiblock approach has a number of additional advantages,
including a smooth
excision boundary~\cite{Calabrese2003:excision-and-summation-by-parts,
Thornburg2004:multipatch-BH-excision, Szilagyi-etal-2006a1}, 
a smooth spherical outer boundary suitable for radiating boundary conditions, 
a spherical wave propagation zone, and absence of coordinate singularities such
as those associated with spherical or cylindrical coordinates. 
Also, multiblock systems such as displayed on Fig.~\ref{F:MyPatchsystems}
allow at low cost to extend the computational domain far outwards in
order to e.g.\ causally disconnect the system dynamics from the outer
boundary and to allow more accurate extraction of gravitational
waves~\cite{Pazos07, Reisswig10b}.

\subsection{Hydrodynamics evolution}
\label{S:HydroEvolution}

The evolution equations of a relativistic fluid are derived from the
covariant equations of conservation of rest mass $\nabla_a(\rho u^a)=0$ and
energy-momentum $\nabla_b T^{ab}=0$, where $T^{ab}$ has the following
form for an ideal fluid~\cite{Landau-Lifshitz2, Zink08}:
$$
T^{ab} = (\rho+u+P) u^a u^b + Pg^{ab},
$$
\noindent where $P$ is the fluid pressure, $\rho$ is the rest mass density,
$u$ is the internal energy density~\footnote{It is also common to define
a specific internal energy $\varepsilon=u/\rho$~\cite{Wilson72, lrr-2003-4}},
$u^a$ the fluid's four-velocity,
and $g^{ab}$ is the spacetime metric in contravariant form.
The quantities $\{\rho, u^i, u\}$ form a set of \emph{primitive variables}
that uniquely determine the state of a single-component relativistic fluid at
every point in space.

Our evolution equations read:
\begin{align*}
 \pd_t(\sqrt{-g}\rho u^t) + \pd_i(\sqrt{-g}\rho u^i) &= 0, \\
 \pd_t(\sqrt{-g}{T^t}_a) + \pd_i(\sqrt{-g}{T^i}_a) &=
 \sqrt{-g}{T^c}_d{\Gamma^d}_{ac},
\end{align*}
\noindent where $t$ is the time coordinate, $g\equiv\det(g_{ab})$ is the
determinant of the spacetime metric and ${\Gamma^d}_{ac}$ are the Christoffel
symbols associated with this metric.
After introducing a set of \emph{conserved variables}
$\{D\equiv\sqrt{-g}\rho u^t, Q_a\equiv\sqrt{-g}T^t_a\}$, the equations can be
cast into a flux-conservative form (as in~\cite{Banyuls97,Gammie03}; see
details specific to our particular scheme in~\cite{Zink08}):
\begin{align*}
 \pd_t D + \pd_i D^i &= 0, \\
 \pd_t Q_a + \pd_i Q_a^i &= S_a,
\end{align*}
\noindent where $D^i$ and $Q_a^i$ are the fluxes of the conserved variables
$D$ and $Q_a$, while $S_a$ are the source terms for $Q_a$.

These equations are solved on each block using a finite volume
cell-centered scheme in the local coordinate basis of that block.
The reconstruction of the primitive variables on the cell interfaces
is performed
using the piecewise-parabolic monotonous (PPM) 
method~\cite{ColellaWoodward84, MartiMuller96}, while
the fluxes through the interfaces are calculated using a Harten-Lax-van Leer
(HLL) Riemann solver~\cite{Harten83}.
In order to compute fluxes, source terms, and the stress-energy tensor
$T^{ab}$, the conservative variables need to be converted into primitive ones
at every timestep.
This is done using a Newton-Rhapson iterative $2D_W$ solver~\cite{Noble2006}
with the non-isentropic $\Gamma$-law EOS\@.
If at a particular cell on the grid the procedure of primitive variables
recovery fails to produce physically meaningful values (which can happen for
number of reasons; see, e.g.,~\cite{Noble2006}), we use a $1D_P$ solver with an
isentropic polytropic EOS\@.
The vacuum region outside the disk is approximated by a low-density
artificial atmosphere, whose density is chosen to be $10^{-6}-10^{-7}$
of the initial maximum density in the system.

The boundary conditions for the hydrodynamics variables are imposed on the
interblock boundaries using interpolation from neighboring blocks. 
The overlap regions where interpolation is performed are created by adding extra
layers of grid points on each block face. At the outer boundary and
the inner excision boundary, we impose outflow boundary
conditions.
Further details on this numerical method of combining multiple blocks can be
found in~\cite{Zink08}.

\subsection{Spacetime evolution}
\label{S:GREvolution}

To evolve the spacetime metric $g_{ab}$ we use a generalized harmonic
(GH) formulation of the Einstein equations in the first-order representation developed
by Lindblom et al.~\cite{LSKOR06}. In this formulation, the coordinate conditions
are specified using a set of four gauge source functions $H_a$, which need to be
prescribed a priori. In our simulations, we have chosen the so-called
``stationary gauge''~\cite{LSKOR06,LindblomSzilagyi09}, in which the gauge
source functions stay frozen at their initial values, 
$H_a(t,x^i) = H_a(t_0,x^i)$. 
Such a choice of the gauge is convenient for quasi-stationary spacetimes, such as
perturbed BHs~\cite{Pazos07} or accretion disks around BHs.

The first-order representation that we use is linearly
degenerate, symmetric hyperbolic~\cite{Secchi96a,Secchi96b,Rauch85}, and
constraint damping. 
Linear degeneracy guarantees that the system will not develop gauge shocks
during the evolution~\cite{Liu79}. 
Symmetric hyperbolicity with boundary conditions imposed on incoming
characteristic fields is the necessary condition for well-posedness
of initial boundary value problems (see, e.g.,\ Chapter 6 of Gustafsson et
al.~\cite{Gustafsson95}).

Being constraint damping, which is one of the most important advantages
of the chosen formulation, means that the constraints
on the variables are
included into the evolution equations in such a way that they are
exponentially damped during evolution. 
Such a property not only reduces the error in the solution, but also eliminates
a variety of numerical instabilities associated with unbounded growth of the
constraints.

The spacetime metric evolution equations are discretized using finite
differences on multiple grid blocks that cover the computational
domain~\cite{Schnetter:2006pg,Zink08}. 
This means that the spacetime grid blocks are only subsets of the
hydrodynamics grid blocks; the overlap regions where the hydrodynamics
variables are interpolated are absent in the spacetime grid blocks.
The advantages of using multiple blocks were outlined in the
Section~\ref{SS:Multiblock} above.
The grids on any two neighboring blocks in the system are designed to
share a 2D interface grid, where so-called penalty boundary conditions are
imposed~\cite{lehner-2005-22}.
For the spatial numerical differentiation, we employ high-order convergent
finite differencing operators that satisfy summation by parts (SBP)
property~\cite{diener-2007-32}.
The SBP property together with penalty boundary conditions guarantees strict
linear numerical stability~\cite{lehner-2005-22}. 
All simulations of self-gravitating disks with dynamical treatment of
relativistic gravity in the paper were performed with the finite
differencing operators of
8th order convergence in the bulk of the grid, and 4th order convergence at
the boundaries. 
Time integration is performed with a 3rd order accurate Runge-Kutta method,
that satisfies a total variation diminishing (TVD)
property~\cite{NeilsenChoptuik00}.

\subsection{Code tests}
\label{S:CodeTests}

Numerical methods for solving the general relativistic hydrodynamics
equations and the spacetime equations are inherently complex, and codes
need to be thoroughly tested before they can be successfully applied to
physical problems (see e.g.~\cite{Font01,Montero08}). 
The hydrodynamics code \Thor{} was tested for fixed spacetimes by Zink et
al.\ in~\cite{Zink08}, where it was demonstrated that the code can handle
situations encountered in many astrophysical scenarios, including
relativistic shocks, rotating polytropes and equilibrium tori around
BHs~\cite{AJS78:AccretingDisks}. The spacetime evolution code \Quilt{}
was tested in~\cite{Pazos07, Korobkin08}.

We present results of our tests of the coupled hydrodynamics and
spacetime evolution codes in
Appendices~\ref{A:RotatingPolytrope}-\ref{A:TOVFreq}.
In particular, in Appendix~\ref{A:RotatingPolytrope} we report stable
convergent evolutions of a rapidly rotating polytropic star, and in
Appendix~\ref{A:TOVFreq}, we demonstrate that our code faithfully reproduces
fundamental frequencies of a non-rotating polytropic star.
Additionally, recently Zink et al.~\cite{Zink10} have used the coupled
\Thor{}+\Quilt{} code to measure the frequencies of
$f$-modes and to study their neutral points in the context of the
Chandrasekhar-Friedman-Schutz instability.

\subsection{Constraint damping}
\label{S:ConstrDamping}

The constraint damping scheme for the spacetime evolution mentioned
above depends on two freely-specifiable parameters,
$\kappa$ and $\gamma_2$. These parameters can be freely chosen as a
function of space and time,
i.e.\ the system of equations does not directly depend on their spatial or
temporal derivatives. Here we describe how these parameters are
specified in our simulations. 

In the generalized harmonic formulation, the set 
$u^{\alpha} = \left\{g_{ab},\Phi_{iab},\Pi_{ab}\right\}$
of dynamical fields consists of the metric $g_{ab}$ and linear combinations of
its derivatives $\Phi_{iab}=\pd_i g_{ab}$, $\Pi_{ab}=-n^c\pd_c g_{ab}$, where
$n^c$ is a normal to a $t=\textrm{const.}$ hypersurface. The evolution equations
for the fields can be written in the following form:
\begin{align}
\label{Eq:CDamping}
\pd_t g_{ab} &= G_{ab}(x^a,u^{\alpha}), \\
\pd_t \Phi_{iab} &= \mathcal{F}_{iab}(x^a,u^{\alpha}) - \gamma_2 C_{iab}, \\
\pd_t \Pi_{ab} &= P_{ab}(x^a,u^{\alpha}) - \kappa C_{ab} - \gamma_2 \beta^i C_{iab},
\end{align}

\noindent where $C_{ab}$, $C_{iab}$ are the constraints, $\beta^i$ is the
shift vector, and $G_{ab}$, $\mathcal{F}_{iab}$, $P_{ab}$ are right-hand sides
of the formulation \emph{without} the constraint damping terms.
In the continuum limit, the
constraints are zero, and the system~\eqref{Eq:CDamping} reduces to the original
Einstein equations. At the discrete level, the constraints can be non-zero, and
the constraint damping terms provide a non-vanishing contribution to the
right-hand sides of the system~\eqref{Eq:CDamping}. Since constraint damping
terms should act only as small corrections to the evolution system, their
contribution should not exceed those of functions $G_{ab}$, $\mathcal{F}_{iab}$,
$P_{ab}$. Otherwise, the evolution of the system will be dominated by the
numerical constraint violations.

We notice in our simulations that the functions $\mathcal{F}_{iab}$, $P_{ab}$
on the right-hand sides of~\eqref{Eq:CDamping} fall off as $1/r^\alpha$ with
$\alpha\sim4$ beyond $r>r_{\textrm{disk}}$, where $r_{\textrm{disk}}$ is the
approximate outer radius of the disk. At the same time, the constraint
violations $C_{ab}$, $C_{iab}$ fall off as $1/r^\beta$ with $\beta\sim1$. This
means that, if we use constant values for $\gamma_2$ and $\kappa$, the
constraint damping terms will dominate the dynamics of the system for
sufficiently large $r$. In order to avoid this situation, the functions $\kappa$
and $\gamma_2$ must fall off with the radius as $\sim1/r^3$.

We find that the following radial profiles of $\kappa$ and $\gamma_2$ for
our simulations lead to satisfactory results:
\begin{align}
\kappa(r) &= \kappa_*\left[1 - \frac{2}{\pi}
             \left(\frac{r_*}{1+r_*^2}+\arctan{r_*}\right)\right] \\
\gamma_2(r) &= \gamma_*\left[1 - \frac{2}{\pi}
               \left(\frac{r_*}{1+r_*^2}+\arctan{r_*}\right)\right]
\end{align}

\noindent where $r_* = (r-r_0)/\sigma$, and $\kappa_*$, $\gamma_*$, $r_0$,
$\sigma$ are (positive) constants. This radial profile approaches a constant
value of $\sim\kappa_*$  (or $\sim\gamma_*$) for $r<r_0$, and falls off as
$\sim1/r^3$ for $r\gg r_0$. The parameter $\sigma$ determines the extent of the
smooth transition region between these two regimes.
We use $\kappa_*=\gamma_*=4$, $r_0=12$, and $\sigma=8$ in our simulations.
Such profiles of the constraint damping coefficients allows imposing strong
constraint damping near the BH and the disk without introducing
spurious dynamics far away from the origin.

\section{Initial setup}
\label{S:IDSetup}

  \begin{figure}[!htp]
  \begin{center}
  \includegraphics[width=0.45\textwidth]{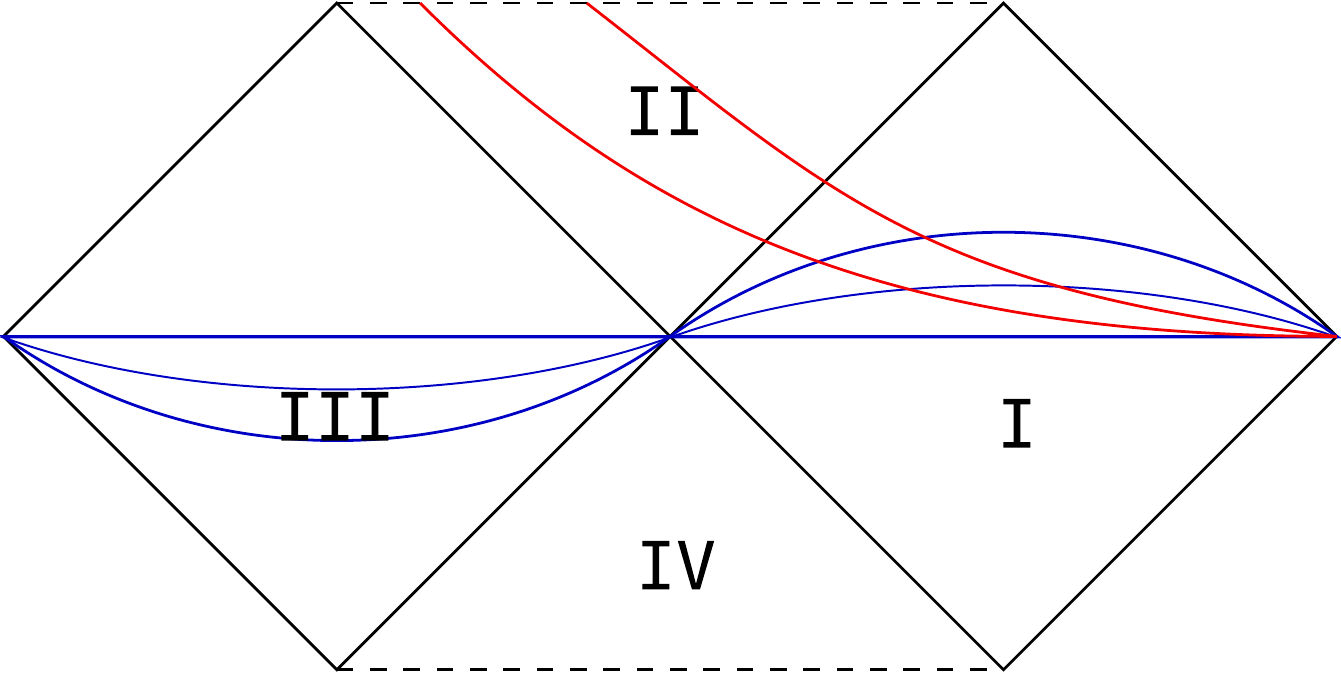} \\
  \caption{
    Conformal Penrose diagram of an axisymmetric spacetime, consisting
    of a non-rotating BH, distorted by a massive stationary disk around
    it.  Each point on the diagram corresponds to a spheroid, located at
    a given geodesic distance from the BH horizon.  Blue lines represent
    a quasi-isotropic foliation of the spacetime, while red lines show a
    horizon-penetrating foliation.
  }
  \label{F:ConformalPicture}
  \end{center}
  \end{figure}

We set up initial equilibrium disk configurations by solving Einstein
constraints using a version of the \rnsid{} solver~\cite{StF95:RNSModels}
adapted to the problem of equilibrium tori.
The method of solution is similar to the one used in~\cite{NishidaEriguchi94}.
The spacetime is assumed to be stationary, axisymmetric, asymptotically flat
and symmetric with respect to reflections in equatorial plane.
The metric is a general axisymmetric metric in quasi-isotropic coordinates:
\begin{align}
\label{Eq:AxisymMetric}
  ds^2 &= -\lambda^2 dt^2
        + e^{2\alpha} (dr_*^2 + r_*^2 d\theta^2)\\
       &+ B^2/\lambda^2 r_*^2 \sin^2\theta (d\ji - \omega dt)^2.
\end{align}

\noindent where $(t,r_*,\theta,\ji)$ are the coordinates, and
$\lambda$, $\alpha$, $B$ and $\omega$ are metric potentials which depend only
on $r_*$ and $\theta$.
The \rnsid{} code implements the KEH(SF) method~\cite{KEH89,StF95:RNSModels},
in which the Einstein equations for the metric potentials $\lambda$, $B$
and $\omega$ are transformed into integral equations~\cite{NishidaEriguchi94}
using Green's functions for the elliptical differential operators.
The remaining metric potential $\alpha$ can then be found by integrating an
ordinary differential equation, once the rest of the potentials are
known.
The KEH(SF) method uses compactified radial coordinate $s$ which maps
the region $[0,\infty)$ into a segment $[0,1]$:
$$
  s \equiv \frac{r}{r+r_+}
$$
\noindent where $r_+$ is the outer radius of the disk.
The boundary conditions are imposed at symmetry interfaces and
at the event horizon.
The latter can always be transformed to a sphere of compactified radius $s_h$,
while preserving the form of the metric given above~\cite{NishidaEriguchi94}.
At the horizon, we impose the boundary conditions for a non-rotating
BH: $\lambda=B=\omega=0$, and we set $B/\lambda=e^{\alpha}$ at the
symmetry axis.
The corresponding integral equations are then solved using Newton-Raphson
iterations~\cite{StF95:RNSModels} in the upper quadrant $\theta\in [0,\pi/2]$,
$s\in [0,1]$ of the meridional plane.

The resulting quasi-isotropic metric is degenerate at the event horizon, which
is very problematic for the evolution with excision of the BH interior using
the generalized harmonic formulation, since this method requires coordinates
without pathologies at the event horizon.
This situation is best illustrated by a conformal picture of the complete
spacetime, shown in Fig.~\ref{F:ConformalPicture}.
All quasi-isotropic slices meet the horizon of the BH at its throat, which
makes the metric degenerate at the horizon. 
Regions II and IV are not covered by the quasi-isotropic foliation.
It is necessary for our time evolution methods to have a
time-independent foliation which penetrates the horizon and continues in
region II rather than region III.

There are several options to address this issue:
\begin{enumerate}
\item[(1)] solve the complete system of equations in horizon-penetrating
	   coordinates rather than quasi-isotropic ones;
\item[(2)] use puncture initial data, as developed in~\cite{Shibata07c}, and       
	   choose such a gauge for the evolution that after some time the
	   spatial slices move from region III to region II, similarly to
	   what happens with punctures in the BSSN formulation and $1+\log$
	   slicing~\cite{brown-2007};
\item[(3)] apply a spacetime coordinate transformation from quasi-isotropic
	   to horizon-penetrating coordinates. Since there is still no
	   solution provided in region II, it needs to be extrapolated
	   into that region.
\end{enumerate}

We use option (3), since it is easier to implement in the context of
the generalized harmonic system, and fits more naturally in the gauge
choice employed by \rnsid{}.
Details on the spacetime transformation that we apply to the initial data can
be found in Appendix~\ref{A:Transform}.

\subsection{Blending numerical and analytical metrics}
\label{SS:Blending}

We could not extrapolate the initial data produced by the elliptic solver to
the region II inside the horizon, because the data does not have enough
smoothness near the horizon.
The problem with initial data at the horizon seems to be similar to the
problem with Gibbs phenomena, when solutions exhibit an oscillatory behavior
and a lower order of convergence near a stellar surface.
To handle this problem, we use an approximation, in which we replace the
numerical metric in the region where it is not accurate with an analytic
Kerr-Schild solution of the same BH mass.
We blend the numerical metric $g^{num}_{ab}$ and the Kerr-Schild metric
$g^{KS}_{ab}$ using the following prescription:
$$
  g_{ab} = (1-w(r)) g^{num}_{ab} + w(r) g^{KS}_{ab}
$$
where the weight function $w(r)$ is defined as:
\begin{align*}
w(r) = 
\begin{cases}
1, & \textrm{if\ }r<b_1,   \\
\cos^2{\frac{\pi(r-b_1)}{2(b_2-b_1)}}, & \textrm{if\ }r\in[b_1,b_2],   \\
0, & \textrm{if\ }r>b_2,   \\
\end{cases}
\end{align*}
\noindent and the segment $r\in[b_1,b_2]$ determines a finite-size blending zone
between the two metrics. 
The weight function $w(r)$ is non-constant only in a narrow spherical layer
outside the horizon.

Blending two metrics introduces constraint violations at the continuum level,
but they subside rapidly in time due to the constraint damping
property of our evolution scheme. 
The constraint damping scheme, however, does not necessarily satisfy the
conservation of mass, so after the constraint violations are
suppressed, the system arrives at a different state, which can be
characterized as a close equilibrium configuration with some
axisymmetric gravitational perturbation. 

The location and size of the blending layer can be adjusted to minimize the
initial unphysical oscillation in a BH mass. For the evolutions presented
below, we used blending in the range $r\in[1.05\ r_g, 1.15\ r_g]$, where
$r_g$ is the radius of the BH event horizon. 
Such a choice results in the initial oscillation of the BH mass within
$\approx12\%$, which then settles down to a stationary value of 
$97.5\pm0.5\%$ during the first orbital period (see 
Fig.~\ref{F:DiskDynamics}b and related discussion in Section~\ref{S:TimeEv}).

\subsection{Initial disk models}
\label{SS:DiskModels}

  \begin{figure}[!htp]
  \begin{center}
  \begin{tabular}{c}
  \includegraphics[width=0.4\textwidth]{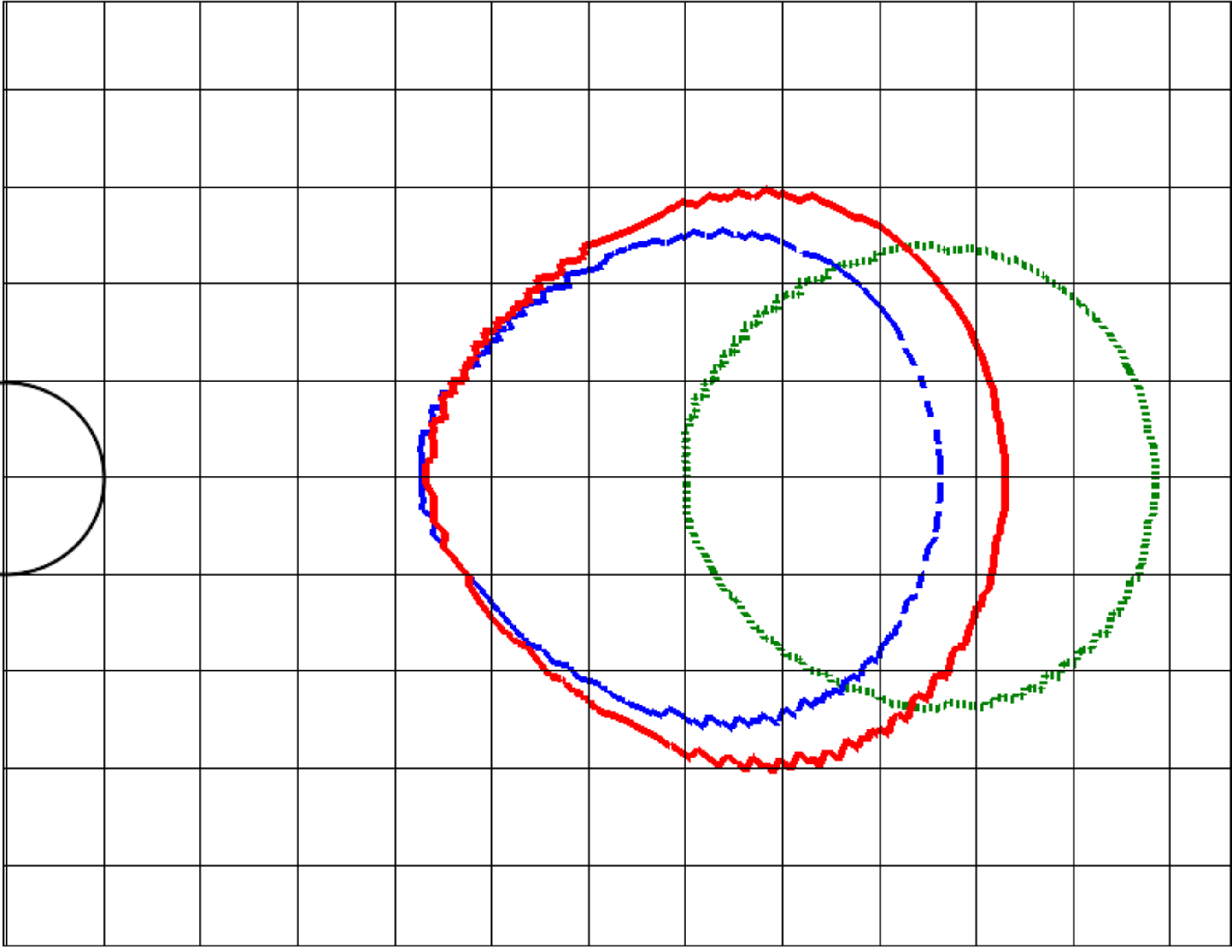} \\
  \end{tabular}
  \caption{
   Contours of the disk surfaces for models A, B and C, and the
   location of BH horizon in the meridional plane.
  }
  \label{F:DiskModels}
  \end{center}
  \end{figure}
  \begin{table}
  \begin{center}
  {\footnotesize
  \begin{tabular}{l|c|c|c}
  \hline
  Model                                                      &   A   &   B   &   C   \\
  \hline
  Specific angular momentum $\ell \ [M_{BH}]$                & 4.50  & 4.32  & 4.87  \\
  Polytropic constant $K \ [10^{14}\cm^3\gr^{-1/3}\sec^{-2}]$& 1.28  & 1.04  & 0.519 \\
  Maximum density $\rho_c \ [10^{13}\gr/\cm^3]$              & 1.23  & 1.17  & 0.755 \\
  Disk-to-BH mass ratio $M_D/M_{BH}$                         & 0.235 & 0.174 & 0.108 \\
  Kinetic to potential energy $T/|W|$                        & 0.479 & 0.497 & 0.499 \\
  Central radius $r_c/r_g $                                  & 6.51  & 5.58  & 8.23  \\
  Ratio of inner to central radius $r_-/r_c$                 & 0.655 & 0.655 & 0.756 \\
  Ratio of inner to outer radius $r_-/r_+$                   & 0.385 & 0.407 & 0.600 \\
  Orbital frequency at $r_c$, $\Omega_c \ [\sec^{-1}]$       & 1713  & 1912  & 1121  \\
  Self-gravity parameter $\tau=4\pi G\rho/\Omega_c^2$        & 3.52  & 2.68  & 5.04  \\
  \hline
  \end{tabular}
  }
  \end{center}
  \caption{Physical parameters of the self-gravitating initial disk models,
  used in our simulations.
  }
  \label{T:SGTParams}
  \end{table}

We have constructed three initial disk models with disk-to-BH mass ratios
$M_D/M_{BH}=0.235$, $0.174$ and $0.108$, labeled A, B and C.
Model C is slender, which means that its width is much smaller than
the radius of the torus, i.e.\ $r_c\gg (r_+-r_-)$.
Models A and B are moderately slender, i.e.\ for these models 
$r_c\approx(r_+-r_-)$. 
Figure~\ref{F:DiskModels} shows the contours of the disk surfaces for our
models and the location of BH horizon in the meridional plane.
All models are constructed using a polytropic EOS with polytropic
index $\Gamma=4/3$, constant specific entropy, and a constant specific angular
momentum distribution.

Table~\ref{T:SGTParams} lists physical and geometrical parameters of the disk,
including the ratio of disk-to-BH mass $M_D/M_{BH}$, the ratio of kinetic
energy $T$ to potential energy $W$ and the self-gravity parameter $\tau$.
The latter can be defined as $\rho_c/\rho_{sph}$, where
$\rho_{sph}\equiv\Omega_c^2/4\pi G$ is the density of a uniform sphere with
radius $r_c$ that creates equivalent gravity at that radius.
Notice that $T/|W|$ correlates to the ``slenderness'' of the disk $r_-/r_+$.
The value of specific angular momentum $\ell$ is given in units of the BH
mass.
These parameters allow us to make qualitative and quantitative comparisons
between our models and the models studied in previous works~\cite{WTH94,
Christodoulou93, GoodmanNarayan88a, ChristodoulouNarayan92, Andalib97}.

\subsection{Adapted curvilinear grid}
\label{S:AdMesh}
   
\begin{figure*}[!htp]
  \begin{center}
  \begin{tabular}{cc}
  \includegraphics[width=0.42\textwidth]{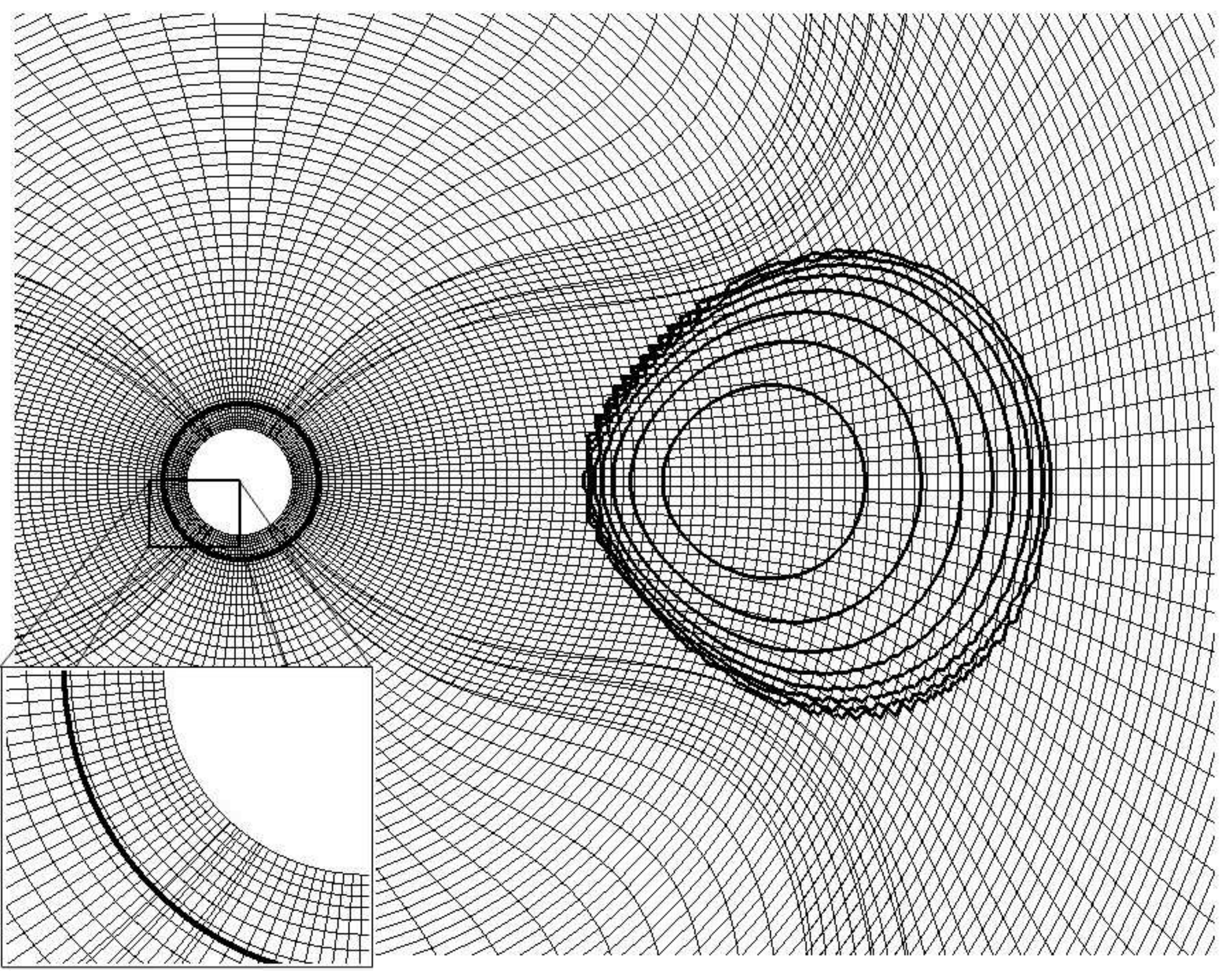} &
  \includegraphics[width=0.48\textwidth]{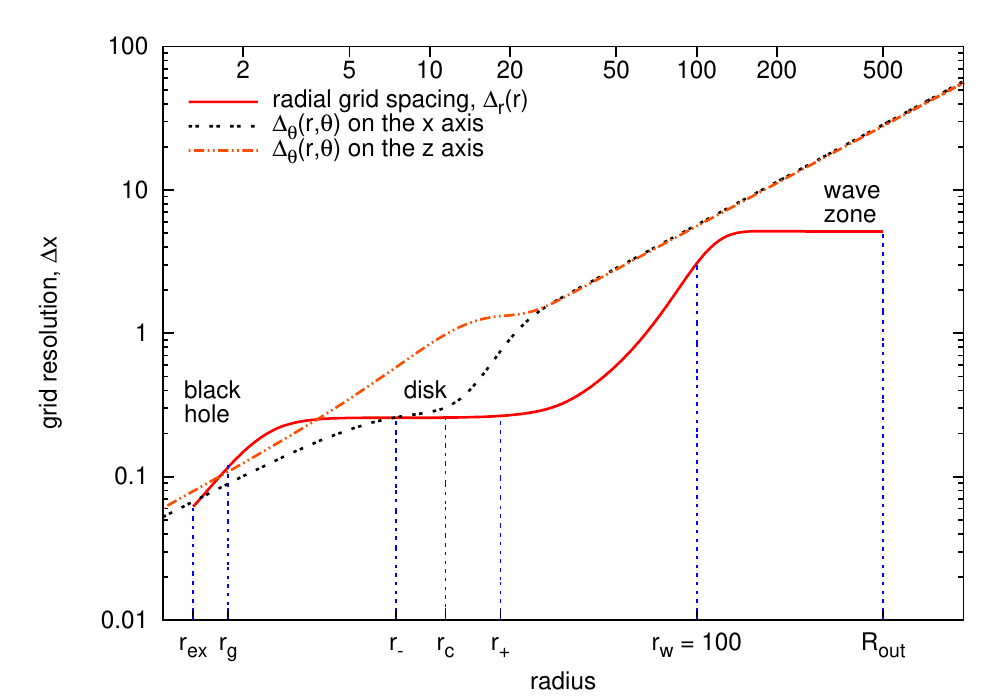} \\
  (a) & (b)
  \end{tabular}
  \caption{
    The left panel (a): meridional cut of an adapted curvilinear grid used
    in one of our simulations, combined with the logarithmic density contours
    of the disk at $t=0$.
    The intersecting radial coordinate lines belong to the neighboring
    blocks in the overlapping (interpolated) regions.
    The thick black circle marks a location of an apparent horizon of
    the BH.
    The inset in the lower left corner of the plot shows the
    high-resolution grid around the BH, adapted in such a way that the
    resolution is uniformly high in every direction.
    The right panel (b): radial profile of the resolution in the $r$ (red
    solid line) and the $\theta$ (dotted lines) directions for the
    grid shown in the top panel. 
    The radial coordinate is plotted in logarithmic scale.
    The top horizontal axis shows the values of the radius in numerical
    units.
    The marks on the lower axis are:
    $r_{ex}$   is the radius of the BH excision sphere;
    $r_g$      is the gravitational radius of the BH;
    $r_-, r_+$ is the inner and outer radii of the disk;
    $r_c$      is the radius of the density maximum;
    $r_W$      is the ``wave extraction'' radius;
    $R_{out}$  is the outer radius of the domain.
    The black dotted line is the resolution in the $\theta$-direction on the $x$
    axis, and the red dashed-and-dotted line is the resolution in the
    $\theta$-direction on the $z$ axis.
   }
   \label{F:Mesh}
   \end{center}
   \end{figure*}

In order to accurately resolve both the disk and the BH while minimizing the
computational cost, we have designed a series of curvilinear multiblock grids
adapted to each of the disk models.
To obtain such a grid, we start from a six-block system (displayed on
Fig.~\ref{F:MyPatchsystems}c) that was previously used in \Quilt{} for
computationally efficient and numerically accurate simulations of perturbed
BHs~\cite{Pazos07}.
We apply a radial stretching and an angular distortion to the six-block system
so as to create a uniformly high resolution near the BH, nearly cylindrical
grid near the disk, and approach a regular six-block spherical grid in the
wave zone.
These mappings are described in detail below.
Figure~\ref{F:MyPatchsystems}d shows an illustration of the grid distortion,
while Fig.~\ref{F:Mesh}a shows the actual curvilinear grid used in some of our
simulations.

A regular six-block system consists of two polar blocks (near the $z$-axis)
and four equatorial blocks (near the $xy$-plane).
We can assign quasi-spherical coordinates $\{r,\theta,\ji\}$ and
$\{r,\theta_1,\theta_2\}$ to the equatorial and polar blocks, respectively.
They can be related to the Cartesian coordinates $\{x,y,z\}$ by the
following transformation:
\begin{enumerate}
\item[(a)] for an equatorial block in the neighborhood of the positive
  $x$ axis:
$$
\begin{array}{ll}
x &= r/\sqrt{1+\tan{\ji}^2+\tan{\theta}^2}, \\
y &= x \tan{\ji}, \\
z &= x \tan{\theta},
\end{array}
$$

\item[(b)] while for a polar block in the neighborhood of the positive
  $z$ axis:
$$
\begin{array}{ll}
x &= z \tan{\theta_1}, \\
y &= z \tan{\theta_2}, \\
z &= r/\sqrt{1+\tan{\theta_1}^2+\tan{\theta_2}^2} .
\end{array}
$$
\item[(c)] The remaining blocks are obtained by applying symmetry
transformations.
\end{enumerate}

We set the coordinate ranges for the polar blocks to be $r\in[R_{min},R_{max}]$,
$\theta_1, \theta_2\in[-\theta_*,\theta_*]$, where the value of $\theta_*$
controls an opening angle of the polar blocks.
For equatorial blocks, $r\in[R_{min},R_{max}]$,
$\theta\in\left[-\frac{\pi}{2}+\theta_*,\frac{\pi}{2}-\theta_*\right]$
and $\ji\in\left[-\frac{\pi}{4},\frac{\pi}{4}\right]$.
The size of the numerical grid for each block is fixed by three numbers:
$N_r$, $N_{\ji}$ and $N_{\theta}$. The equatorial blocks have $N_r\times
N_{\ji}\times N_{\theta}$ cells, and the polar ones have 
$N_r\times N_{\theta}\times N_{\theta}$ cells.

In order to obtain a variable radial resolution,
we apply a smooth one-dimensional radial stretching 
$S: r\to\bar{r}$ that yields the desired resolution profile $\Delta_r(r)$.
This profile is chosen based on several stringent requirements imposed by an
accuracy and available computational resources.
First, there need to be at least $8$ grid points between the excision radius and the
BH horizon in order to prevent constraint violations from leaving the interior
of the BH (this question is considered in some detail in~\cite{Brown:2008sb}),
as well as to allow for some (restricted) BH movement.  
Second, our convergence tests show that near the disk, the grid needs to allow
for at least $40$ points across the disk in order to achieve a global
convergent regime in hydrodynamical evolutions.
For this purpose, the radial resolution profile is adapted to have
sufficiently high resolution near the disk as well.
Third, in the wave zone, there is no need to maintain very high radial
resolution.
However, this resolution needs to be uniform (rather than, for example,
exponentially decreasing) to be able to carry the radiation accurately without
dissipation. 
All these requirements result in the radial resolution profile
$\Delta_r(r)$ shown on Fig.~\ref{F:Mesh}b.

The resolution in the $\theta$ direction $\Delta_{\theta}$ near the disk also
needs to have at least $40$ points across the disk for convergence.
However, simple increase of $N_{\theta}$ in the six-block system leads to
a very small minimal grid step $\Delta_{min}$ at the excision radius
$R_{min}$ that is too restrictive for the time step due to CFL condition.
To increase the grid resolution in the $\theta$-direction near the
disk only, we apply a radial-dependent distortion to the angular coordinates
$\theta$, $\theta_1$ and $\theta_2$:
$$
\begin{array}{lll}
\theta(\bar{\theta},r) &= \bar{\theta}
                          (1 - \beta(r)
                               \frac{\cos{2\theta_*}}{\pi-2\theta_*}), \\
\theta_i(\bar{\theta}_i,r) &= \bar{\theta_i} 
                              (1 + \beta(r)
                                   \frac{\sin{2\theta_*}}{2\theta_*}), 
                           &  \textrm{i = 1,2}
\end{array}
$$

\noindent where the function $\beta(r)$ is the amplitude of the distortion,
chosen to have a Gaussian profile $\beta(r) =
\beta_*\exp(-(r-r_0)^2/\sigma^2)$,
in which the parameters $\beta_*$, $r_0$ and $\sigma$ are chosen to satisfy
the above requirements.
This distortion bends diverging radial coordinate lines towards equatorial
plane around the radius $r_0$, making the grid resemble a cylindrical
shape near that radius (as illustrated by Fig.~\ref{F:Mesh}a).
Figure~\ref{F:Mesh}b shows the dependence of $\Delta_{\theta}$ on $r$ along
the $x$-axis (dotted black line) and along the $z$-axis (dash-dotted red line).
The radial-dependent $\theta$-distortion increases $\Delta_{\theta}$ on the
$x$-axis at the expense of $\Delta_{\theta}$ on the $z$-axis near the disk.
Away from the disk, the distortion vanishes and $\Delta_{\theta}$ approaches
the linear dependence $\Delta_{\theta}(r)\propto r$. 

Figure~\ref{F:Mesh} shows an example of the resulting curvilinear grid.
For this example, $\theta_*=30^o$, $N_{\theta}=49$ and $N_{\ji}=25$.
These parameters make the grid spacing approximately uniform in angular direction
near the BH and at large $r$. 
The inner (excision) radius is $R_{min}=1.3$ and the outer one is
$R_{max}=500$, while
the minimum grid spacing is $\Delta_{min}=\Delta_{\theta}(r_{min})\approx0.06$. 
The apparent horizon has radius of $\approx1.7$, so that as many as 8 grid points
can be placed inside the horizon without decreasing the minimum grid spacing. 
Distortion parameters of the Gaussian have values $r_0=12$,
$\sigma=8$, and $\beta_*=1.3$, which is sufficient to concentrate
about 40 grid points across the disk in the vertical direction.
The radial resolution is adapted to be $(\Delta_r)_{BH}\approx0.06$ at the excision
sphere, $(\Delta_r)_{disk}\approx0.25$ around the disk, and
$(\Delta_r)_{WZ}\approx5.2$ in the wave zone.

All of the time evolutions described in the next two
Sections~\ref{S:TimeEv} and~\ref{S:Nonaxisym} use adapted six-block
curvilinear grids. 
Table~\ref{T:Models} summarizes dimensions and resolutions of the grids
for each of the simulations used in these two sections.
The first column lists simulation names, which consist of two or
three symbols.  
Models K1-K6 are considered in Section~\ref{SS:CompKojima}.  
In the rest of the model names, the first letter denotes the initial disk
model (A, B or C), the second letter signifies whether the simulation is
evolved in full GR (F) or Cowling (C) approximation (see corresponding
Sections~\ref{SS:DynamicalBG} and~\ref{SS:FixedBG}).
The next number (if present) is the azimuthal number $m$ of an added
non-axisymmetric perturbation (as explained in Section~\ref{S:Nonaxisym}). 
Simulations AFc, AF and AFf represent disk model A evolved in full GR using
coarse, medium and fine resolution grids.  
These models are used in Section~\ref{S:TimeEv} for convergence studies.
Finally, note that the adapted curvilinear grid example
above (displayed in Fig.~\ref{F:Mesh}) corresponds to the the grid used in
simulation AF, and the grids of simulations AFf and AFc are obtained by
increasing and decreasing the resolutions in the AF grid by a factor
of $3/2$.

  \begin{table*}
  \begin{center}
  {\small
  \begin{tabular}{l|ccc|c|c|cccc}
  \thickhline
   Model & 
   $r_g$ &
   $R_{min}$ &
   $R_{max}$ &
   $N_x\times N_z\times N_r$  &  
   $\Delta_{min}$ &
   $(\Delta_r)_{disk}$ &
   $(\Delta_{\theta})_{r_c,x}$ &
   $(\Delta_{\theta})_{r_c,z}$ &
   $(\Delta_r)_{WZ}$ \\
   \hline 
   K1                 & $2.0$    & $4.0$  & $30$ & $25\times49\times96$   & 0.07 & 0.25 & 0.32 & 0.72 & --  \\  
   K2                 & $2.0$    & $4.0$  & $25$ & $25\times49\times96$   & 0.07 & 0.25 & 0.32 & 0.72 & --  \\  
   K3                 & $2.0$    & $4.0$  & $22$ & $25\times49\times96$   & 0.09 & 0.20 & 0.17 & 1.28 & --  \\  
   K4                 & $2.0$    & $6.0$  & $17$ & $25\times49\times96$   & 0.08 & 0.18 & 0.13 & 1.68 & --  \\  
   K5                 & $2.0$    & $7.0$  & $16$ & $25\times49\times96$   & 0.03 & 0.10 & 0.05 & 0.50 & --  \\  
   K6                 & $2.0$    & $8.0$  & $14$ & $25\times49\times96$   & 0.02 & 0.08 & 0.03 & 0.53 & --  \\  
   \hline                        
   AC, AC1, AC2       & $1.760$  & $1.7$ & $25$  & $37\times73\times144$  & 0.05 & 0.17 & 0.21 & 0.48 & --  \\
   BC, BC1, BC2       & $1.806$  & $1.7$ & $25$  & $37\times73\times144$  & 0.05 & 0.17 & 0.21 & 0.48 & --  \\
   CC, CC1, CC2, CC3  & $1.812$  & $1.7$ & $25$  & $37\times73\times144$  & 0.05 & 0.17 & 0.21 & 0.48 & --  \\
   \hline                        
   AF, AF1, AF2       & $1.760$  & $1.3$ & $500$  & $25\times49\times280$ & 0.06 & 0.25 & 0.22 & 0.95 & 5.2 \\
   AFc                & $1.760$  & $1.3$ & $1000$ & $17\times33\times180$ & 0.09 & 0.37 & 0.33 & 1.43 & 7.8 \\
   AFf                & $1.760$  & $1.3$ & $1000$ & $37\times73\times420$ & 0.04 & 0.17 & 0.15 & 0.62 & 3.4 \\
   BF, BF1, BF2       & $1.806$  & $1.3$ & $1000$ & $25\times49\times280$ & 0.06 & 0.25 & 0.22 & 0.95 & 5.2 \\
   CF, CF1, CF2, CF3  & $1.812$  & $1.3$ & $1000$ & $25\times49\times280$ & 0.06 & 0.22 & 0.22 & 1.01 & 5.2 \\
  \thickhline

  \end{tabular}}
  \end{center}
  \caption{Parameters of numerical grids for all the simulations of accretion
	   disks used in this study. 
	   The naming convention of the simulations is explained in the main
	   text in Section~\ref{S:AdMesh} on adapted curvilinear grids.
	   All resolutions and linear dimensions are given in computational
           units.
           The first column, which lists the values of the BH
	   gravitational radius $r_g$, allows to convert all quantities from
           computational to CGS units.
	   The remaining columns contain:
	   $R_{min},R_{max}$ are radial extents of the computational domain;
	   $N_x$ is the number of grid points in the horizontal $x$- or $y$-direction;
	   $N_z$ is the number of grid points in the vertical direction;
	   $N_r$ is the number of grid points in the radial direction;
	   $\Delta_{min}$ is the minimal grid step size, in $\theta$ direction at the excision sphere;
	   $(\Delta_r)_{disk}$ is the resolution in the $r$-direction at radius $r_c$;
	   $(\Delta_{\theta})_{r_c,x}$ is the resolution in the $\theta$-direction
	   on the $x$-axis at radius $r_c$;
	   $(\Delta_{\theta})_{r_c,z}$ is the resolution in the $\theta$-direction
	   on the $z$-axis at radius $r_c$;
	   $(\Delta_r)_{WZ}$ is the radial resolution in the wave zone (not used
	   for simulations on fixed background).}
  \label{T:Models}
  \end{table*}

\subsection{Data analysis}
\label{SS:DataAnalysis}

To identify and characterize non-axisymmetric instabilities, we will
adopt an approach commonly used in linear perturbative studies of accretion
disks (e.g. in~\cite{Kojima86b,WTH94}). Namely, we analyze first few terms in
Fourier expansion in angle $\ji$ of the disk density $\rho(r,\ji)$ on a
sequence of concentric circles in the equatorial plane:
$$
\rho(t,r,\ji) = \bar{\rho}(t,r) \left(1 + \sum_{m=1}^{\infty} D_m
e^{-i(\omega_m t - m\ji)} \right).
$$
where $\bar{\rho}$ is a $\ji$-averaged density at a given radius.
The quantity $D_m$ represents the (complex) amplitude of an azimuthal
mode $m$, the real part of the quantity $\omega_m$ determines a mode pattern
speed, while its imaginary part determines a mode growth rate.
Following~\cite{WTH94, WilliamsTohlineAPJ87}, we quantify the growth rate and
the pattern speed of a non-axisymmetric mode by two dimensionless parameters
$y_1$ and $y_2$, defined as
$$
y_1(m) = \frac{\re(\omega_m)}{\Omega_{orb}} - m, \qquad
y_2(m) = \frac{\im(\omega_m)}{\Omega_{orb}}.
$$
We calculate a value of the parameter $y_2(m)$ from a slope of
$\log{|D_m|}$ versus $t$ line at an arbitrary radius, while $y_1(m)$ is
obtained from a slope of the mode phase angle $\ji_m=\ji_m(t)$.
Notice that because the modes that we consider are global, their growth rates
and pattern speeds do not depend on a radius. 

We use the parameter $y_1(m)$ to obtain the value of a corotation radius
$r_{cr}$ for a given mode: using the mode pattern speed,
$$
\Omega_p = \Omega_{orb}\left(1+\frac{y_1(m)}{m}\right),
$$
and the radial profile $\Omega=\Omega(r)$ of a fluid angular velocity in the
disk, we can calculate $r_{cr}$ from the equation $\Omega(r=r_{cr})=\Omega_p$.

In those simulations where the disk is oscillating radially, the values of the
mode amplitude $D_m$ at a given radius oscillate due to disk oscillations,
which makes it more difficult to extract the mode growth rates from
$D_m$.
In such cases, we have found that more accurate growth rates can be obtained
if we use normalized root mean squared (RMS) mode amplitudes $G_m$,
defined as: 
$$
   G_m = \left<D_m\right>_2 / \left<D_0\right>_2
$$
\noindent where the angle brackets $\left<\dots\right>_2$ denote an RMS value
over radii from $r_-$ to $r_+$:
$$
   \left<D_m\right>_2 \equiv \frac{1}{r_{{}_+}-r_{{}_-}}
			     \left(\int_{r_-}^{r_+}|D_m|^2 dr\right)^{1/2}
$$

\section{Time evolution}
\label{S:TimeEv}

In this section, we present the results of the fully general
relativistic time evolution of the initial data for the reference model A,
constructed as described above in Section~\ref{S:IDSetup}. 
Overall, the dynamics of model B is qualitatively similar to that of
A, while model C exhibits a qualitatively different time
evolution.
The BH initial mass in all of our models is $2.5\ \MSun$, as in some
of the previous works~\cite{FontDaigne02, Zanotti03a, DaigneFont04}, and
the disk rotational period $t_c=3.667\ \ms$ is used as a unit of time
in all of the plots in this section.

At the beginning of time evolution, the metric blending procedure (described
earlier in Section~\ref{SS:Blending}) introduces axisymmetric constraint-violating
perturbation to the spacetime near the BH, causing an unphysical oscillation
in the BH mass, which damps out in about one orbital period of the disk. 
We discard the first orbital period when analyzing simulation data and drawing
physical conclusions about the system dynamics.

In the meanwhile, the axisymmetric perturbation propagates outwards and
triggers axisymmetric disk oscillations.
Disk oscillations lead to formation of shock waves, which transform
kinetic energy of the shock to thermal energy, resulting in damping of the
oscillations. We discuss different aspects of the dynamics of the disk in more
detail below.

\begin{figure}[!htp]
  \begin{center}
  \begin{tabular}{c}
  \includegraphics[width=0.45\textwidth]{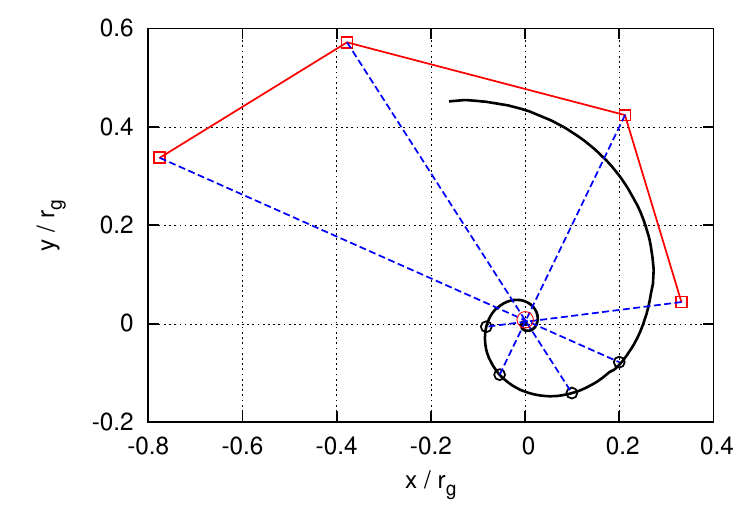}
  \end{tabular}
  \caption{Trajectories of the center of mass of the accretion disk (red
    line) and the BH (black line). Dashed lines show four consecutive
    simultaneous locations of the two centers of mass, and a small red
    circle at the origin marks the location of their common center of
    mass.
    The spiral motion of the BH is caused by the development of the
    non-axisymmetric $m=1$ mode in the disk.
  }  
  \label{F:GRInst_a}
  \end{center}
  \end{figure}

After about three orbital periods, the disk develops an $m=1$
non-axisymmetric mode, which we have identified as Papaloizou-Pringle
(PP) type instability \cite{PPI, PPII}
(discussed in more detail below in Section~\ref{S:Nonaxisym}). 
The same type of mode develops in model B, while the more slender model C
develops an $m=2$ mode of an intermediate type (see
Section~\ref{SS:DynamicalBG}).
As the $m=1$ mode grows, the center of mass of the disk drifts away from its
initial position along a spiral-like trajectory, as shown in
Fig.~\ref{F:GRInst_a} (red squares).
As a result of gravitational interaction between the deformed
disk and the BH, the latter also starts spiraling away, mirroring the
motion of the disk center of mass, as shown in Fig.~\ref{F:GRInst_a}
(black line). This plot also shows dashed lines that connect the
positions of the center of mass of the BH and the disk at different
moments of time. As we can see, all of the these lines intersect with
each other at one point at the initial location of the center of mass
of the disk-BH system, implying that the center of mass of the system
does not move, as should be the case for BH motion caused by physical
interaction with the disk (but not gauge effects).   

\begin{figure}[!htp]
  \begin{center}
  \begin{tabular}{c}
  \includegraphics[width=0.45\textwidth]{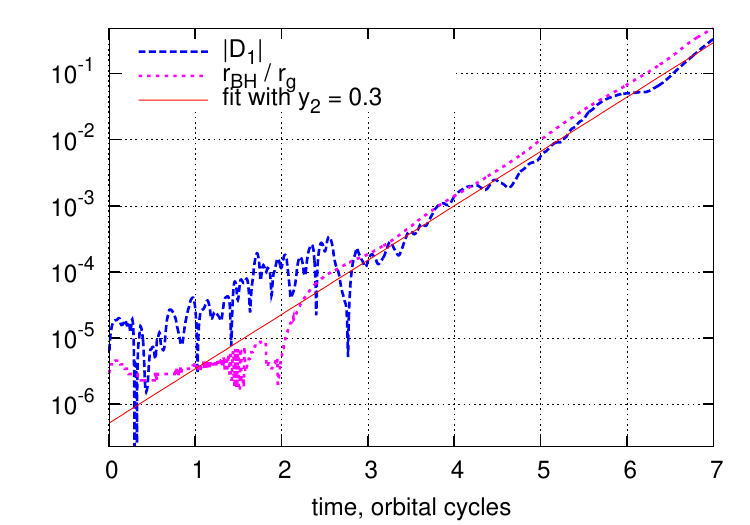}
  \end{tabular}
  \caption{Time evolution of the amplitude $|D_1|$ of the
    non-axisymmetric $m=1$ mode in the disk and the length of the BH
    position vector $ r_{BH} / r_{g} $. 
    The growth of these quantities is correlated, i.e.\ they develop at
    the same time and with the same rate.}
  \label{F:GRInst_b}
  \end{center}
  \end{figure}

Figure~\ref{F:GRInst_b} shows the time evolution of the amplitude
$|D_1|$ of the $ m=1 $ PP non-axisymmetric mode and the distance $
r_\mathrm{BH} $ (normalized to $r_\mathrm{g}$) from the BH center to
its initial position. As we can see, both $|D_1|$ and $ r_\mathrm{BH}
$ have the same growth rate. This feature provides another evidence
that the BH motion is a result of the physical interaction with the
$m=1$ disk deformation, but not due to gauge effects. 

  \begin{figure}[!htp]
  \begin{center}
  \includegraphics[width=0.45\textwidth]{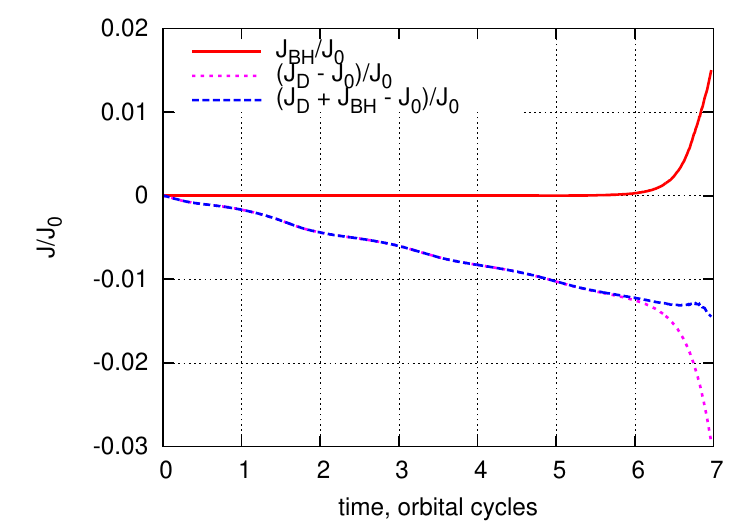}
  \caption{
    Illustration of the angular momentum transfer from the disk to the
    BH.
    The solid (red) line shows the orbital angular momentum of the BH
    $J_{BH}$, the dotted (magenta) line shows the decrease in the
    angular momentum of the disk $J_D-J_0$, and the dashed (blue) line
    shows the total decrease of the angular momentum of disk+BH 
    $(J_D+J_{BH})-J_0$.
    All quantities are divided by the initial value of angular momentum
    of the disk $J_0$.
    The noticeable angular momentum transfer can only be seen in the
    last orbital period, when the $m=1$ distortion reaches significant
    amplitude.
    The gradual decrease in total angular momentum is due to the mass
    loss at the interpolation boundaries. 
    This is a numerical artifact which converges away with resolution.
  }
  \label{F:JBH}
  \end{center}
  \end{figure}

We point out that, as a result of the interaction of the $m=1$
deformed disk with the BH, the latter acquires significant orbital
angular momentum from the disk. Fig.~\ref{F:JBH} illustrates how
angular momentum of the disk ($J_\mathrm{D}$), BH
($J_\mathrm{BH}$)\footnote{We calculate the total angular momentum of
  the disk using expression $ J_D = \int d^3x \sqrt{-g}
  T^t_{\ji}$ \cite{Misner73}, while the orbital angular momentum
  of the BH is calculated using a simple Newtonian estimate: using the
  BH speed $r\dot{\ji}$ and its distance from the origin $r$, we
  get $ J_{BH} \simeq M_{BH} r^2 \dot{\ji}$.} and BH+disk
system ($J_\mathrm{BH} + J_\mathrm{D}$) changes with time with respect
to initial disk angular momentum ($J_\mathrm{0}$). The total angular
momentum of the disk+BH system decreases by $ \sim 1.5 \% $ 
in $ \sim 7 $ orbital periods due to numerical
errors such as interpolation at the block boundaries and evaporation
to the artificial atmosphere in the case of medium resolution.
As a result of angular momentum transfer,
$J_\mathrm{D} / J_\mathrm{0} $ additionally decreases by $ \sim 1.5
\%$, which is completely compensated by the $ \sim 1.5 \% $ increase
of $J_\mathrm{BH} / J_\mathrm{0} $.    

Unfortunately, the continued outspiraling motion of the BH ultimately
leads to the intersection of the apparent horizon with the excision
boundary. At this point, the inner excision boundary conditions become
ill-posed, and we have to terminate our simulations. 

Similarly to the findings of~\cite{Montero10,Rezzolla10b}, we did
not observe runaway instability in all three models.
We also do not expect this instability to occur at a later time. 
In the most likely scenario of the subsequent evolution, the development of
non-axisymmetric instabilities will redistribute disk angular momentum and
lead to a profile of specific angular momentum that increases
outwards~\cite{Zurek86a,Christodoulou93}.
Such angular momentum profile was shown to strongly disfavor runaway
instability~\cite{Daigne97a}. 
Moreover, the damping of radial disk oscillations in models A and B reduces
and eventually completely terminates mass transfer from the disk to the BH,
preventing onset of runaway instability.

  \begin{figure}[!htp]
  \begin{center}
  \begin{tabular}{c}
  \includegraphics[width=0.45\textwidth]{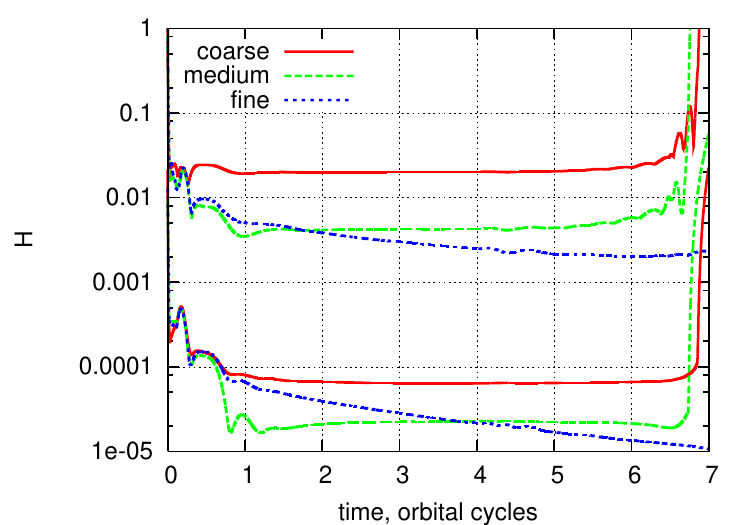} \\
  (a) \\
  \includegraphics[width=0.45\textwidth]{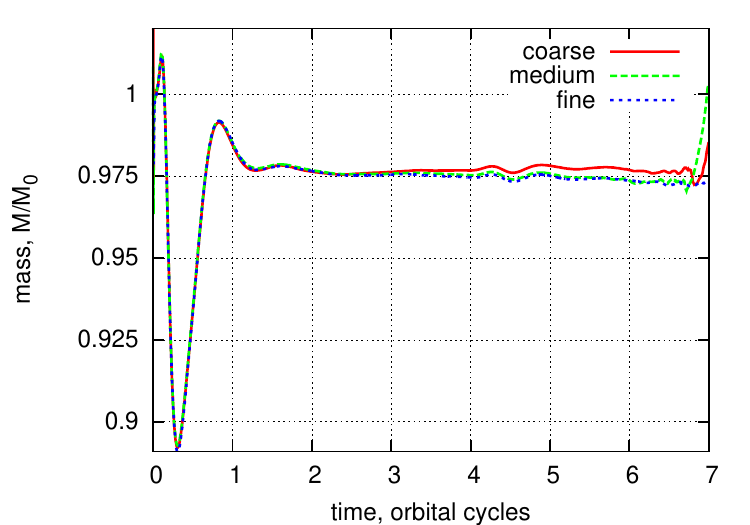} \\
  (b)
  \end{tabular}
  \caption{
    Time evolution of the constraints and the BH mass for simulations
    with the coarse, medium and fine resolution grids.
    (a) $L_\infty$- and $L_1$-norms of the Hamiltonian constraints as
    a function of time.
    For the coarse and medium resolution cases, the plot shows that
    the constraints are reduced down to the discretization error level
    during the first orbital period (see the main text for more
    detailed discussion). 
    (b) Time evolution of the BH mass (measured from the area of
    apparent horizon, normalized by its initial value).
    After the initial transitional oscillatory phase, the BH mass
    settles down to $97.5\pm0.5\%$ of its initial value. 
    The oscillation in the first orbital period is caused by the
    constraint violations due to blending procedure in the construction
    of our initial data (see Section~\ref{SS:Blending}).
  }
  \label{F:DiskDynamics}
  \end{center}
  \end{figure}

To determine how much our results depend on numerical resolution, we have
performed simulations of this model for three different resolutions with grid
cell size scaling as $1:1.5:1.5^2$. We refer to these as the coarse, medium
and fine resolutions hereafter. The curvilinear geometry of the blocks was
chosen to be the same for all of the three resolutions. We list the parameters
of the resulting coarse, medium and fine (denoted as AFc, AF and AFf,
respectively) grid models in Table~\ref{T:Models}.

Figure~\ref{F:DiskDynamics}a shows the $L_1$- and
$L_{\infty}$-norms of Hamiltonian constraint violation for the coarse, medium
and fine resolution simulations. This plot shows that there are two distinct 
regimes in the evolution of the constraints: initial exponential decrease and
subsequent steady plateau. 
The latter is due to the constraint damping mechanism of our evolution
scheme, in which the rate of production of discretization errors is
balanced by the rate of constraint damping. 
Since the discretization error depends on grid cell size, the value of the
plateau decreases with increasing resolution. 
In the case with the highest resolution, the ``plateau regime'' is not
reached during the time span of the
simulation. Fig.~\ref{F:DiskDynamics}a also shows that there is a
small region of a rapid growth of the constraints at the very end of
the simulations. This increase is caused by the approach of the BH
apparent horizon too close to the inner excision boundary as a result 
of the interaction of the BH with the $m=1$ deformation of the disk
described above.   

Figure~\ref{F:DiskDynamics}b shows the time evolution of the BH
mass~\footnote{
We measure irreducible BH mass $M_{BH}$ using the area of apparent
horizon~\cite{Schnetter03a}.}.
Because the constraint violations due to the metric blending are introduced
at the continuum limit, the BH mass shows an unphysical oscillation
that does not converge away with resolution, but which completely
damps out in about one orbital period. The BH mass then stabilizes at
$97.5\pm0.5\%$ of its initial value and remains near this value until
the end of the simulation. 

  \begin{figure}[!htp]
  \begin{center}
  \begin{tabular}{c}
  \includegraphics[width=0.45\textwidth]{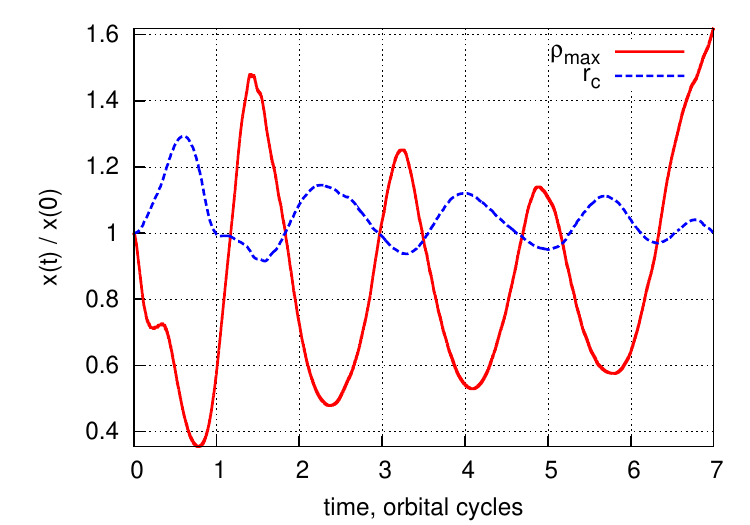} 
  \end{tabular}
  \caption{
    Time evolution of the maximum rest-mass density and the location of 
    a rest-mass density maximum $r_c=r(\rho_{max})$.
  }
  \label{F:DiskOscillations}
  \end{center}
  \end{figure}

  \begin{figure*}[!htp]
  \begin{center}
  \begin{tabular}{cc}
  \includegraphics[width=0.45\textwidth]{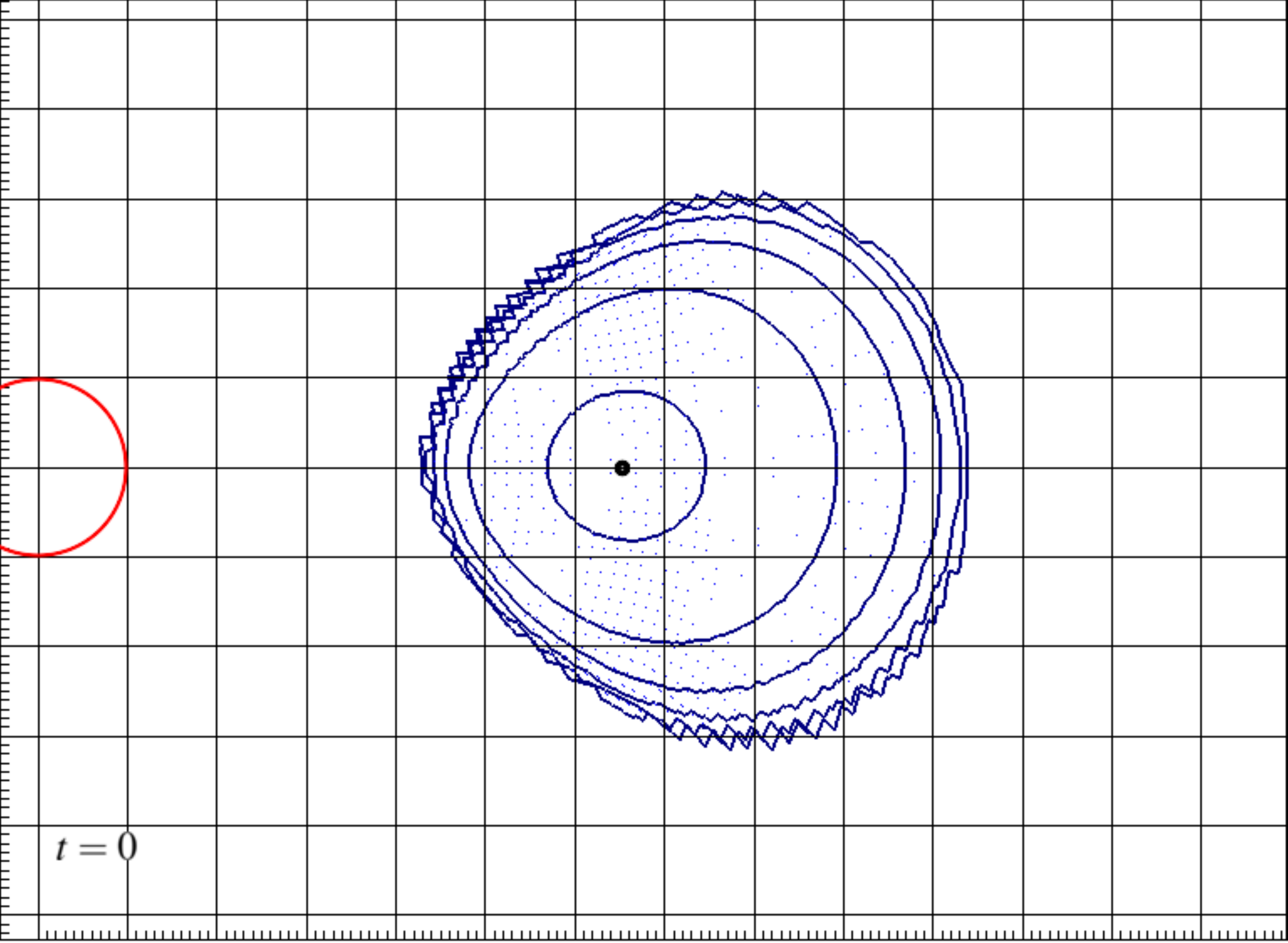} &
  \includegraphics[width=0.45\textwidth]{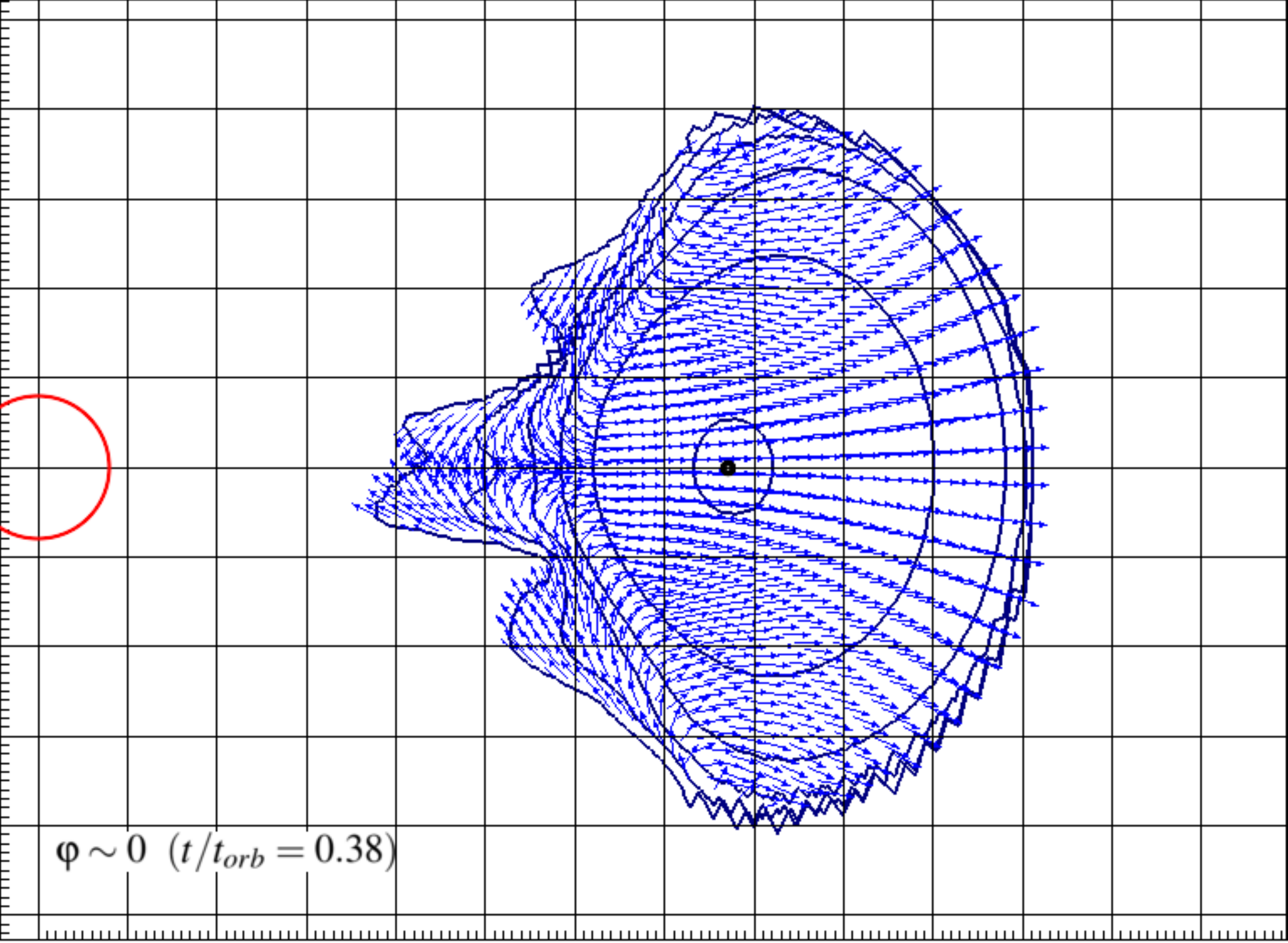}
  \\
  \includegraphics[width=0.45\textwidth]{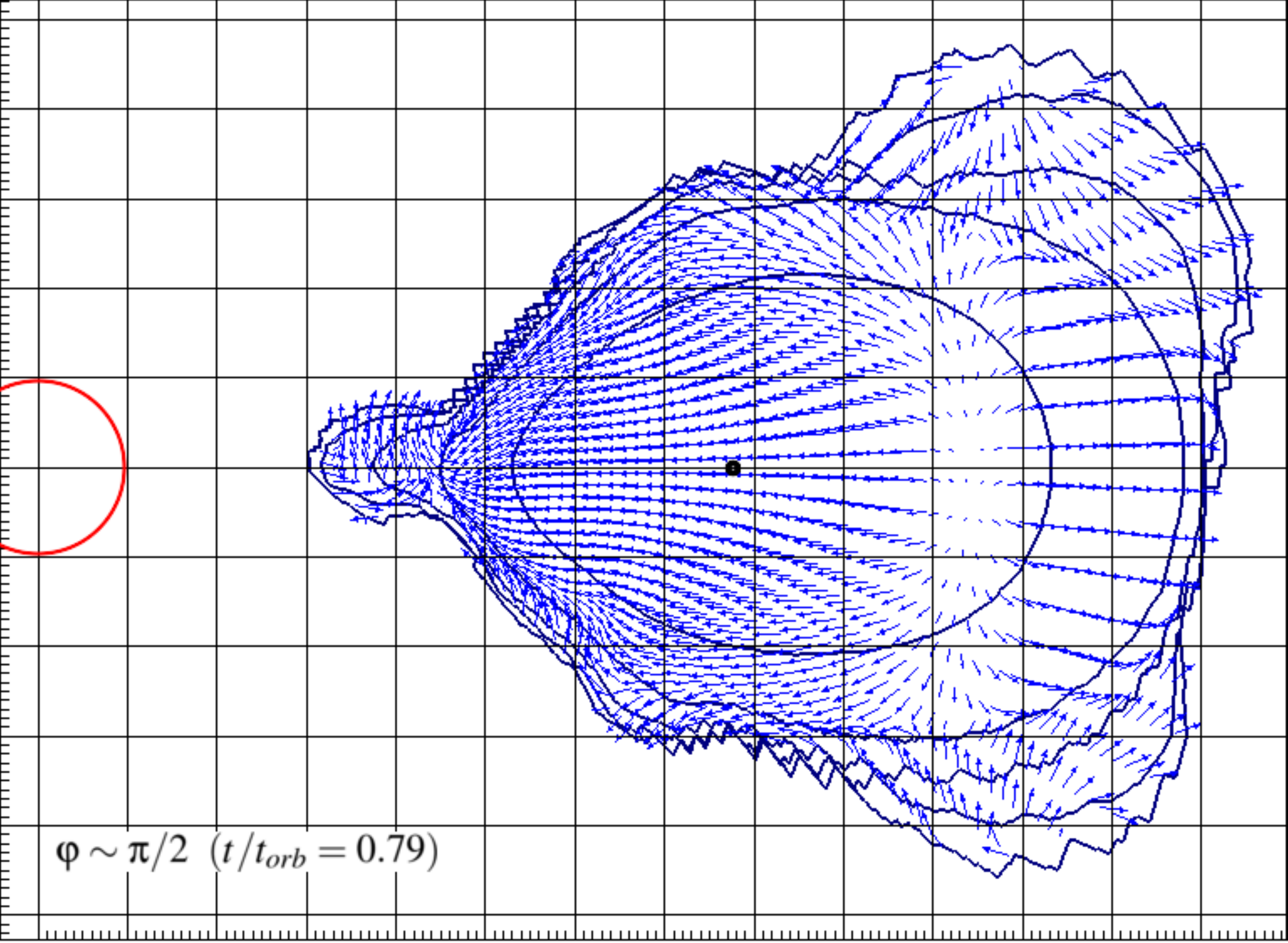} &
  \includegraphics[width=0.45\textwidth]{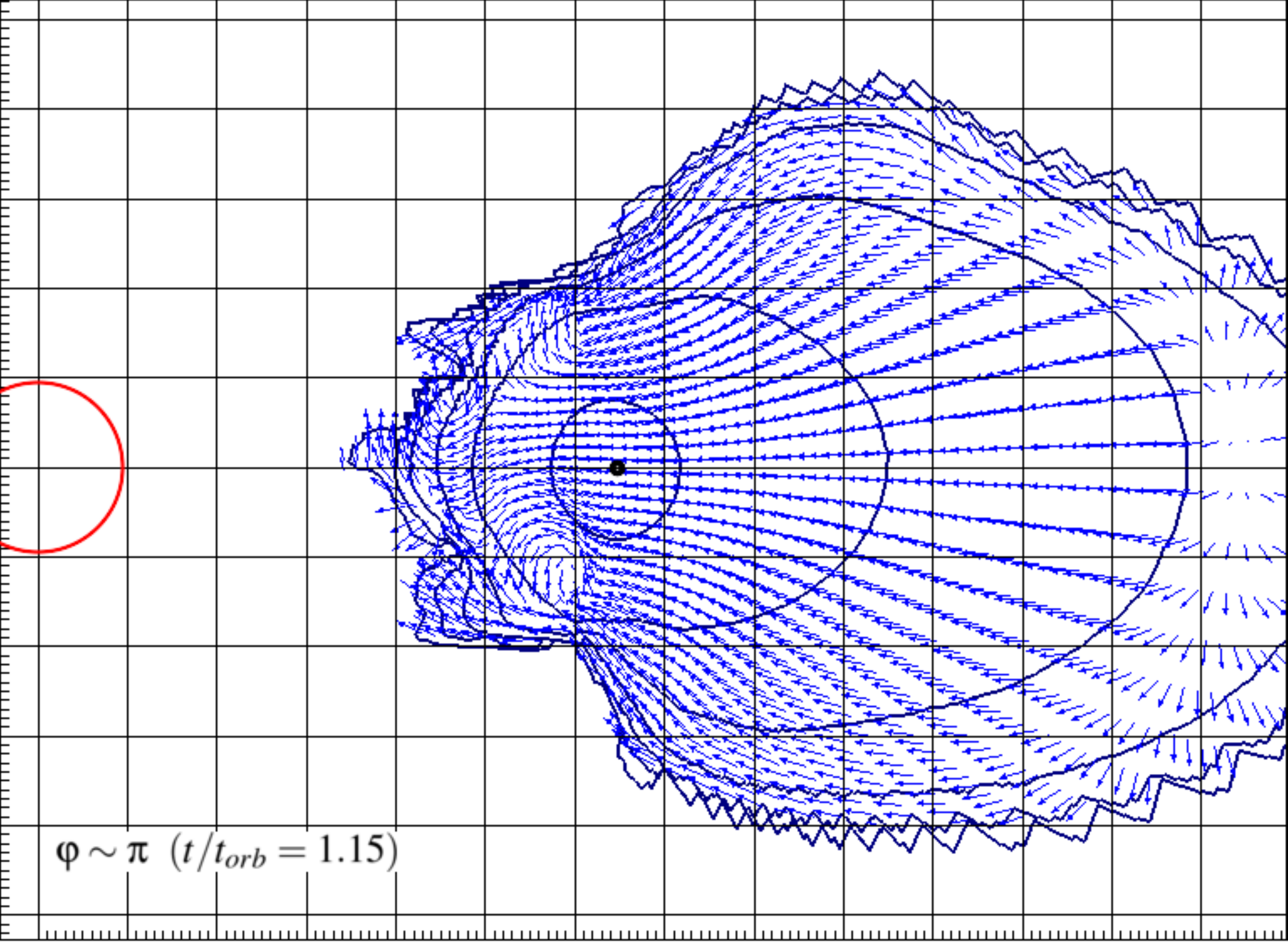}
  \\
  \includegraphics[width=0.45\textwidth]{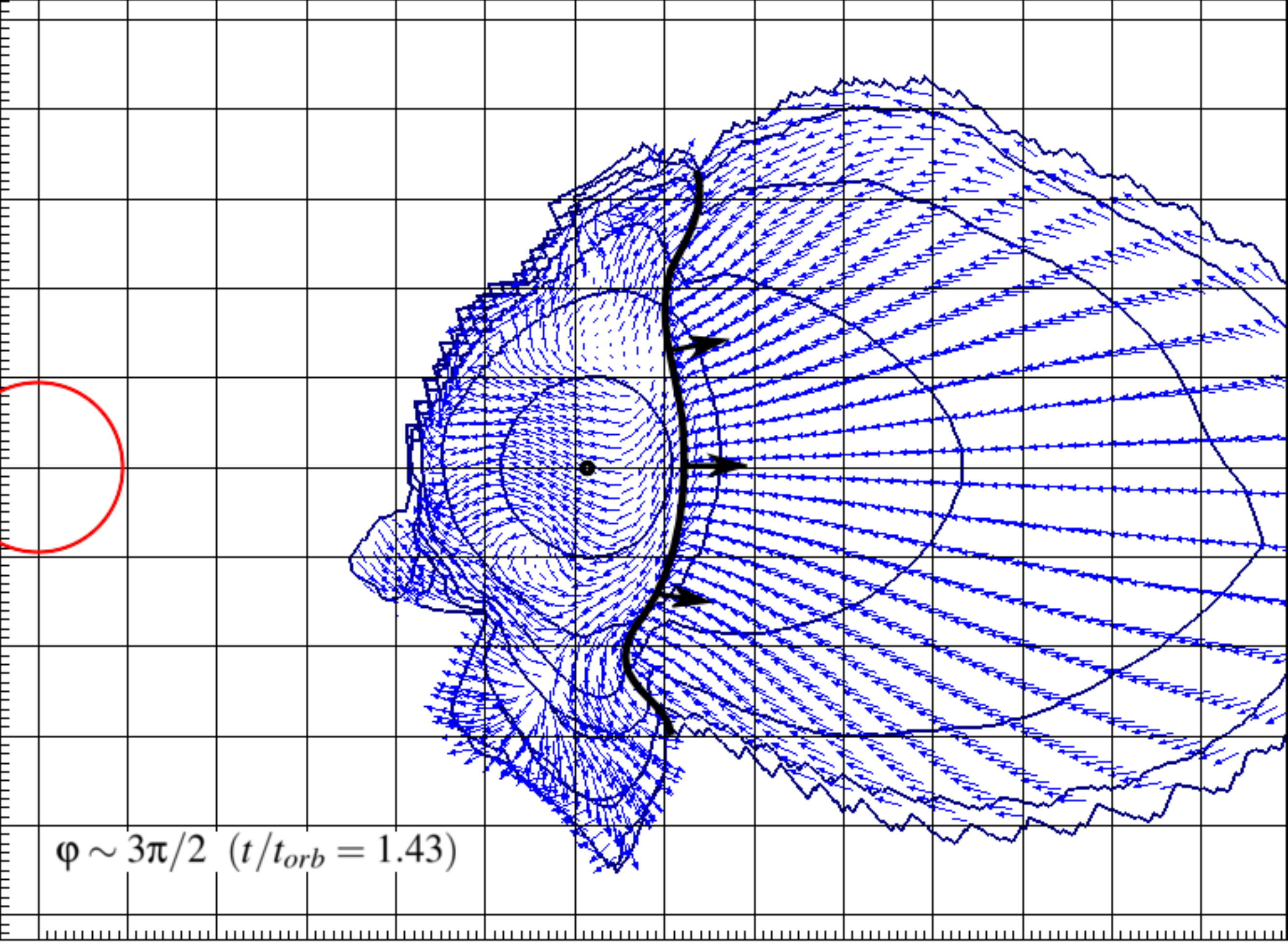} &
  \includegraphics[width=0.45\textwidth]{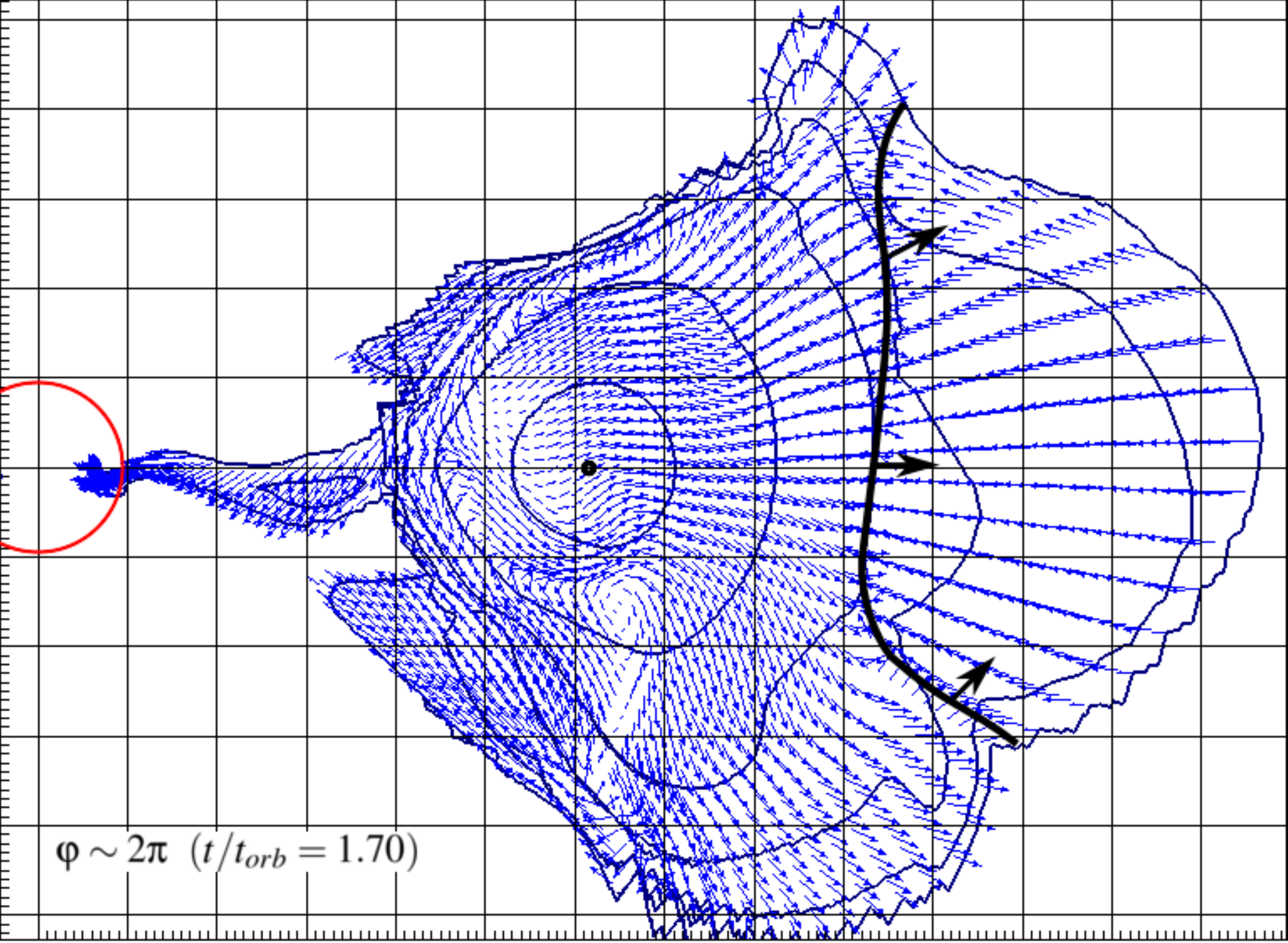}
  \\
  \end{tabular}
  \caption{
    Meridional cuts of the disk and the BH horizon in the $xz$-plane at
    several different phases $\ji \in (0-2\pi)$ of the radial
    oscillation. 
    Each frame shows six density contours, equally spaced
    in logarithmic scale between the initial maximum density
    $\rho_{\max}(0)$ and $10^{-6}\rho_{\max}(0)$.
    Also shown are the velocity field in the meridional plane and the
    location of the density maximum.
    The bottom two frames also show the position of the shock wave
    that is propagating outwards (thick black line).
    The last frame shows a brief episode of accretion.
  }
  \label{F:PolarCuts}
  \end{center}
  \end{figure*}

We now proceed to the analysis of the axisymmetric disk oscillations
induced by the metric blending procedure (described in
Section~\ref{SS:Blending}).
Fig.~\ref{F:DiskOscillations} shows the time evolution of the maximum disk
density $\rho_{max}$ (solid line) and the radial position of the disk center
$r_c$~\footnote{
  We define the radius of the disk center $r_c$ to be the radius, where the
  $\ji$-averaged disk density reaches its maximum.
} (dashed line), normalized to their values at $t=0$.
As can be seen from this plot, the evolution of both $\rho_{max}$ and $r_c$ is
dominated by a single mode of oscillation. 
The disk expands while moving away from the BH, and contracts while moving
towards the BH. 
This can also be seen in Fig.~\ref{F:PolarCuts}, which shows meridional cuts
of the disk in the $xz$-plane in different phases of the oscillation.

The radial oscillation of the disk has frequency of $\sim164\ \Hz$, a
value that is slightly smaller than the epicyclic frequency
$\kappa=201\ \Hz$ at $r=r_c$ for a Schwarzschild BH of 
the same mass~\footnote{The relativistic epicyclic frequency for the
  Schwarzschild metric is given by
  $\kappa=\frac{1}{2\pi}\sqrt{\frac{GM}{r^3}\left(1-\frac{3r_g}{r}\right)}$~\cite{Okazaki87}.}. 
Rezzolla et al.~\cite{RYZ03} studied relativistic axisymmetric
oscillations of accretion disks in Cowling approximation using disk
models that have the same $\Gamma$ ($=4/3$) and (constant) specific
angular momentum profile as the ones studied here. They found that the
disks with $r_c=3.860\ r_g$ have the radial oscillation frequency of
$\simeq 261\ \Hz$ ($=0.02017$ for their c2 model in normalized
units~\cite{RYZ03}), which is higher than the $\sim 164\ \Hz$
frequency that we obtain in our case. Part of this difference stems from
the difference in the disk models: while model of \cite{RYZ03} has
$r_c=3.860\ r_g$, our disk model A has $r_c=6.51\ r_g$ (cf.
Table~\ref{T:SGTParams}).
Rezzolla et al.~\cite{RYZ03} has found that the oscillation frequency
$f$ depends on the disk extent $L$ in a particular
way. Namely, $f$ decreases with increasing $L$, and in the limit of
very thin disks ($L\to0$), $f$ approaches the local value of the
epicyclic frequency $\kappa$ (see Fig.~4 of~\cite{RYZ03}). If we
assume this dependence $f(L)$ to hold for any $r_c$, we can obtain the
value of radial oscillation frequency for a disk with the same $L$ and
$r_c$ as our model. This frequency turns out to be $\sim168\ \Hz$,
which is very close to $f=164\ \Hz$ that we obtain in our
simulations. This result might indicate that for the disks that are
similar to the ones considered here, the disk self-gravity does not
significantly affect the frequencies of the radial oscillations. Note
that in the case of NSs, the frequency of the fundamental quasi-radial
modes was found to differ by a factor of $\sim2$ between Cowling and
full GR simulations~\cite{Font01,Dimmelmeier06a}, perhaps implying that the
self-gravity plays a more important role in the case of NSs. However,
since it is unclear whether such dependence of $f$ on $L$ holds
exactly for other values of $r_c$, this result should be taken with
caution. We will revisit this issue in a future publication. 

The disk oscillations are damped due to formation of shock waves that convert
the kinetic energy of the oscillations into the thermal one. At the
end of the contraction phase of the disk, the high-density inner part
of the disk bounces back earlier than the lower-density outer
part. 
The collision of the former with the still-infalling lower-density outer
material leads to formation of an outward-propagating shock wave (see two
bottom panels on Fig.~\ref{F:PolarCuts}). 
The shock accelerates during propagation and reached relativistic velocities
of $ \sim 0.3 c$ in the rarefied outer disk shells.
In order to quantify the amount of the shock dissipation, we evolve this model
also with isentropic polytropic EOS with the same $\Gamma=4/3$ (in addition to
the evolution with the non-isentropic $\Gamma$-law EOS), which does not allow
entropy changes and thus shock heating. 
In this case, the oscillations exhibit little damping, while in the case of
evolutions with the $\Gamma$-law EOS, the amplitude of $\rho_\mathrm{max}$
decreases by a factor of $ \sim 2 $ in $ \sim 4 $ orbital periods. 
Note that dissipation of oscillation kinetic energy into heat by
shocks has  been found to operate in many other astrophysical
scenarios, including damping of NS oscillations in a migration from an
unstable to a stable branch~\cite{Font01b,Baiotti04}, in 
phase-transition-induced collapse of NSs~\cite{Abdikamalov09a}, as
well as damping of ring-down oscillations of nascent proto-NSs formed
in core-collapse supernovae (e.g.,~\cite{Ott09a}) and
accretion-induced collapse of white-dwarfs
(e.g.,~\cite{Abdikamalov10}).

\section{Non-axisymmetric instabilities}
\label{S:Nonaxisym}

In this section, we discuss non-axisymmetric instabilities in the disks.
First, we test our method by reproducing results obtained by
Kojima~\cite{Kojima86b} for thick tori with negligible self-gravity on a
Schwarzschild background. In the next two sections we analyze the
non-axisymmetric instabilities in our models first on fixed background, and
then with fully dynamical general relativistic treatment.

In various astrophysical scenarios, the disk might be formed with
different non-axisymmetric structures, leading to preferential
excitation of specific unstable non-axisymmetric modes.
In order to account for these possible scenarios, in our work we
evolve initial disk models with added small non-axisymmetric
perturbations at $t=0$.

We expect~\cite{Kojima86a,NGG87,WTH94} our thick disk models A and B to be
dominated by non-axisymmetric unstable modes with azimuthal numbers $m=1$ and
$m=2$, while for more slender model C instabilities with $m=3$ and $m=4$
should also play an important role.
In our simulations without initial perturbations, these modes are triggered
by numerical errors and start to grow at a random moment in time.
In order to explore evolutionary scenario in which a particular mode is
excited initially, we add a small density perturbation of the form
$\tilde{\rho} = \rho[1 + A\cos m(\ji - \ji_0)]$
to our initial disk models.
We use a perturbation amplitude $A=0.001$, which we have found to be both
large enough to trigger the instability, and sufficiently small to remain in
the linear regime.
The amount of constraint violations due to these artificial perturbations,
which is quickly suppressed by our constraint damping scheme, is too small to
significantly affect the subsequent evolution of the disk.

For each initial disk model A, B and C we have completed a sequence of
evolutions both in Cowling and in full GR treatment.
As explained in the end of Section~\ref{S:AdMesh}, corresponding simulations
are referred to by two- or three-letter notation (such as AC, AC1, AC2, AF,
etc.), in which the first letter denotes the initial disk model (A, B or C),
the second letter is either 'C' for Cowling or 'F' for full GR treatment, and
the third one is the azimuthal number $m$ of the initial perturbation (absent
if evolved without initial perturbation).

\subsection{Comparison with previous work}
\label{SS:CompKojima}

  \begin{table}
  \begin{center}
  {\small
  \begin{tabular}{l|cccc}
  \hline \hline
   Model & 
   $r_c/r_g$ & 
   $r_{-}/r_{c}$ & 
   $r_{+}/r_{-}$ &
   $\rho_{\max}$ \\
   \hline   
   K1 & 5.236 & 0.60 & 3.952 & $1.715\cdot10^{-4}$ \\
   K2 & 5.236 & 0.65 & 2.956 & $5.321\cdot10^{-5}$ \\
   K3 & 5.236 & 0.70 & 2.351 & $1.472\cdot10^{-5}$ \\
   K4 & 5.236 & 0.75 & 1.939 & $3.522\cdot10^{-6}$ \\
   K5 & 5.236 & 0.80 & 1.640 & $5.809\cdot10^{-7}$ \\
   K6 & 5.236 & 0.85 & 1.419 & $8.682\cdot10^{-8}$ \\
   \hline \hline
  \end{tabular}}
  \end{center}
  \caption{Physical parameters of the initial disk models, used for comparison
           with calculations by Kojima~\cite{Kojima86b}. Here,
           $r_c/r_g$ is the location of the density maximum in units of the BH
           gravitational radius $r_g$, 
           $r_{-}$ ($r_{+}$) is the inner (outer) radius of the torus, and
           $\rho_{max}$ is the maximum density.
           For all models, the mass of the BH is $\MSun$, specific
           angular momentum has constant value $\ell=4.0$, and
           polytropic constant $K=0.06$ (the latter two quantities are given
           in the normalized system of units, in which $G=c=\MSun=1$).
 }
  \label{T:KojimaIDModels}
  \end{table}

Below we present the growth rates of PP instability in evolutions of
equilibrium tori in the Cowling approximation and compare them to the results
obtained by Kojima~\cite{Kojima86b}, who studied PP instability in thick disks
around Schwarzschild BHs using a linear perturbative approach. Kojima analyzed
tori with constant distribution of specific angular momentum, constructed in
a relativistic framework using the Abramowicz-Sikora-Jaroszynski (AJS)
prescription~\cite{AJS78:AccretingDisks}. Kojima calculated the
$m=1$ mode growth parameter $y_2$ for several sequences of disk models with
different values of specific angular momentum $\ell$ \ ($=3.8,\ 4.0,\ 4.2$).
In particular, \cite{Kojima86b} found that, due to relativistic redshift
effects, $y_2$ for GR models is generally smaller than in the Newtonian
case~\cite{Kojima86a}.

We have constructed a sequence of AJS tori of varying radial extent with the
same physical parameters as in~\cite{Kojima86b}. 
All these models have polytropic index $\Gamma$ of $4/3$ and specific angular
momentum $\ell$ of $4.0$. 
The mass of the BH is set to $M_{BH}=1$. 
The models in this sequence are labelled as K1-K6, and their parameters are
listed in Table~\ref{T:KojimaIDModels}. 
To excite the $m=1$ mode, we add a small non-axisymmetric perturbation as
described above, and evolve the disk in the Cowling approximation, using
a curvilinear grid (as described in Section~\ref{S:AdMesh}), adapted to each
of the models.
Parameters of the curvilinear grids for each model are listed in
Table~\ref{T:Models}.
We then measure the growth rate $y_2$ as described in
Section~\ref{SS:DataAnalysis} above and compare it with the results of Kojima.

  \begin{table}
  \begin{center}
  {\small
  \begin{tabular}{l|cc|c|cccc||c|cc|cc|c}
  \hline \hline
   Model & $y_1$  & $y_2$ & $\Omega_p/\Omega_c$ & $r_{cr}/r_c$ & Type \\
   \hline                                                                                           
   K1 & -0.174 & 0.097(6) & 0.826 & 1.136 & PP\\
   K2 & -0.147 & 0.124(3) & 0.853 & 1.112 & PP\\
   K3 & -0.113 & 0.153(3) & 0.887 & 1.083 & PP\\
   K4 & -0.084 & 0.145(4) & 0.916 & 1.041 & PP\\
   K5 & -0.048 & 0.120(3) & 0.952 & 1.060 & PP\\
   K6 & -0.011 & 0.093(5) & 0.989 & 1.007 & PP\\
   \hline \hline
  \end{tabular}}
  \end{center}
  \caption{Parameters of the $m=1$ PP instability, measured for the dynamical
           evolutions of Kojima disk models K1-K6.
           $y_1$ and $y_2$ are the pattern speed and growth rate parameters,
           $\Omega_p$ is the mode pattern speed, and $r_{cr}$ is the mode
           corotation radius.}
  \label{T:KojimaModes}
  \end{table}

  \begin{figure}[!htp]
  \begin{center}
  \includegraphics[width=0.45\textwidth]{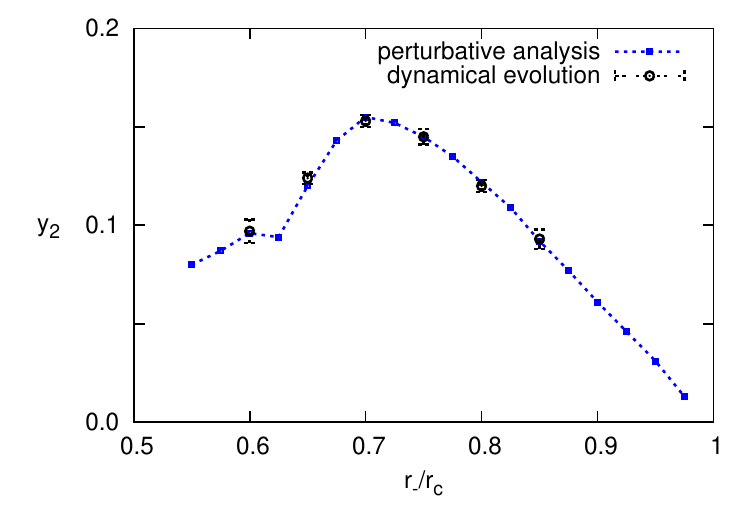} \\
  \caption{
    Comparison of the $m=1$ PP mode growth rates $y_2$ obtained by
    Kojima~\cite{Kojima86b} in linear perturbative approach with the values
    measured from the evolutions of the same initial data models in Cowling
    approximation.
    The abscissa represents the ratio of the inner radius of the disk  
    $r_-$ to the radius of the maximum disk density $r_c$.
    Note that the error bars of the measured growth rates originate from the
    uncertainty in determining the time span of a clear exponential 
    growth of the mode.
  }
  \label{F:Kojima86b}
  \end{center}
  \end{figure}

Figure~\ref{F:Kojima86b} shows the values of $y_2$ as a function of the disk
radial extent for our models and for those of Kojima. As we can see, $y_2$ for
our models are within estimated error bars from the values, calculated by
Kojima in~\cite{Kojima86b}, as it should be the case.

\subsection{Fixed background}
\label{SS:FixedBG}

In this section, we analyze non-axisymmetric instabilities which develop
when our initial disk models are evolved in Cowling approximation.
We find that all of our models develop the PP instability. 
More specifically, for models A and B the fastest growing
mode is $m=2$, while for more slender model C it is $m=3$.
This is expected from the Newtonian
considerations~\cite{Kojima86a,NGG87,WTH94}.
Below, we first describe instabilities in models A and B, after which we focus
on model C.
As mentioned above, we evolve our models with and without artificial density
perturbations.
For models A and B, we add $m=1,2$ and for model C we add $m=1,2,3$
perturbations.
Notice that all simulations contain a spurious $m=4$ perturbation which is
a numerical artifact of interpolation at the boundaries between four blocks
near the equatorial plane.
The evolutions without artificial perturbation are therefore similar to the
ones in which an $m=4$ density perturbation is added.

To analyze the non-axisymmetric modes, we adopt an approach from Woodward et
al~\cite{WTH94}.
Namely, we expand the disk density in the equatorial plane in a Fourier series, as
explained in Section~\ref{SS:DataAnalysis} above. Then we construct and
analyze the following \emph{four diagrams}:

\begin{itemize}
\item[(a)] $D_m-t$ diagram shows logarithm of the normalized mode
  amplitude $D_m$ as a function of time at some radial location close
  to the disk density maximum $r_c$. The slope of this curve yields
  the growth rate $y_2$.  
\item[(b)] $D_m-r$ diagram represents a radial profile of the
  normalized mode amplitude $D_m$. This diagram will be helpful in
  identifying the type of non-axisymmetric
  instabilities~\cite{WTH94}. 
\item[(c)] $\ji_m-t$ diagram displays a phase angle of the
  non-axisymmetric mode $m$ as a function of time at a specified
  radius. The slope of this function determines the mode pattern speed
  and parameter $y_1$ that is related to it.  
\item[(d)] $\ji_m-r$ diagram represents mode phase angle as a function
  of radius. This diagram also provides a convenient way to identify
  the type of the mode~\cite{WTH94}. 
  In all our $\ji_m-r$ diagrams, the disk rotates counterclockwise.
\end{itemize}

  \begin{table*}
  \begin{center}
  {\small
  \begin{tabular}{l|cc|cc|cc||l|cc|cc|cc}
  \thickhline
   \multicolumn{7}{c||}{Cowling} & \multicolumn{7}{c}{Full GR} \\
  \hline                                                 
   Model & m & type & $y_1$  & $y_2$ & $\Omega_p/\Omega_0$ & $r_{cr}/r_c$ &
   Model & m & type & $y_1$  & $y_2$ & $\Omega_p/\Omega_0$ & $r_{cr}/r_c$ \\
  \thickhline
   AC  & 2 & PP & -0.10(5) & 0.21(1) & 0.89(2) & 1.07(1) & AF  & 1 & PP & -0.17(5) & 0.300(8) & 0.83(5) & 1.12(4) \\
   AC1 & 1 & PP & -0.08(4) & 0.17(1) & 0.93(2) & 1.04(1) & AF1 & 1 & PP & -0.18(5) & 0.294(8) & 0.82(5) & 1.11(4) \\
   AC2 &   &    &          &         &         &         & AF2 & 1 & PP & -0.17(3) & 0.30(1)  & 0.83(3) & 1.12(3) \\
       & 2 & PP & -0.09(5) & 0.22(1) & 0.95(2) & 1.03(1) &     & 2 & I  & -0.6(1)  & 0.17(3)  & 0.68(5) & 1.26(6) \\
  \hline                                                                                                        
   BC  & 1 & PP & -0.04(5) & 0.16(1) & 0.96(5) & 1.03(3) & BF  & 1 & PP & -0.16(5) & 0.28(3)  & 0.84(5) & 1.12(4) \\
       & 2 & PP & -0.12(6) & 0.18(1) & 0.94(3) & 1.04(2) &     &   &    &          &          &         &         \\
   BC1 & 1 & PP & -0.06(5) & 0.16(1) & 0.94(5) & 1.04(3) & BF1 & 1 & PP & -0.12(5) & 0.270(8) & 0.88(5) & 1.08(4) \\
   BC2 & 1 & PP & -0.04(5) & 0.17(1) & 0.96(5) & 1.03(3) & BF2 & 1 & PP & -0.13(5) & 0.29(2)  & 0.87(5) & 1.09(4) \\
       & 2 & PP & -0.08(6) & 0.17(1) & 0.96(3) & 1.03(2) &     & 2 & I  & -0.5(1)  & 0.11(2)  & 0.75(5) & 1.19(5) \\
  \hline                                                                                                        
   CC  & 3 & PP & -0.06(5) & 0.21(2) & 0.98(2) & 1.01(1) & CF  &   &    &          &          &         &         \\
       & 4 & PP & -0.04(5) & 0.14(1) & 0.99(2) & 1.00(1) &     & 4 & I  & -0.84(8) & 0.16(1)  & 0.79(2) & 1.14(2) \\
   CC1 & 3 & PP & -0.04(5) & 0.24(1) & 0.99(2) & 1.00(1) & CF1 & 1 & PP${}^*$?& ??       & ??       & --      & --      \\
       & 4 & PP & -0.04(5) & 0.16(2) & 0.99(2) & 1.00(1) &     & 4 & I  & -0.84(8) & 0.15(1)  & 0.79(2) & 1.14(2) \\
   CC2 & 2 & PP & -0.04(5) & 0.20(1) & 0.99(2) & 1.00(1) & CF2 & 2 & I  & -0.6(1)  & 0.279(7) & 0.66(5) & 1.26(5) \\
       & 3 & PP & -0.06(5) & 0.22(1) & 0.98(2) & 1.01(1) &     & 4 & I  & -0.8(2)  & 0.16(1)  & 0.80(5) & 1.14(4) \\
   CC3 & 3 & PP & -0.07(5) & 0.23(1) & 0.98(2) & 1.01(1) & CF3 & 3 & I  & -0.7(1)  & 0.318(8) & 0.76(3) & 1.17(3) \\
       & 4 & PP & -0.04(5) & 0.15(2) & 0.99(2) & 1.00(1) &     & 4 & I  & -0.80(8) & 0.16(2)  & 0.80(2) & 1.14(2) \\
  \thickhline
   \multicolumn{14}{p{0.75\textwidth}}{
    \begin{footnotesize}
      ${}^*$For the simulation CF1, it was not possible to accurately determine the growth rate and pattern 
      speed of the $m=1$ mode. We classified this mode as a PP-type due to the character of its $\ji_m-r$ 
      and $D_m-r$ diagrams, which are typical for the PP-modes. 
    \end{footnotesize} 
   }
  \end{tabular}}

  \end{center}
  \caption{
    Quantitative characteristics and types of the non-axisymmetric modes
    for the simulations studied in
    Section~\ref{S:Nonaxisym}.
    For each of the simulations, the table lists one or two dominant unstable
    modes.
  }
  \label{T:ABCModes}
  \end{table*}

  \begin{figure*}[!htp]
  \begin{center}
   \begin{tabular}{cc}
   \includegraphics[width=0.45\textwidth]{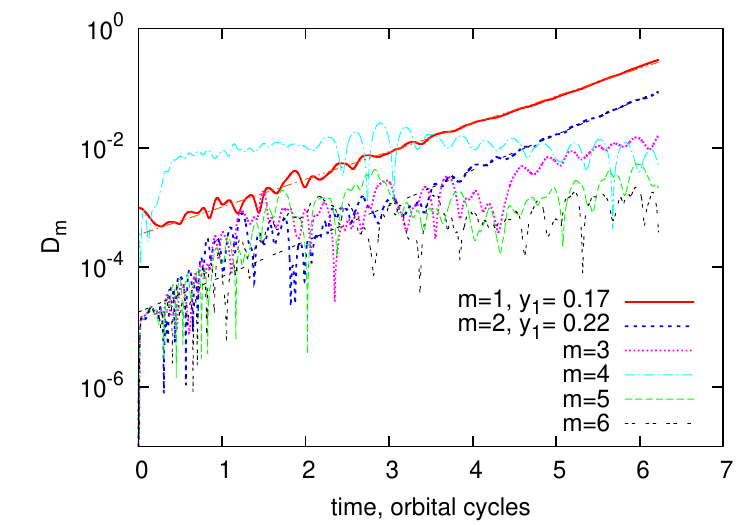} &
   \includegraphics[width=0.45\textwidth]{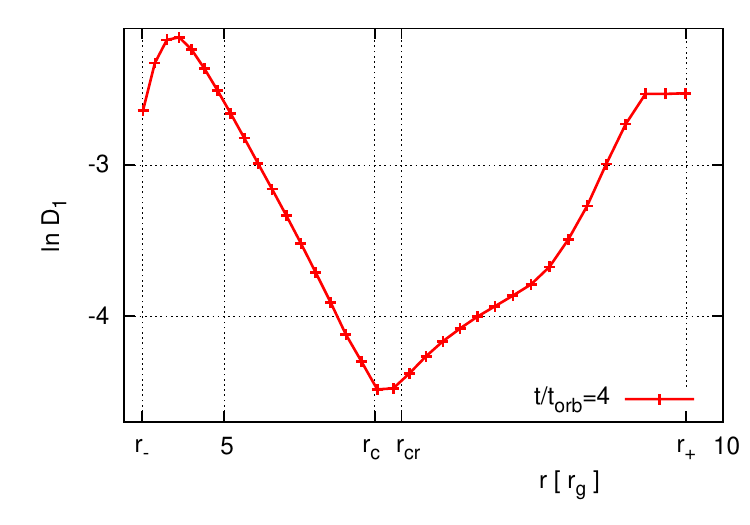} 
   \\
   (a) & (b)
   \\
   \includegraphics[width=0.45\textwidth]{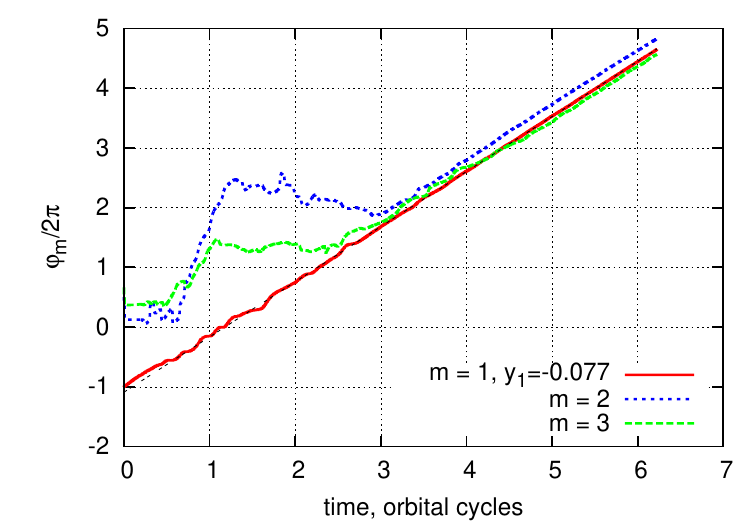} &
   \includegraphics[width=0.45\textwidth]{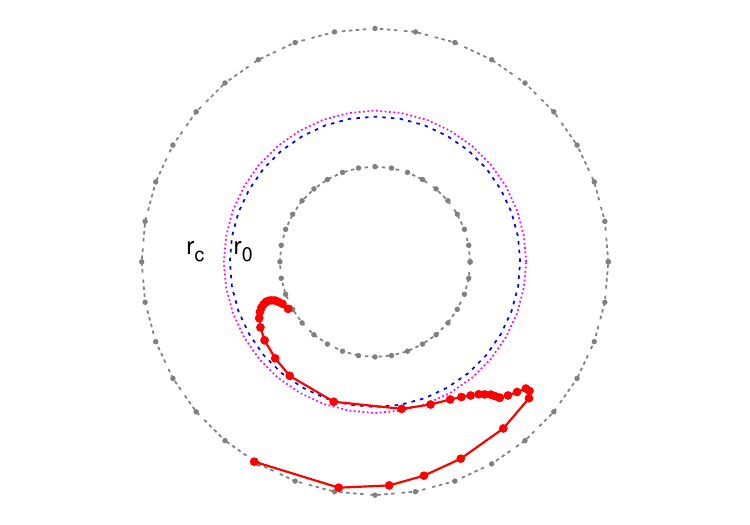} 
   \\
   (c) & (d)
   \\
   \end{tabular}
  \caption{
    The behavior of the dominant non-axisymmetric modes and the radial
    character of the $m=1$ (0,0) PP-mode in the simulation AC1:
    (a) $D_m-t$ diagram shows the time evolution of the mode
    amplitudes $D_m$ for $m=1-6$;
    (b) $D_m-r$ diagram shows the radial character of the $m=1$ mode
    amplitude (in logarithmic scale);
    (c) $\ji_m-t$ diagram shows the time evolution of the Fourier angle
    $\ji_m$, which allows to determine a pattern speed and a corotation
    radius for each mode;
    (d) $\ji_m-r$ diagram shows the dependence of the mode Fourier
    angle from the radius $r$ in equatorial plane at
    $t/t_{orb}\simeq4$.
    The $D_m-r$ and $\ji_m-r$ diagrams on panels (b) and (d) also show
    the locations of the mode corotation radius.
  }
  \label{F:PPStruct1}
  \end{center}
  \end{figure*}

We found that instabilities which develop during evolutions of models A and B
are very similar.
Therefore we present the properties of these instabilities on example of
model A, while the case of model C will be described
separately.

Figure~\ref{F:PPStruct1} shows the four mode diagrams for the case of model
AC1, which represents time evolution of initial disk model A in Cowling
approximation with an added $m=1$ non-axisymmetric density perturbation.
Panel (a) shows the $D_m-t$ diagram for the first six non-axisymmetric modes
with $m=1,2,\dots,6$.
Three lowest-$m$ modes exhibit clear exponential growth, with $m=1$ being the
dominant mode throughout the time span of the simulation (first $\approx6.5$
orbital periods).
This is the case because an $m=1$ perturbation is artificially added from the
beginning and has more time to grow and to remain the dominant mode.
The mode $m=2$ has higher growth rate, but appears subdominant, since it is
triggered later than the $m=1$ mode and has less time to develop.
It may eventually overshoot the $m=1$ mode at a later time, when both modes
reach nonlinear regime (not covered in this work, this will be a subject
of our future publication).
Notice that the relatively high values of the $m=4$ mode amplitude are due to
the effect of interpolation errors on four interblock boundaries of the grid
near the equatorial plane.

Panels (b) and (d) of Fig.~\ref{F:PPStruct1} demonstrate the radial structure
of the $m=1$ mode at $t=4\ t_{orb}$, when it is sufficiently developed.
The $D_m-r$ diagram on Fig.~\ref{F:PPStruct1}b represents the radial profile
of the amplitude of the mode.
This amplitude is highest near the edges of the disk, it does not have nodes
(does not become zero), and it reaches its minimum near the radius of
corotation.
Previous works on PP instability in Newtonian gravity~\cite{PPI,NGG87} suggest
that such radial behavior is a characteristic of the principal PP mode, or a
mode of (0,0)-type in the classification of Blaes and
Hawley~\cite{BlaesHawley88}.
The $\ji_m-r$ diagram on Fig.~\ref{F:PPStruct1}d has a specific S-shaped
structure, which is also a well-known feature of the PP instability,
discovered in previous Newtonian works~\cite{WTH94,NGG87}.

Finally, panel (c) of Fig.~\ref{F:PPStruct1} shows the $\ji_m-t$ diagram
for the $m=1-3$ modes. 
It shows that while all modes initially have arbitrary
phases and pattern speeds, they eventually settle to the pattern speed of the
dominant $m=1$ mode possibly due to nonlinear interaction between the modes.
The pattern speed of the $m=1$ mode is slightly below the speed of the disk at
$r_c$, which means that the mode corotation radius $r_{cr}$ lies just outside
$r_c$.
The close proximity of the mode corotation radius to the radius of the disk
density maximum is also typical for PP non-axisymmetric instabilities, as
discovered in previous Newtonian works~\cite{PPI,GGN86,WTH94}.
All these features allow us to conclude that the observed $m=1$ mode is indeed
the PP instability.

\begin{figure*}[!htp]
  \begin{center}
   \begin{tabular}{cc}
   \includegraphics[width=0.45\textwidth]{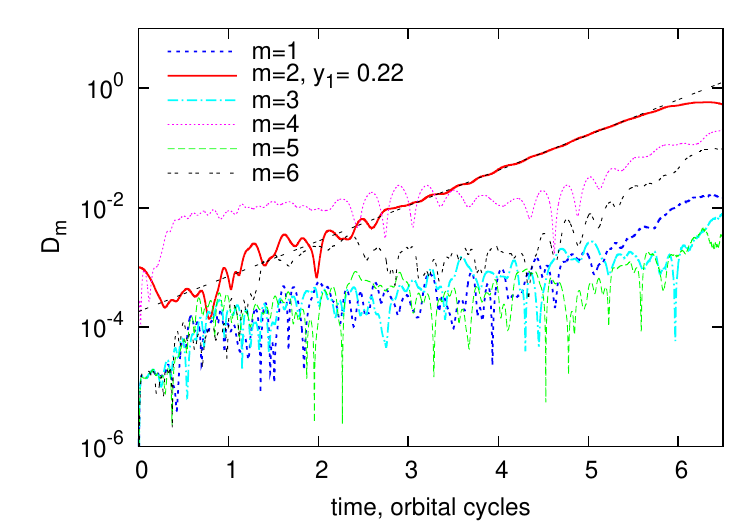} &
   \includegraphics[width=0.45\textwidth]{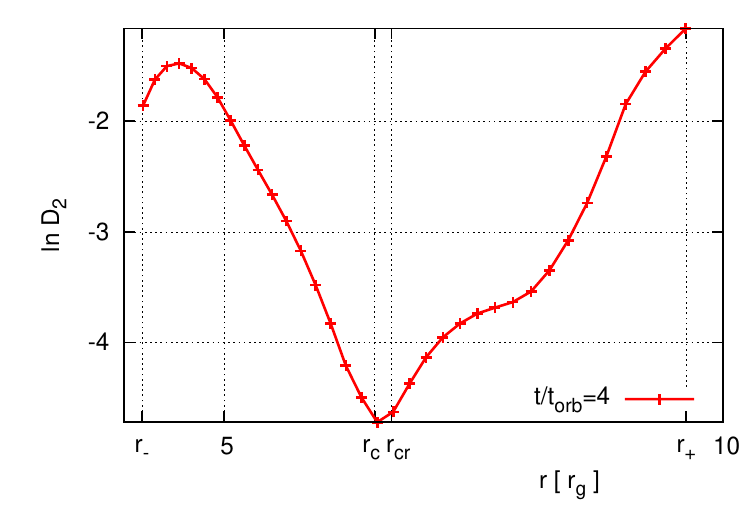} 
   \\
   (a) & (b)
   \\
   \includegraphics[width=0.45\textwidth]{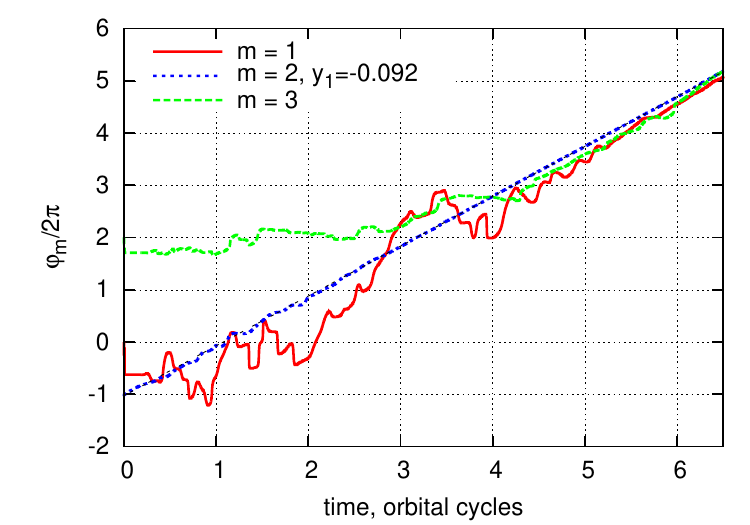} &
   \includegraphics[width=0.45\textwidth]{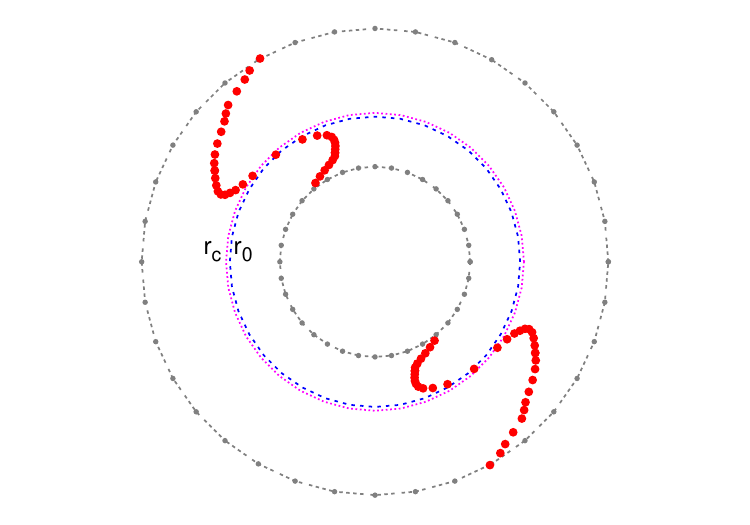} 
   \\
   (c) & (d)
   \\
   \end{tabular}
  \caption{
    The behavior of the dominant non-axisymmetric modes and the radial
    character of the $m=2$ (0,0) PP-mode in the simulation AC2
    (see captions to Fig.~\ref{F:PPStruct1} for details).
  }
  \label{F:PPStruct2}
  \end{center}
  \end{figure*}

Figure~\ref{F:PPStruct2} shows the set of four diagrams for simulation AC2,
in which an $m=2$ density perturbation is added initially.
Panel (a) represents $D_m-t$ diagram for the modes with $m=1-6$. 
In this case, only the mode with $m=2$ exhibits pronounced exponential growth. 
The rest of the modes remain either stable or are not excited, showing rapid
growth only in the very end of the simulation, when the amplitude of $m=2$
mode reaches nonlinear regime and coupling between the modes becomes
important. Panels (b) and (d) of~Fig.\ref{F:PPStruct2} show the radial profile
and azimuthal shape of the mode. These are again typical for the principal
(0,0)-type PP instability with $m=2$.
Fig.~\ref{F:PPStruct2}c shows the $\ji_m-t$ diagram for the $m=1-3$ modes.
The dominant $m=2$ mode rotates uniformly in the same direction with the disk,
while other two modes do not exhibit clear rotation pattern until after
$\sim4$ orbital periods, when they align in phase with the $m=2$ mode.
The pattern speed of the dominant mode inferred from this plot corresponds to
the corotation radius just outside of $r_c$.
Similarly to the simulation AC1 above, all these features are typical of the PP
instability, studied in previous works in the Newtonian approximation~\cite{PPI,
GGN86, WTH94}.

While unstable modes in model B are very similar to those of model A, in model
C higher-order PP modes become dominant.
This is expected~\cite{Kojima86a} since model C is more slender than models A
and B.
Indeed we observe that the most unstable mode for the model C is the $m=3$ PP
mode, while $m=2$ and $m=4$ have comparable but smaller growth rates.
The four diagrams for $m=3,4$ modes are not qualitatively
different from those for the $m=1,2$ modes in simulations AC1 and AC2 above.
They also exhibit all the features typical of the PP instability described
above.

We can therefore conclude that the PP instability with various values of
azimuthal number $m$ is observed in all of our disk models in Cowling
approximation.
For reader's reference, the parameters of unstable modes calculated for all
of our simulations are summarized in Table~\ref{T:ABCModes}.

\subsection{Dynamical background}
\label{SS:DynamicalBG}

We now turn to the analysis of non-axisymmetric instabilities, which develop
when the disks are evolved in a fully dynamical general relativistic framework.
We again consider evolutions with and without artificial density
perturbation, adding $m=1,2$ perturbations for models A and B and $m=1,2,3$
perturbations for model C. 
While we observe PP type instabilities in the Cowling case, in the fully
dynamical GR case we observe two distinct types of instabilities: the
Papaloizou-Pringle (PP) type and a GR analog of the so-called
intermediate type (I-type) instability~\cite{GoodmanNarayan88a,
Christodoulou93}.
Similarly to the Cowling case, instabilities in the moderately slender models
A and B have very similar properties, so it suffices to present the result only
for the case of model A.
Model C is more slender and therefore favors instabilities with higher
azimuthal numbers than those of models A and B, so we consider this model
separately.

  \begin{figure*}[!htp]
  \begin{center}
   \begin{tabular}{cc}
   \includegraphics[width=0.45\textwidth]{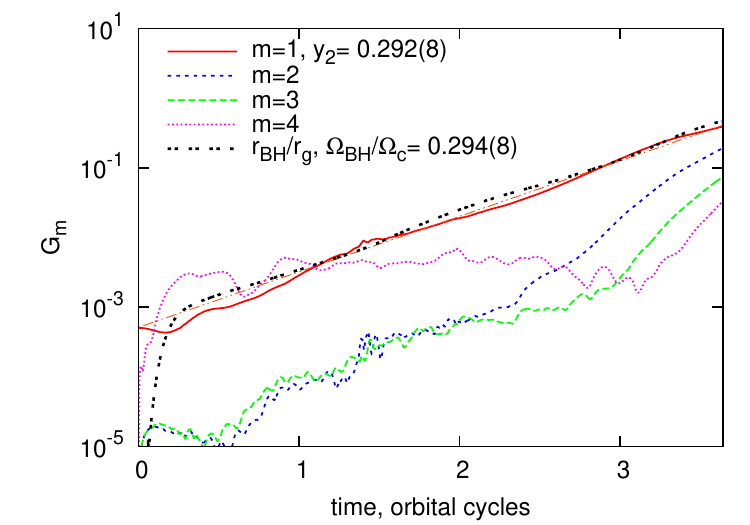} &
   \includegraphics[width=0.45\textwidth]{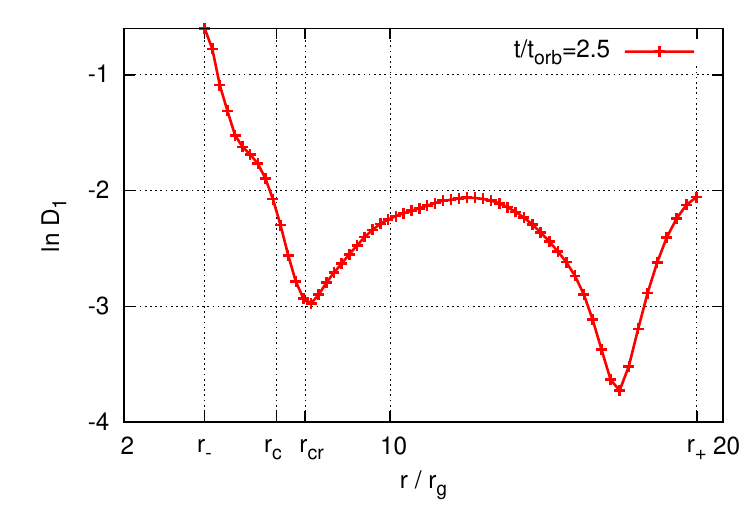} 
   \\
   (a) & (b)
   \\
   \includegraphics[width=0.45\textwidth]{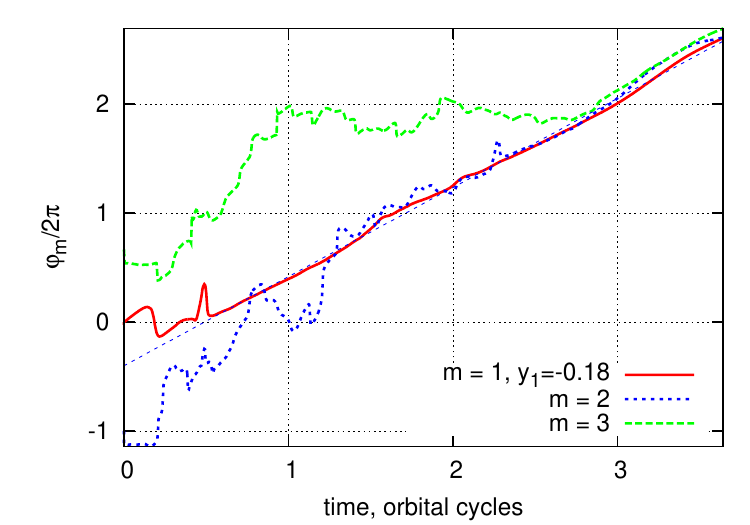} &
   \includegraphics[width=0.33\textwidth]{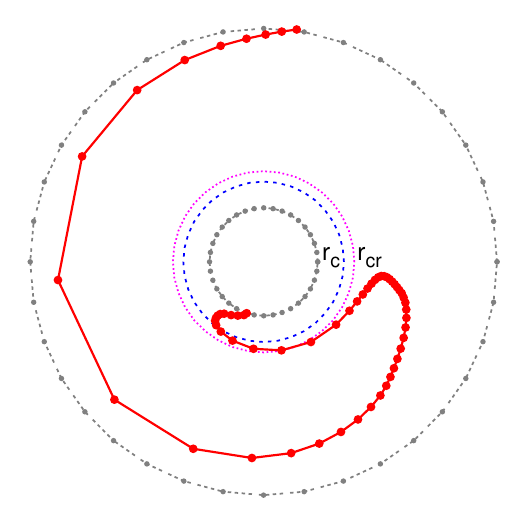} 
   \\
   (c) & (d)
   \\
   \end{tabular}
  \caption{
    The behavior of the dominant non-axisymmetric modes and the radial
    character of the $m=1$ (0,0) PP-mode in the simulation AF1
    (see captions to Fig.~\ref{F:PPStruct1} for details).
  }
  \label{F:StructAF1}
  \end{center}
  \end{figure*}

For the analysis of non-axisymmetric modes we adopt the same approach as in
Section~\ref{SS:FixedBG} above for simulations on a fixed background,
with one exception: for evaluating mode growth rates $y_2$, instead of $D_m$,
we use quantities $G_m$ introduced in Section~\ref{SS:DataAnalysis} above.
This is necessary because the values of the mode amplitudes $D_m$ at a fixed
radial location oscillate due to disk oscillations, making it hard to infer
accurate growth rates of the modes.
We have found that mode growth rate $y_2$ can be calculated more accurately
from a time behavior of $G_m$,
because it is expressed in terms of integrals over the radius and as
such it is much less affected by radial oscillations.
With this exception, we follow the same approach as described in
Section~\ref{SS:FixedBG} to determine mode types, growth rates and pattern
speeds.
These quantities for all models are tabulated in Table~\ref{T:ABCModes} for
reference.

In most of the simulations with fully dynamical GR, the BH responds
to the excitation of the $m=1$ mode by developing an outspiraling motion, as
described earlier in section~\ref{S:TimeEv}.
The position vector $\bar{r}_{BH}$ of the BH starts to rotate with approximately
constant angular velocity $\Omega_{BH}$, while the length of this vector grows
exponentially.
In order to characterize this motion and study it in the context of the
development of non-axisymmetric modes, we plot the time evolution of the BH
position vector length $r_{BH}/r_g$ and phase angle $\ji_{BH}$ on the $G_m-t$
and $\ji_m-t$ diagrams, respectively.
From these plots we can calculate the quantities $\Omega_{BH}/\Omega_0$ and
$y_2(BH)$ that can be directly compared to those of non-axisymmetric modes.
These quantities are also listed in the Table~\ref{T:ABCModes}.

Figure~\ref{F:StructAF1} shows the time evolution and radial profiles of the
amplitudes and Fourier angles of the dominant unstable modes for the
simulation AF1, which represents fully dynamical GR evolution of the initial
disk model A with an added $m=1$ density perturbation.
The top left panel contains the time evolution of $G_m$ for $m=1-4$ and the
normalized coordinate length of the BH position vector $r_{BH}/r_g$.
The diagram shows that the $m=1$ mode is the dominant one.
It also shows that the BH responds to the growth of the $m=1$ mode and the
distance from the BH to the origin grows exponentially at the same rate as
the dominant $m=1$ mode.

Panel (c) of Fig.~\ref{F:StructAF1} shows the time evolution of the Fourier
phase angles $\ji_m$ for $m=1,2,3$, measured at a fixed radial coordinate
location near the inner edge of the disk $r_-$.
The phase of the dominant $m=1$ mode after short initial
readjustment exhibits almost uniform linear growth. 
Readjustment of the mode happens because the added artificial perturbation
initially does not have the right shape of the mode and needs some time (less
than one orbital period) to readjust itself.
Similar behavior is observed in the corresponding Cowling simulation, but it
is less pronounced (see the $m=1$ phase angle during the first half orbit on
Fig.~\ref{F:PPStruct1}c).
Phases of the rest of the modes initially do not show uniform linear growth, but
as the amplitude of the $m=1$ mode increases, the higher order modes start to
align in phase with the dominant $m=1$ mode, most likely due to nonlinear
modes interaction.
Notice that the pattern speed of the $m=1$ mode calculated from this diagram
is lower compared to the Cowling case.

Panels (b) and (d) of Fig.~\ref{F:StructAF1} show the radial and angular
profiles of the dominant $m=1$ mode at $t=2.5\ t_{orb}$.
The location of $r_-$, $r_+$ and $r_c$ shown on the plot refers to the same
time.
Because the disk undergoes radial oscillations, special care must be taken when 
calculating the mode corotation radius $r_{cr}$, which is an important quantity
that characterizes non-axisymmetric modes.
To find $r_{cr}$, we solve an equation between the mode pattern speed
$\Omega_p$ and the disk angular velocity: $\Omega(r_{cr})=\Omega_p$.
The latter changes during the evolution of the disk, so strictly speaking the
value of $r_{cr}$ will also depend on time.
However, we have found that in all our simulations the change of the
profile of $\Omega$ in the course of the evolution is very small, therefore within
the measured accuracy the corotation radius is independent of time.
The values of $r_{cr}/r_c$, calculated in this way for all simulations on
a dynamical background are also listed in Table~\ref{T:ABCModes}.
For simulation AF1, $r_{cr}$ lies outside of $r_c$ with
$r_{cr}/r_c\approx1.17$.
Such a value of $r_{cr}/r_c$ is typical for a PP instability and comparable to
the values observed in previous Newtonian studies with a moving central object
and a massive self-gravitating disk~\cite{WTH94}.

  \begin{figure*}[!htp]
  \begin{center}
   \begin{tabular}{cc}
   \includegraphics[width=0.45\textwidth]{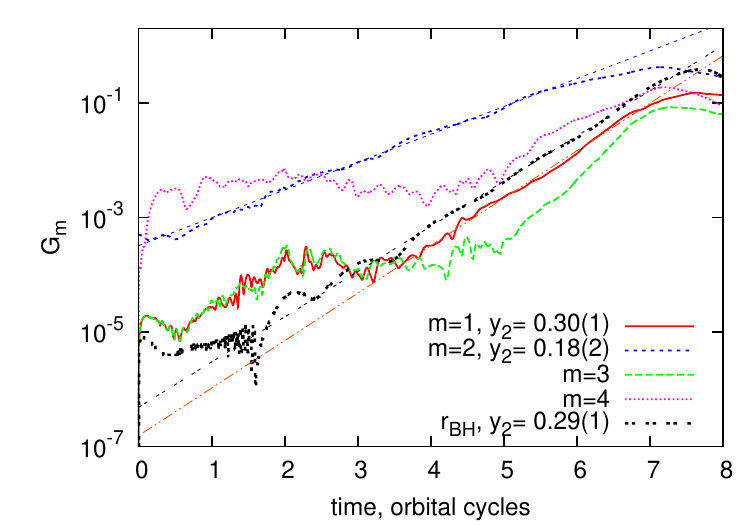} &
   \includegraphics[width=0.45\textwidth]{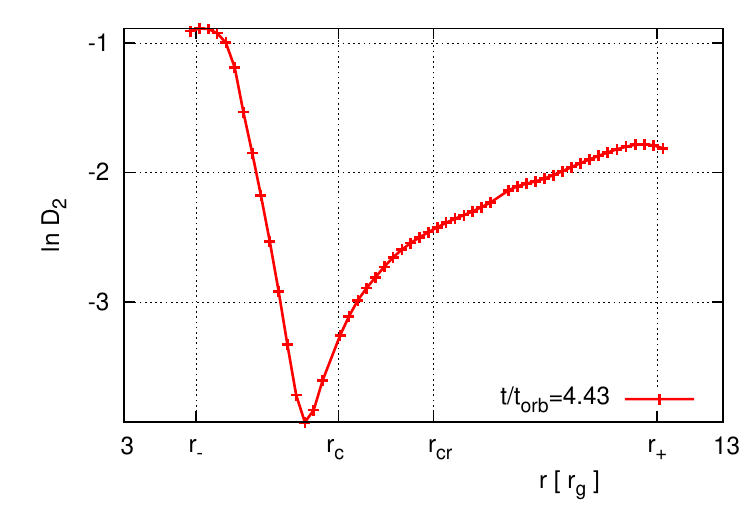} 
   \\
   (a) & (b)
   \\
   \includegraphics[width=0.45\textwidth]{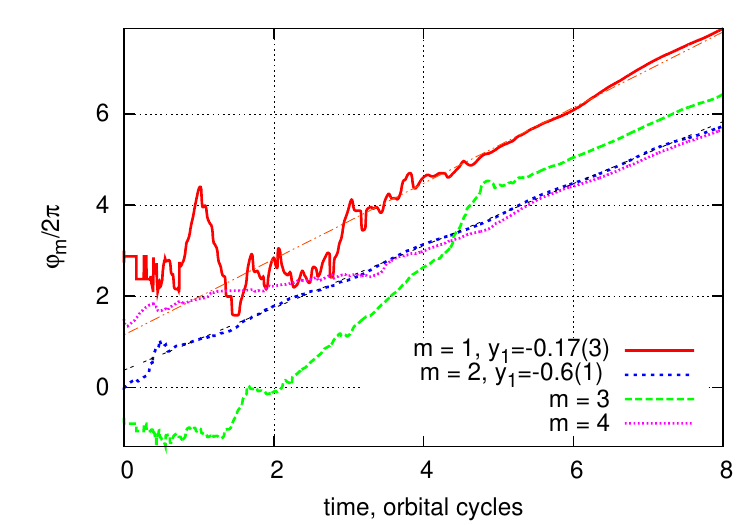} &
   \includegraphics[width=0.35\textwidth]{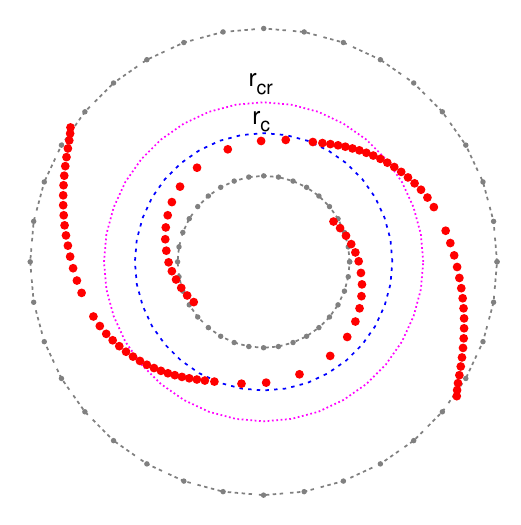}
   \\
   (c) & (d)
   \\
   \end{tabular}
  \caption{
    The behavior of the dominant non-axisymmetric modes and the radial
    character of the $m=2$ I-mode in the simulation AF2
    (see captions to Fig.~\ref{F:PPStruct1} for details).
  }
  \label{F:StructAF2}
  \end{center}
  \end{figure*}

Similar to the Cowling case, the radial profiles of $D_m$ and $\ji_m$ 
show structural features that are typical for a PP instability of (0,0)-type in
the Blaes and Hawley classification~\cite{BlaesHawley88}.
In particular, the mode amplitude displayed on Fig.~\ref{F:StructAF1}b is
higher near the edges of the disk and has a minimum near corotation $r_{cr}$.
The mode amplitude does not vanish anywhere in the disk, which
characterizes the mode as having type (0,0).
The profile of the Fourier angle of the mode, shown on
Fig.~\ref{F:StructAF1}d, has a specific S-shaped structure, that consists of a
trailing spiral pattern outside the corotation radius~\footnote{
We remind the reader that the disk rotates counterclockwise on all $\ji_m-r$
diagrams.}, a leading spiral pattern inside the corotation radius and a
short segment near $r_{cr}$ that connects the two spiral patterns.

However, compared to the Cowling approximation, the growth rate of the $m=1$ mode
is amplified by almost a factor of~$\sim1.5$, and there are reasons to believe
that the outspiraling motion of the BH is responsible for this amplification.
The mechanism which drives the unstable outspiraling motion of the BH is
similar to the one described in~\cite{Heemskerk92}.
The disk creates a hilltop potential, which has a maximum at the origin, so the
BH initially is located at the point of unstable equilibrium.
The BH can reduce its potential energy by converting it into kinetic energy of
orbital motion around the common center of mass (CM) of the disk+BH system.
Such orbital motion requires angular momentum which can be borrowed from the
disk through the development of a non-axisymmetric $m=1$ mode.
Because the orbital motion of the BH requires a compensating displacement of the
disk CM, removing angular momentum from the disk increases the amplitude of
the $m=1$ mode, which in this case is the PP-mode.

Next we consider the dominant non-axisymmetric modes that develop in the
simulation AF2, in which an $m=2$ density perturbation was added.
Figure~\ref{F:StructAF2} presents the time evolution and radial profiles of
$G_m$ and $\ji_m$ of these modes.
Comparing Fig.~\ref{F:StructAF2} to Fig.~\ref{F:PPStruct1}, which presents the
same set of diagrams for the Cowling simulation AC2, we can see that the type
of the $m=2$ mode in this simulation is quite different from the PP one
observed in the simulation AC2.
First, on the $D_m-r$ diagram in Fig.~\ref{F:StructAF2}b, the minimum
of $D_m$ lies inside $r_c$, unlike the case of the PP mode where such minimum is
located close to the mode corotation radius $r_{cr}$.
Second, the $\ji_m-r$ diagram of the $m=2$ mode in Fig.~\ref{F:StructAF2}d
consists only of trailing spirals and does not have a leading spiral pattern
inside corotation, as it would be the case for PP modes.
Finally, as Fig.~\ref{F:StructAF2}d illustrates, the mode phase angle makes an
abrupt turn by $\pi/2$ radians near $r_c$, which would be the case when the
disk were subjected to an elliptic (bar-like) deformation.

\begin{figure*}[!htp]
  \begin{center}
   \begin{tabular}{cc}
   \includegraphics[width=0.45\textwidth]{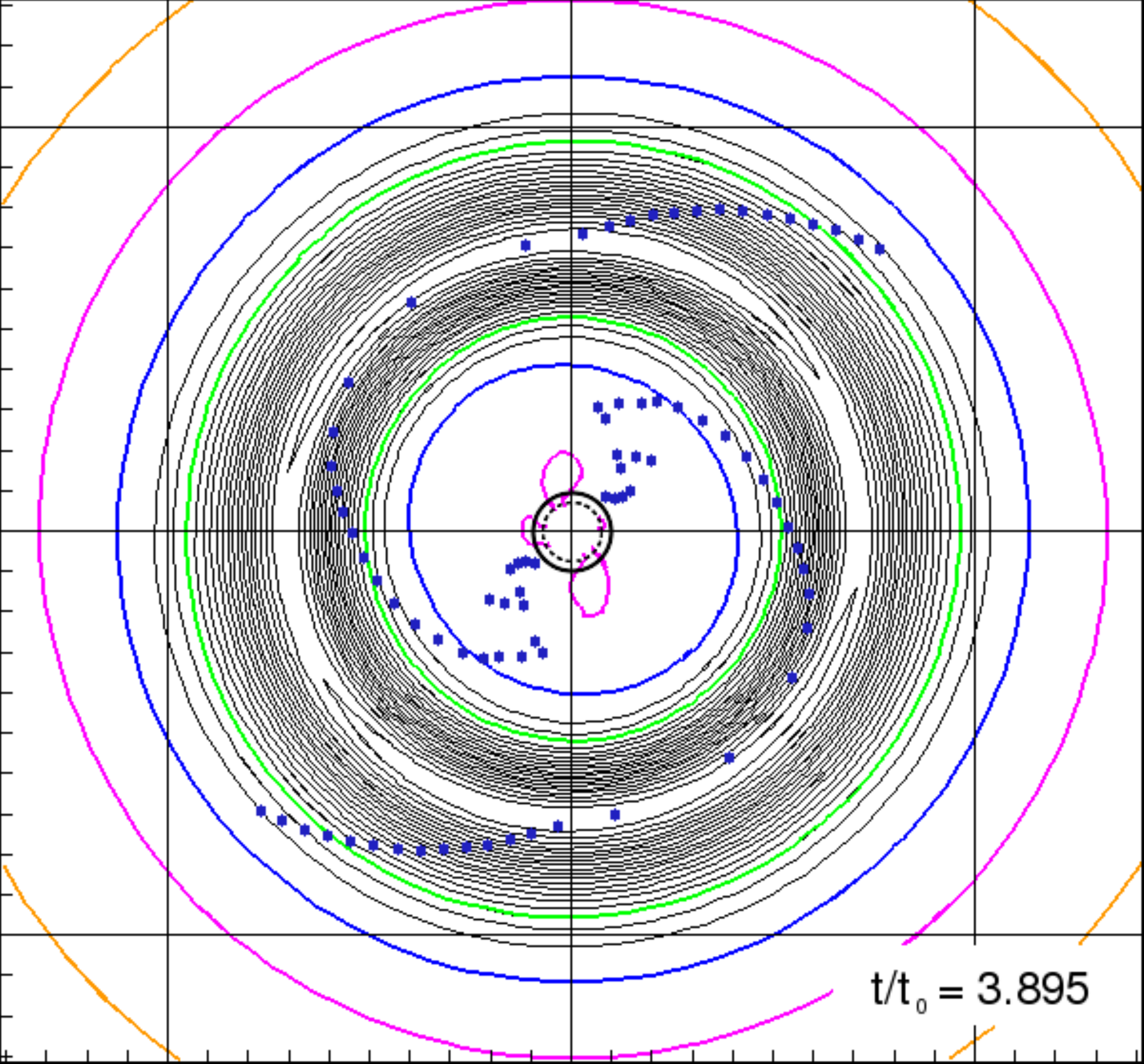} &
   \includegraphics[width=0.45\textwidth]{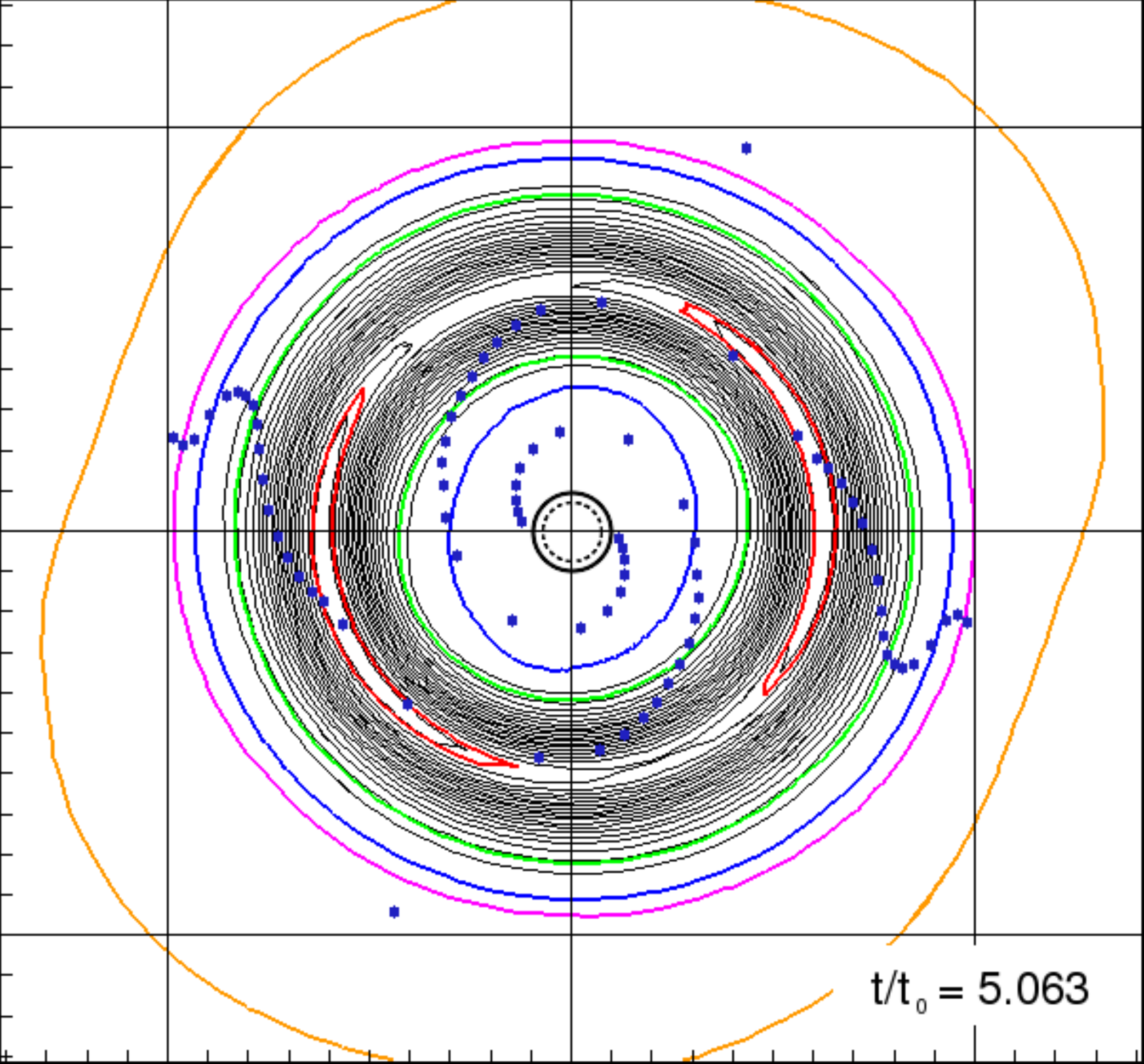} 
   \\
   \includegraphics[width=0.45\textwidth]{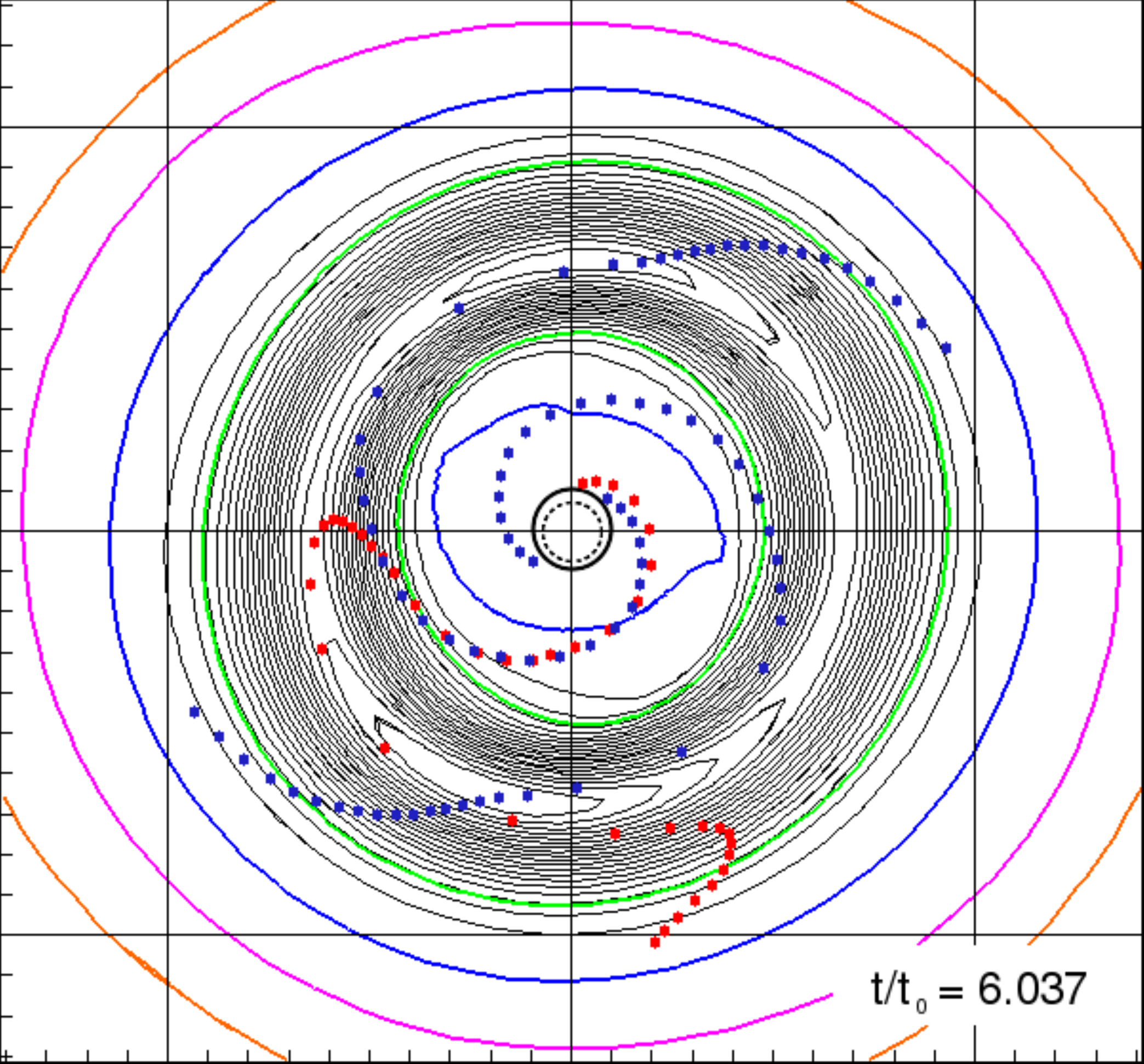} &
   \includegraphics[width=0.45\textwidth]{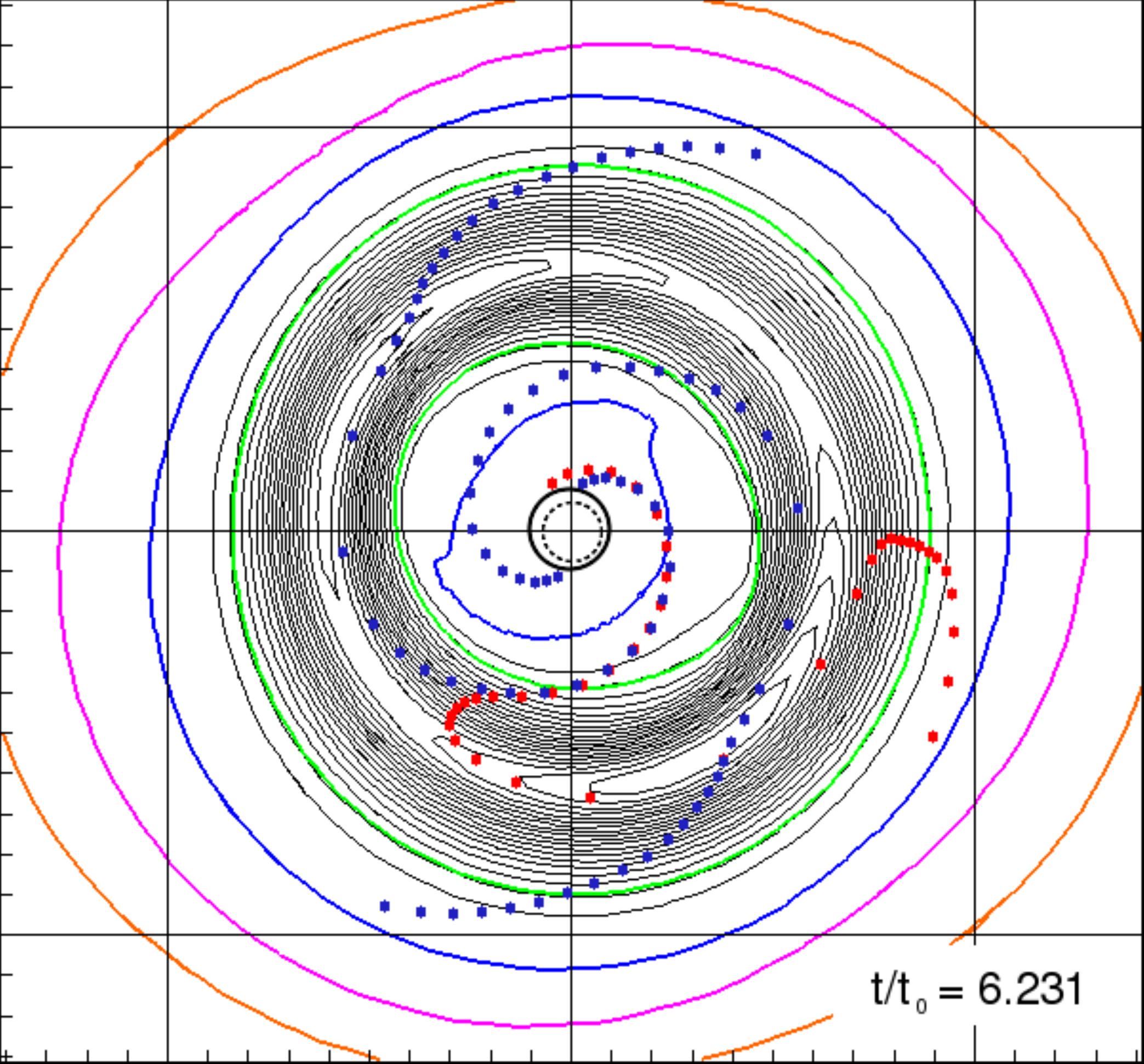} 
   \end{tabular}
  \caption{
    A sequence of four successive snapshots of the disk density in the
    equatorial plane, combined with the corresponding $\ji_m-r$
    diagrams for the $m=1$ and $m=2$ modes for the simulation AF2.
    The radial character of the $m=1$ and $m=2$ modes is represented by
    a sequence of red and blue dots resp.
  }
  \label{F:GRm2cnt}
  \end{center}
  \end{figure*}

Overall, this mode looks very similar to the so-called intermediate type
(I-type) modes found in earlier studies using Newtonian
gravity (see~\cite{GoodmanNarayan88a, ChristodoulouNarayan92,
  Christodoulou93, WTH94}). 
In particular,~\cite{WTH94} observed the I-modes in their 3D
Newtonian simulations of self-gravitating disks with various
disk-to-central object mass ratios $M_D/M_c$ and values of parameter
$T/|W|$ (see Section $4$ in~\cite{WTH94}). 
A subset of their models that develop the I-mode instability (namely E31 and
E32) have parameters $M_D/M_c=0.2$, $T/|W|\sim0.47$ and $4\pi
G\rho/\Omega_c\sim4$, which are comparable to those of our models A or B
(listed in Table~\ref{T:SGTParams}).
Notice that in this simulation the $m=1$ mode is also excited, as can be
inferred from Fig.~\ref{F:StructAF2}a.
The growth rate and pattern speed of this mode is the same as in simulation
AF1 (cf. Table~\ref{T:ABCModes}).
The $m=1$ mode development is again correlated in time with outspiraling
motion of the BH, which is apparent in Fig.~\ref{F:StructAF2}a.
The analysis of the mode character, similar to the one performed above
confirms that in this simulation the $m=1$ mode has the same (0,0) PP type as
in the simulation AF1.
It grows faster than $m=2$ mode, so that both modes become of comparable
amplitude by the end of the simulation.
The radial character of the two co-existing modes is also depicted on
Fig.~\ref{F:GRm2cnt}, which shows a sequence of snapshots of the $\ji_m-r$
diagrams at different times, superimposed with the disk density in equatorial
plane.

\begin{figure*}[!htp]
  \begin{center}
   \begin{tabular}{cc}
   \includegraphics[width=0.45\textwidth]{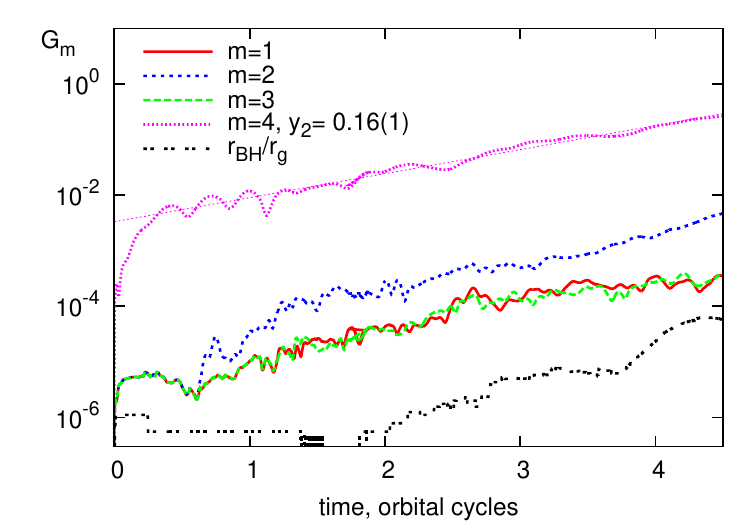} &
   \includegraphics[width=0.45\textwidth]{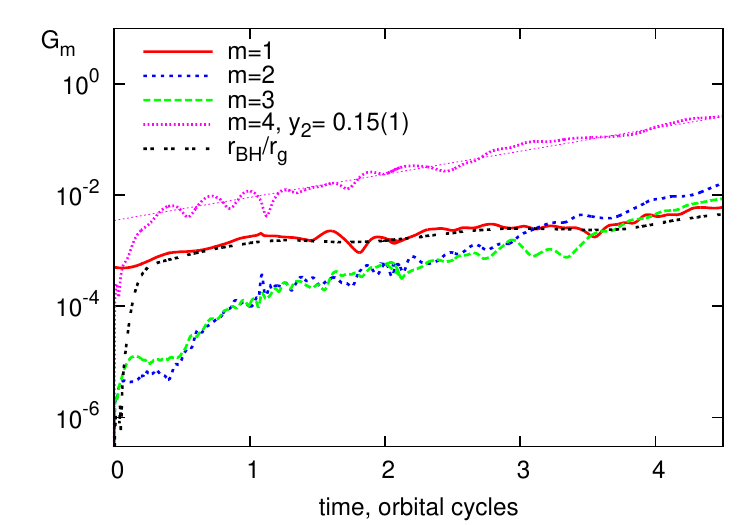} 
   \\
   CC & CC1 
   \\
   \includegraphics[width=0.45\textwidth]{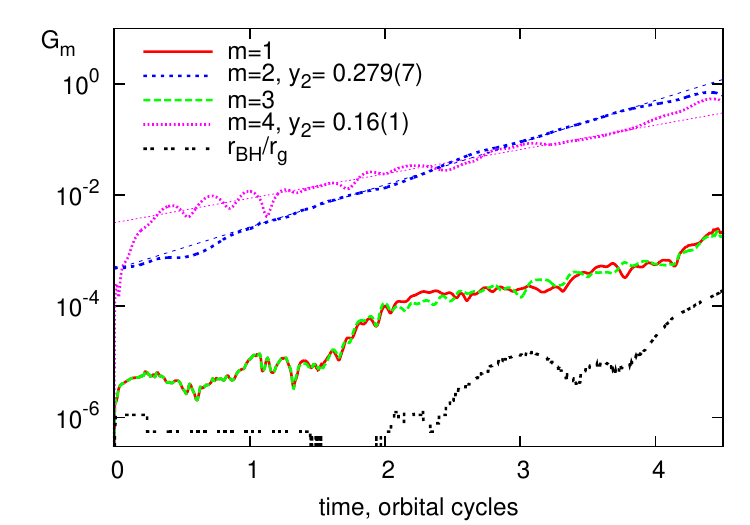} &
   \includegraphics[width=0.45\textwidth]{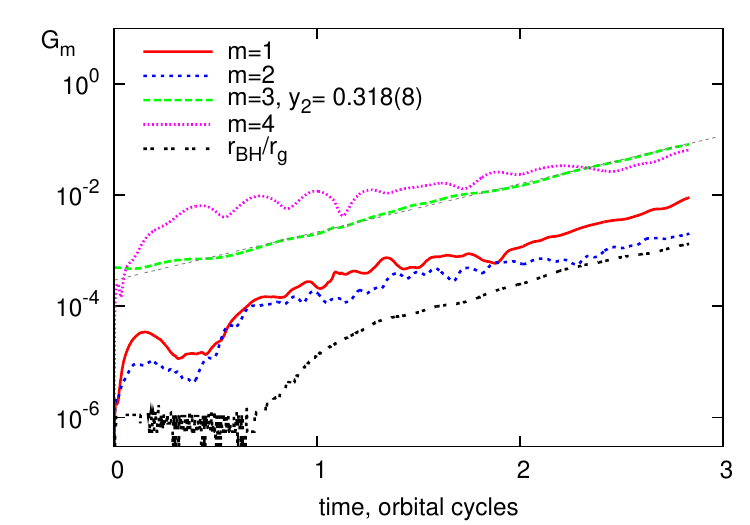} 
   \\
   CC2 & CC3 
   \end{tabular}
  \caption{
    $G_m-t$ diagrams for modes with $m=1-4$ in simulations with disk model C
  }
  \label{F:GmtC}
  \end{center}
  \end{figure*}

In the case of model C, we used artificial density perturbations with $m=1,2,3$
and observed the development of four unstable modes.
The same analysis that we have done before for models A and B reveals that
only the $m=1$ mode has PP type, while the modes with $m=2,3,4$ have I-type.
The I-modes with $m=3$ and $m=4$ represent triangular and square
deformations of the disk.
Such modes were previously observed in Newtonian simulations of narrow
self-gravitating annuli~\cite{ChristodoulouNarayan92}.\ Figure~\ref{F:GmtC} 
shows the time behavior of $G_m$ and $r_{BH}/r_g$
for simulations of model C with different added perturbations.
As can be seen from these plots, the fastest growing mode is the I-mode with
$m=3$, while the slowest one is the $m=1$ PP mode 
(cf.\ Table~\ref{T:ABCModes}).
The latter remains subdominant in all simulations, its growth does not
show clear exponential behavior and it is poorly correlated with the
motion of the BH.
This happens because the $m=1$ mode does not form a global coherent pattern.
In this case, the quantity $G_m$ does not correspond to the amplitude of a
global mode, but rather to some combination of local $m=1$ Fourier harmonics.
However, as soon as the amplitude of the $m=1$ mode reaches
$G_m\sim10^{-4}$, it shows a better correlation with the motion of the BH,
as can be seen on Fig.~\ref{F:GmtC} for simulations CC1 and CC3.

In model C, the $m=1$ mode starts growing when the other modes with
higher $m$ have already reached significant amplitudes ($G_m\sim
0.1$). Moreover, the growth rate of the $m=1$ mode has different values
depending on which of the higher-$m$ modes dominates the dynamics of
the disk. For example, in models CF1, CF2, and CF3, in which the modes
with $m=4$, $m=2$ and $m=3$ reach the largest amplitudes, the $m=1$
mode has the growth rates of $0.11$, $0.23$ and $0.28$,
respectively. Such behavior of the growth rates is likely to be a
result of non-linear interaction between these modes. In other words,
$m=1$ mode is never linearly independent from the $m>1$ modes, making
it hard to explore unambiguously the influence of the BH motion on the
growth rate of the $m=1$ mode.

Finally, we need to point out that the comparisons with the Cowling
simulations presented in this section should be taken with a grain of salt.
The initial perturbation introduced by the blending of two metrics in the
initial data changes the BH mass by a small amount ($\approx2.5\%$), which not
only drives the disk out of equilibrium, but also changes the
equilibrium configuration itself.
Therefore, strictly speaking, the disks in Cowling and full GR represent
different objects and cannot be directly compared.
As a result, these disks will have different evolutionary paths not only
due to dynamical GR effects, but also due to differences in BH masses.
However, because the latter is small, we believe that the differences in
evolution are mainly caused by the former, while the latter should not affect
the time evolution significantly.
For example, for models A and B which have different disk-to-BH mass ratios
($0.24$ vs $0.17$), we do not observe qualitative differences in time
evolution, and quantitative differences are small (e.g.\ the differences in
the growth rates are within $6\%$).
Therefore, we conclude that most of the differences between our Cowing and
full GR simulations are caused by the effects of GR.

\subsection{Gravitational Wave Detectability}
\label{S:GWDetectability}

  \begin{table}
  \begin{center}
  {\small
  \begin{tabular}{l|ccc}
  \hline \hline
   Source & LIGO & Adv. LIGO & ET\\
   \hline   
   A at $10$~kpc & $1$     & $1$ & $1$ \\
   C at $10$~kpc & $2$     & $1$ & $1$ \\
   A at $18$~Mpc & $1.7\times10^6$  & 7900  & 40  \\
   C at $18$~Mpc & $6.4\times10^6$  & 80000 & 150 \\
   \hline \hline
  \end{tabular}}
  \end{center}
  \caption{
    Minimal number of wave cycles needed for gravitational waves from
    non-axisymmetric instabilities to be detectable by LIGO, Advanced LIGO
    and Einstein Telescope (ET).
    It is assumed that the value of the amplitude of the instability is at
    least $D_m=0.1$ during this time.
    Estimates are given for the sources with parameters of models A and C,
    located at distances $10$~kpc and $18$~Mpc.
  }
  \label{T:Detect}
  \end{table}

As mentioned before in Section~\ref{S:TimeEv}, all of our disk models
are unstable to non-axisymmetric modes. Once formed, these modes start
growing exponentially until they reach a saturation regime due to
non-liner effects. This process is accompanied by a redistribution of 
the angular momentum of the disk until the profile of the specific 
angular momentum becomes steep enough for the disk to be stable to
non-axisymmetric instabilities (see related discussion
in~\cite{Zurek86a,Christodoulou93}). Before the angular momentum is
redistributed, and even after the disk becomes stable, the amplitude
of non-axisymmetric modes in the disk is likely to remain high
(possibly near the saturation level)~\cite{Zurek86a}. The presence of
non-axisymmetric deformations in the disk leads to emission of
potentially detectable gravitational radiation.  

Below we give estimates of the detectability of the GW signal from
saturated non-axisymmetric instabilities in our disk models. 
We make our estimates based on the Newtonian quadrupole formula for the
initial disk models with added $m=2$ mode with an amplitude $D_m=0.1$.
We calculate an approximate number of cycles that the instability needs to
remain at that amplitude in order for the emitted GW to be detectable.
These numbers are listed in Table~\ref{T:Detect} for an event at a
distance of $10$~kpc (our galaxy) and $18$~Mpc (a distance to the
Virgo cluster), for models A and C. The table shows that an event in
our galaxy will be detectable with the current LIGO detector even with
a single cycle of the non-axisymmetric mode.  
An event in the Virgo cluster, on the other hand, is unlikely to be  
detectable with the current LIGO detector, since it would
require unrealistically large number ($> 10^6$) of cycles. Second and
third generation detectors such as the advanced LIGO and the Einstein
Telescope can detect such events if non-axisymmetric modes persist for
$\sim 10^4-10^5$ and $\sim 40-150$ cycles, respectively. Finally, we
point out that it is currently unclear how long a non-axisymmetric
mode in a given disk model will persist in non-linear regime. This is
likely to depend on the details of non-linear mode properties,
accretion rate, magnetic fields and the thermodynamic state of the
disk matter~\cite{Zurek86a}.

\section{Conclusion}
\label{S:Conclusion}

In this paper we have explored non-axisymmetric instabilities in
self-gravitating disks around black holes (BHs) using
three-dimensional hydrodynamical simulations in full general relativity (GR).
We studied several moderately slender and slender models with disk-to-BH mass
ratio ranging from $0.11$ to $0.24$. The parameters of these models are listed in
Table~\ref{T:SGTParams}.

To obtain a self-consistent equilibrium disk model outside BH, we solve the
coupled system of Einstein constraints and Euler equations using an iterative
Green functions approach, implemented in the \rnsid{}
code~\cite{StF95:RNSModels}.
To avoid coordinate singularities, we transform the stationary initial
data outside the BH horizon from quasi-isotropic to non-singular
horizon-penetrating coordinates.
We set the data inside the BH horizon to the analytic Kerr-Schild solution
and smoothly blend it with the computed data outside the horizon.

We evolve the metric using a first-order form of the generalized harmonic
formulation of the Einstein equations with adaptive constraint damping.
The metric evolution equations are discretized on multiblock grids and solved
using 8th order finite difference operators.
We evolve the matter with relativistic hydrodynamics equations in
flux-conservative form, using a finite volume cell-centered discretization
scheme.
We use a $\Gamma$-law equation of state to model disk matter.
Our numerical approach makes extensive use of the curvilinear mesh adaptation
in order to achieve desired resolutions in different parts of the domain.

We did not observe the runaway instability in our models, which could have
developed due to the disk overfilling its toroidal Roche lobe in the process
of radial oscillations.
Such radial axisymmetric oscillations of the disk around its equilibrium state
are triggered in all of our simulations by an axisymmetric perturbation in the
metric due to the blending of two metrics inside and outside the BH horizon
(see Section~\ref{SS:Blending}).
Although our initial disk models are close to overfilling the toroidal Roche
lobe, the radial oscillations do not lead to the development of the runaway
instability within several initial orbital periods that we have simulated. 
However, this result may be specific to the particular model that we focused
on; we can not exclude the possibility that the runaway instability
develops in models with different initial parameters.

In all models that were explored we observed unstable non-axisymmetric
modes.
We have performed detailed analysis of these modes to determine their types,
growth rates, radial profiles and pattern speeds (see Table~\ref{T:ABCModes}).
For all simulations in the Cowling approximation we observe the development of
the Papaloizou-Pringle (PP) instability with $m=1-4$. 
In this case, the azimuthal number $m$ of the fastest growing mode 
depends on the disk slenderness: for moderately slender models A and B
such mode is $m=2$, while for more slender model C, it is $m=3$.
In the simulations in full GR, we observe two distinct types of
instabilities.
The unstable mode with $m=1$ has PP type, similar to the one observed in
Cowling case.
Unstable modes with $m>1$ become the intermediate modes (I-modes),
representing elliptical, triangular or square deformations of the disk.
In full GR, the fastest growing mode is $m=1$ in models A and B, and
$m=3$ in model C. 

In the full GR case, the development of the $m=1$ PP mode is accompanied by
an outspiraling motion of the BH. The distance from the BH center to
its initial position has the same growth as that of the $m=1$ mode
amplitude. We find that due to this motion, the growth rate of the
$m=1$ mode is amplified by a factor of $\approx1.5$ compared to the
Cowling case for massive models A and B. This amplification makes the
$m=1$ PP mode the fastest growing one in the models A and B, while in
the case of less massive model C, this mechanism is not as
efficient. The overall picture of the unstable modes in full GR is 
qualitatively similar to and consistent with the Newtonian
case~\cite{Kojima86b, WTH94, ChristodoulouNarayan92}.  

Evolution of non-axisymmetric instabilities in non-linear regime
will be associated with the emission of high-frequency gravitational
radiation.
In Table~\ref{T:Detect}, we give rough estimates of the detectability of this
radiation in terms of the number of cycles that a non-axisymmetric deformation
must persist in order to be detectable.
While even a single cycle of gravitational radiation from this deformation is
detectable if occurs in our galaxy, for more reasonable distances
such as Virgo cluster it is only detectable with Advanced LIGO, and only in
the case that the disk deformation persists for thousands of cycles (see
Table~\ref{T:Detect}).
It is currently unclear how long a non-axisymmetric deformation can persist in
non-linear regime.

Finally, we would like to point out limitations of our current simulations.
We use simplified initial disk models and do not include realistic
microphysics, neutrino cooling and magnetic fields.
Future studies of non-axisymmetric instabilities should take into account
these effects, as well as consider larger set of parameters, such as
non-constant angular momentum distribution, various disk sizes, masses, BH
spins etc.
The properties of the disk in the nonlinear regime, such as the persistence of
non-axisymmetric structures in realistic disk models should also be addressed.

\section{Acknowledgements}
\label{S:Acknowl}

The authors would like to thank our colleagues
Eloisa Bentivegna, Peter Diener, Juhan Frank,
Tyler Landis, Frank L\"offler, Luis Lehner, Christian D. Ott, Jorge Pullin,
Jian Tao, Manuel Tiglio, and Joel Tohline for valuable discussions and ideas.
We also thank Yasufumi Kojima for providing his data for comparison with our
results (Fig.~\ref{F:Kojima86b}).
This work is supported by the NSF grants 0721915 (Alpaca), 0904015
(CIGR), and 0905046/0941653 (PetaCactus).
The simulations were performed using the supercomputing resources Ranger and
Lonestar at TACC via the NSF TeraGrid, and Queenbee at LONI\@.
We also used the PetaShare infrastructure to store the data from our
simulations.
N. S. acknowledges the hospitality of the University of T\"ubingen, and O.K.
acknowledges the hospitality of AEI numerical relativity group.

\appendix

\section*{Appendix}\label{Apx}

\subsection{Transforming initial data to horizon-penetrating coordinates}
\label{A:Transform}

Here we present the transformation of the stationary axisymmetric initial data
in quasi-isotropic coordinates to time-independent horizon-penetrating
coordinates which is used in Section~\ref{S:IDSetup} in order to remove 
the degeneracy at the BH horizon.
In our initial data, the metric is given in general form~\eqref{Eq:AxisymMetric}
and represents an axisymmetric deformation of a Schwarzschild BH by a massive
equilibrium torus.
The sought transformation has to satisfy the following requirements:
\begin{itemize}
\item the metric in the new coordinates is time-independent;
\item the metric does not have pathologies (degeneracy or divergence) at the
      event horizon;
\item the three-metric on $t=\textrm{const.}$ foliation is positive definite (i.e.\
      the $t=\textrm{const.}$ foliation is spacelike).
\end{itemize}      

We build our transformation by analogy with the transformation from isotropic
Boyer-Lindquist coordinates~\cite{Brandt94a} to horizon-penetrating
Kerr-Schild coordinates~\cite{KerrSchild09} of Schwarzschild spacetime (see
also~\cite{Takahashi07a} for a general case of rotating BH).
In this case, the line element has the following form:
$$
  ds_{is}^2 = -\left(\frac{1-m/2r_*}{1+m/2r_*}\right)^2 dt^2 
              + \psi^4(dr_*^2 + r_*^2 d\Omega^2) ,
$$
where 
$m$ is the BH mass, 
$r_*$ is the isotropic radius, 
$d\Omega^2 \equiv d\theta^2 + \sin{\theta}^2 d\ji^2$ is the solid angle element, and 
$\psi \equiv 1+\frac{m}{2r_*}$ is the conformal factor.
At the event horizon $r_{*,h} = m/2$ the determinant of the isotropic metric
is zero.

In the horizon-penetrating Kerr-Schild coordinates, the line element will be:
$$
  ds_{ks}^2 = -(1-H) d\bar{t}^2 
              + 2H d\bar{t} dr 
              + (1+H)dr^2 
              + r^2 d\Omega^2
$$
where $H \equiv 2m/r$ and
$r \equiv r_*\left(1+m/2r_*\right)^2$ is the Schwarzschild radial coordinate.

The Jacobian of the transformation from the isotropic $(t, r_{*},\theta,\ji)$ to
the horizon-penetrating coordinates $(\bar{t},r,\theta,\ji)$ has the following
form:
$$
\frac{D(t, r_*,\theta,\ji)}{D(\bar{t},r,\theta,\ji)} = 
\left[\begin{array}{cccc}
1   &  -\frac{H}{1-H} &  0 & 0 \\
0   &   \frac{r_*}{r\sqrt{1-H}} &  0 & 0 \\
0   &   0           &  1           & 0 \\
0   &   0           &  0           & 1 \\
\end{array}\right].
$$

The metric in the new coordinates remains independent of time, which is also true
for an arbitrary transformation with the following Jacobian:
$$
\frac{D(t, r_*,\theta,\ji)}{D(\bar{t},r,\theta,\ji)} = 
\left[\begin{array}{cccc}
1   &   f(r,\theta) &  h(r,\theta) & 0 \\
0   &   g(r,\theta) &  p(r,\theta) & 0 \\
0   &   0           &  1           & 0 \\
0   &   0           &  0           & 1 \\
\end{array}\right],
\qquad \textrm{(A1)}
$$
where the functions $\{f,g,h,p\}$ do not explicitly depend on time and
are only constrained by the regular Jacobian integrability conditions:
$$
\frac{\partial f(r,\theta)}{\partial\theta} = \frac{\partial h(r,\theta)}{\partial r}, 
\qquad                                                                 
\frac{\partial g(r,\theta)}{\partial\theta} = \frac{\partial p(r,\theta)}{\partial r}.
$$

For the case of a general axisymmetric spacetime, the metric in quasi-isotropic coordinates 
is given by:
$$
\bf{g}_{is} \equiv 
  \left[\begin{array}{cccc}
    g_{tt}             &  0           & 0               & -\omega g_{\ji\ji} \\
    0                  &  e^{2\alpha} & 0               & 0 \\
    0                  &  0           & r^2 e^{2\alpha} & 0 \\
    -\omega g_{\ji\ji} &  0           &                 & g_{\ji\ji} \\
  \end{array}\right] 
$$
where 
$g_{tt}\equiv-\lambda^2 + \omega^2 g_{\ji\ji}$ and
$g_{\ji\ji}\equiv B^2\lambda^{-2}r_*^2 \sin{\theta}^2$. 
We can select the set of functions $\{f,g,h,p\}$ in the Jacobian (A1)
above to construct a transformation from the quasi-isotropic to
horizon-penetrating coordinates, which satisfies the above requirements for an
arbitrary stationary axisymmetric deformation of Schwarzschild if we choose:
$$
    f(r,\theta) = 1-1/\lambda^2(r,\theta), \quad
    g(r,\theta) = e^{-\alpha(r,\theta)}/\lambda(r,\theta).
$$
Since the metric potentials $\lambda(r_*,\theta)$ and $\alpha(r_*,\theta)$
are known as functions of $r_*$ and not $r$, we need to express the new radial
coordinate $r$ in terms of $r_*$.
The required relations in the following differential form are obtained by
inverting the Jacobian (A1):
$$
\left(\frac{\partial r_*}{\partial r}\right)_{\theta=\textrm{const.}} = \frac{1}{g(r_*,\theta)}, \quad
\left(\frac{\partial r_*}{\partial \theta}\right)_{r_*=\textrm{const.}} = 0.
$$
These need to be integrated along the radial coordinate from $r_{*,h}$ to $r_*$
for each $\theta$:
$$
 r(r_*,\theta) - r_h = \int_{r_{*,h}}^{r_*}\frac{d\zeta}{g(\zeta,\theta)} 
                     = \int_{r_{*,h}}^{r_*}\lambda(\zeta,\theta)e^{\alpha(\zeta,\theta)}d\zeta.
$$
where $r_h$ is the radius of the event horizon in new coordinates.
The remaining two unknown functions $h$ and $p$ can be calculated by 1D
integration of the Jacobian integrability conditions:
\begin{align*}
  h(r,\theta) &= \int_{r_h}^r dr'
                      \frac{\partial f(r',\theta)}{\partial\theta}
               = 2 \int_{r_{*,h}}^{r_*} 
                  \frac{\lambda'_{\theta}(\zeta,\theta) 
                        e^{\alpha(\zeta,\theta)}}
                       {\lambda^2(\zeta,\theta)}
                  d\zeta,
              \\
  p(r,\theta) &= \int_{r_h}^r dr'
                      \frac{\partial g(r',\theta)}{\partial\theta}
              \\
              &= - \int_{r_{*,h}}^{r_*} \left(
                   \alpha'_{\theta}(\zeta,\theta)
                   +
                   \frac{\lambda'_{\theta}(\zeta,\theta)}
                        {\lambda(\zeta,\theta)}
                 \right) d\zeta.
\end{align*}

After the transformation, the metric in the new horizon-penetrating
coordinates has the following form:
$$
\bf{g}_{ks} = 
  \left[\begin{array}{cccc}
    g_{tt}  &  f g_{tt} & h g_{tt}                                      & -\omega g_{\ji\ji} \\
    \dots   &  e^{2\alpha}g^2+g_{tt}f^2 & f h g_{tt}+e^{2\alpha}g p     & -\omega f g_{\ji\ji} \\
    \dots   &  \dots    & h^2 g_{tt}+e^{2\alpha}(p^2+r_{*}^2)           & -\omega h g_{\ji\ji} \\
    \dots   &  \dots    & \dots                                         & g_{\ji\ji} \\
  \end{array}\right] 
$$
where ellipsis indicate matrix elements which can be filled in by symmetry.
In the limit of $r\to r_h$, the functions $f$ and $g$ tend to infinity, while $h$ and
$p$ vanish. The resulting metric at the horizon remains finite and non-degenerate:
$$
\lim_{r\to r_h}\bf{g}_{ks} = 
  \left[\begin{array}{cccc}
   0 & 1 & 0                  &  0 \\
   1 & 2 & C                  &  0 \\
   0 & C & e^{2\alpha}r_{*}^2 &  0 \\
   0 & 0 & 0                  & g_{\ji\ji} \\
  \end{array}\right]
$$
where $C\equiv\lim_{r\to r_h}{[f h g_{tt} + e^{2\alpha}g p]}$ is a finite
constant.

\subsection{Stable evolution of a uniformly rotating polytrope}
\label{A:RotatingPolytrope}

  \begin{table}
  \begin{center}
  {\small
  \begin{tabular}{l|c|c}
   \hline
                                           & geom.  & CGS \\ 
   \hline                     
   polytropic scale $K$                    & 100    & $1.46\cdot 10^5\;\cm^5\gr^{-1}\sec^{-2}$   \\
   polytropic index $\Gamma$               & 2      & 2                                          \\
   central rest-mass density $\rho_c$      & 0.001  & $6.17\cdot 10^{14}\ \gr\ \cm^{-3}$         \\
   ratio $R_p/R_e$                         & 0.7    & 0.7                                        \\
   ADM mass $M$                            & 1.49   & 1.49                                       \\
   rest mass $M_0$                         & 1.59   & 1.59                                       \\
   equatorial radius $R_e$                 & 12.32  & $1.823\cdot10^6\;\cm$                      \\
   angular momentum $J$                    & 1.32   & $1.16\cdot10^{49}\;\gr\ \cm^2\sec^{-2}$    \\
   normalized ang. mom. $J/M^2$            & 0.59   & 0.59                                       \\
   kinetic / binding en. $T/|W|$           & 0.0748 & 0.0748                                     \\
   angular velocity $\Omega$               & 0.0215 & 4300 $\sec^{-1}$                           \\
   Keplerian angular velocity $\Omega_K$   & 0.0286 & 5801 $\sec^{-1}$                           \\
   rotational period $P$                   & 292.1  & $1.44\cdot10^{-3}\ \sec$                   \\
   \hline
  \end{tabular}}
  \end{center}
  \caption{Physical parameters of the uniformly rotating polytropic star used
           for the code tests, in geometrized and CGS units, where:
           $R_p/R_e$ is the ratio of the polar to equatorial radii of the star,
           $J/M^2$ is its angular momentum, normalized with the square of the 
           ADM mass of the star $M$, and
           $T/|W|$ is the ratio of the kinetic to binding energy of the star.
          }
  \label{T:RNSParams}
  \end{table}

In this appendix, we present results of testing the time evolution of a
uniformly rotating polytropic star for numerical stability and convergence.
In our tests, we use geometrized units based on the solar mass, in which
$G=c=\MSun=1$.\ 
The parameters of the star in the geometrized and CGS
units are summarized in Table~\ref{T:RNSParams}.
We use a thirteen-block cubed sphere system that was described in~\cite{Zink08}
(see Fig.~\ref{F:MyPatchsystems}b and the related discussion in
Section~\ref{SS:Multiblock} above).\ 
For the current setup, we fix the sizes of the blocks by choosing $r_0=2.5$,
$r_1=9$ and $r_2=14$ (see Section~4 of~\cite{Zink08} for definition of $r_0$,
$r_1$ and $r_2$), with each block having an equal number of $N^3$ grid cells.
The sizes of the domain and its blocks are selected in such a way that the
star occupies $\approx90\%$ of the entire domain in radial equatorial
direction, and the inner seven blocks of the system lie inside the star.
This setup allows to test how much the accuracy and convergence of our
numerical scheme are affected by interpolation errors on the interblock
boundaries which thread the bulk of the star.

The stability of the numerical scheme for evolving the spacetime metric
depends on the numerical dissipation parameter
$\epsilon$~\cite{diener-2007-32} and the constraint damping coefficients
$\kappa,\ \gamma_2$ (see Section~\ref{S:ConstrDamping}).
In general, higher values of numerical dissipation restrict the
timestep, while lower values are undesirable because they do not
provide enough suppression of the numerical noise, which needs to be
dissipated for stability~\cite{Calabrese:2003vx}.
For the current setup, we choose $\epsilon = 0.2$ and 
$\kappa = \gamma_2 = 0.1$. 
Values of the constraint damping parameters higher than $\approx0.5$
lead to numerical instabilities in our simulations of stars.

  \begin{figure}[!htp]
  \begin{center}
  \begin{tabular}{c}
  \includegraphics[width=0.45\textwidth]{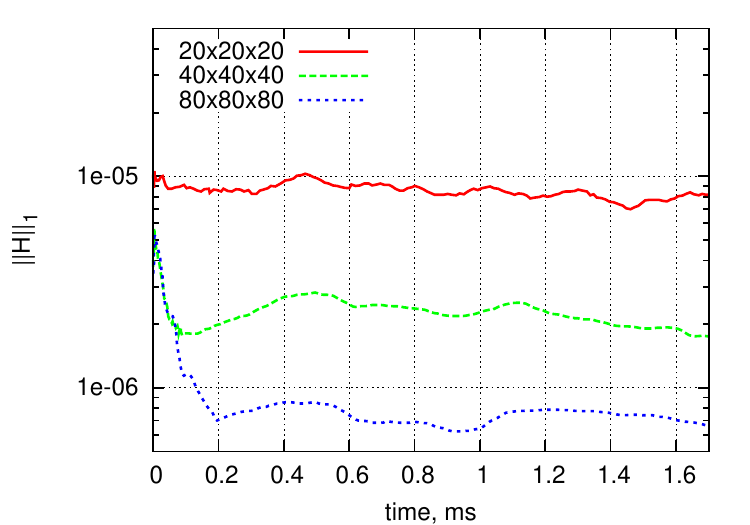}    \\
  \includegraphics[width=0.45\textwidth]{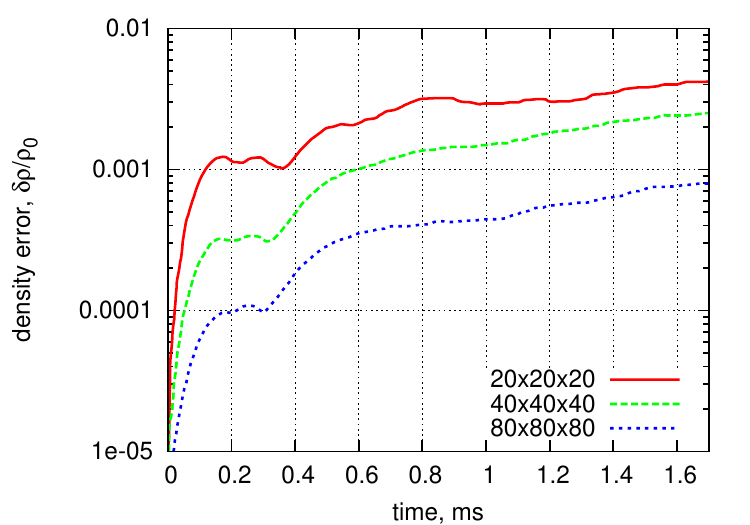} 
  \end{tabular}
  \caption{
    Time evolution of $L_1$ norms of the Hamiltonian constraint (top
    panel) and density solution error $\delta\rho/\rho_{max}(0)$
    (bottom panel) for three different resolutions.
  }
  \label{F:RNSErrorConv}
  \end{center}
  \end{figure}

  \begin{figure}[!htp]
  \begin{center}
  \begin{tabular}{c}
  \includegraphics[width=0.45\textwidth]{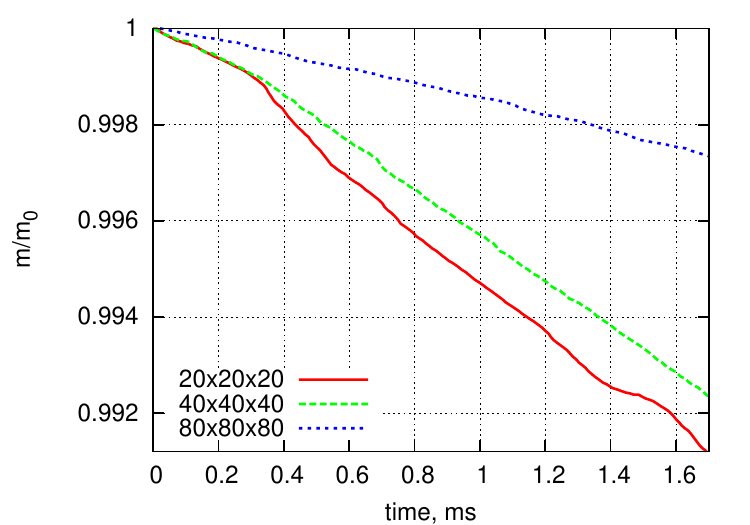} \\
  \includegraphics[width=0.45\textwidth]{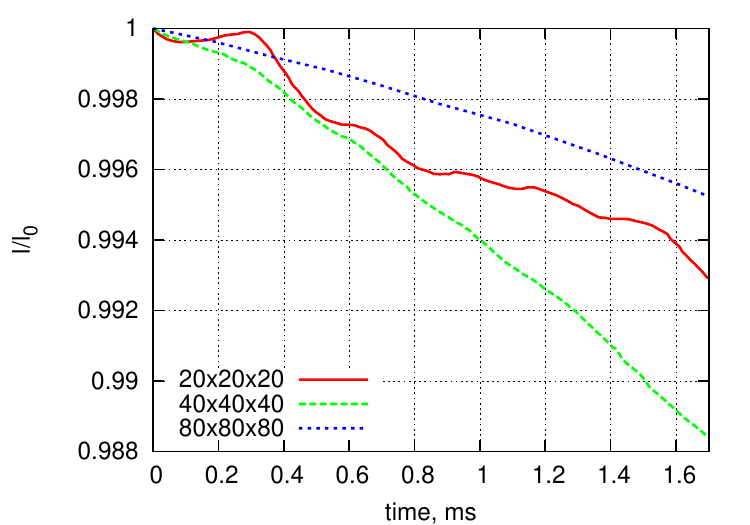}
  \end{tabular}
  \caption{
    Time evolution of the total rest mass (top panel) and the total
    angular momentum (bottom panel) for three different
    resolutions. Both quantities are normalized to their values at
    $t=0$. 
  }
  \label{F:RNSConservedConv}
  \end{center}
  \end{figure}

Initial data for the time evolution is generated by the \rnsid{}
code~\cite{StF95:RNSModels}, which uses the KEH(SF)
method~\cite{KEH89,StF95:RNSModels} to produce equilibrium models of
stationary rotating relativistic stars. 
Since \rnsid{} is a 2D solver which uses its own grid that is different
from the 3D multiblock grid of our time evolution code, we interpolate
the data from 2D grid to the 3D multiblock grid using 4-th order
Lagrange interpolation.
Also, because the variables in the GH formulation contain first derivatives
of the metric and because the resolution on the 2D grid is usually much higher
than on the multiblock 3D grid, we perform numerical differentiation on the 2D
grid. 
The resulting derivatives are then interpolated onto the 3D grid. 
Note that the interpolation procedure is not consistent with the
Einstein constraint equations, and hence produces numerical noise.

The system is evolved up to $t=350$, which corresponds to 1.73 ms, or
$10$ dynamical timescales of the star\footnote{The dynamical time
  $t_D$ is defined as $t_D=R_e\sqrt{R_e/M}$, where $R_e$ is a proper
  equatorial circumferential radius, and $M$ the ADM mass of the
  star. It corresponds to the inverse of the orbital frequency
  $\Omega=\sqrt{M/R_e^3}$ at $R_e$.}. In vacuum regions outside the
star, we use an artificial atmosphere, which has density of 
$\rho_{\textrm{atm}}=10^{-7}\rho_\mathrm{max} (0)$, where $\rho_\mathrm{max}
(0)$ is the maximum density at $t=0$. 
If during the evolution the density in a cell drops down below
a threshold value set to $\rho_{\textrm{thr}} = 2\ \rho_{\textrm{atm}}$,
the density in this cell is reset to the artificial atmospheric value. 
To estimate the accuracy of our code, we
have performed a convergence study using three different resolutions
with $N^3 = 20\times20\times20$, $40\times40\times40$ and
$80\times80\times80$ grid  points in each block. 
We have analyzed various integral norms of the errors in all evolved
variables, including 50 spacetime variables, 5 primitive variables and 5
conserved variables. 
We have also analyzed integral norms of the Hamiltonian and momentum constraints,
as well as the behavior of conserved integral quantities such as total rest
mass and total angular momentum.

In all cases, we observe the expected 2-nd order convergence.
As an example, Fig.~\ref{F:RNSErrorConv} (top panel) shows the time
evolution of the $L_1$ norm of the normalized density deviation
$\delta\rho \equiv \left[\rho(t)-\rho(0)\right]/\rho_{\textrm{max}}(0)$ for
the three resolutions.
Due to accumulation of truncation errors, this deviation exhibits a steady
growth (modulo small variations because of oscillations of the star)
throughout entire evolution. 
The deviation for $N=40$ is larger than that for $N=80$ by a factor of
$\approx4$, which is a clear signature of 2-nd order convergence.
However, the deviation for $N=20$ is larger than that for $N=40$ by a smaller
factor of $\approx1.5$, which means that the resolution $N=20$ is insufficient
for achieving a convergent regime.
A similar convergent behavior is observed for integral norms of the
deviations of all of the rest of the variables.

Figure~\ref{F:RNSErrorConv} (bottom panel) shows the plot of the $L_1$ norm of
the Hamiltonian constraint violation as a function of time.
This quantity is not zero at $t=0$, since initial conditions were
interpolated from the 2D grid and interpolation errors were introduced.
However, because of the constraint damping scheme, the Hamiltonian
constraint violation significantly drops for medium and high
resolutions within the first $0.2\ \ms$. 
During subsequent evolution the value of the Hamiltonian constraint remains
stable and clearly shows 2-nd order convergence with resolution, i.e.\ 
the values of the Hamiltonian constraint for $N=20$, $40$ and $80$ are
in an approximate ratio $16:4:1$.
Momentum constraints show a similar behavior.

Figure~\ref{F:RNSConservedConv} demonstrates time evolution of the total rest
mass (upper panel) and total angular momentum (lower panel) of the star.
In our numerical simulations these quantities are not conserved mostly due to
interpolation errors on the interblock boundaries that pass through the bulk
of the star.
By the end of the simulation, for $N=20$, $40$ and $80$, the total rest mass
decreases by $0.88$, $0.84$ and $0.25$ percent, while the total angular
momentum decreases by $0.07$, $0.12$ and $0.05$ percent.
These numbers show that the smallest necessary resolution for the convergent
regime is $N=40$, which amounts to $\approx70-100$ points across the star.

\subsection{Fundamental modes of a TOV star}
\label{A:TOVFreq}

  \begin{figure*}[!htp]
  \begin{center}
  \begin{tabular}{cc}
  \includegraphics[width=0.45\textwidth]{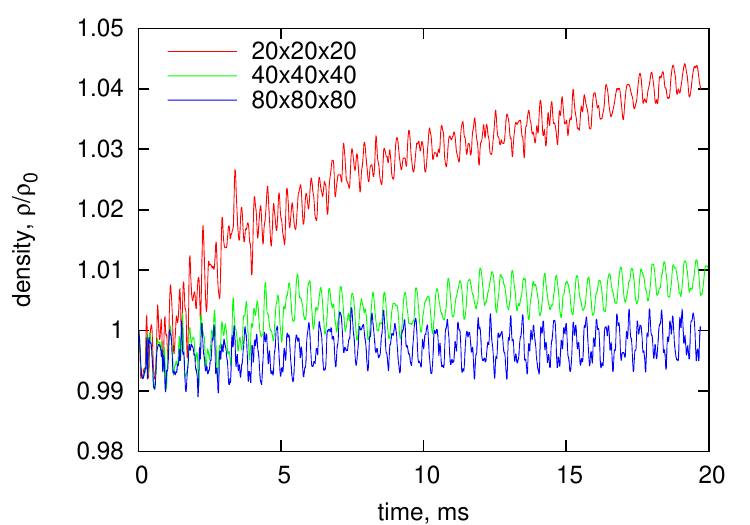}  &
  \includegraphics[width=0.45\textwidth]{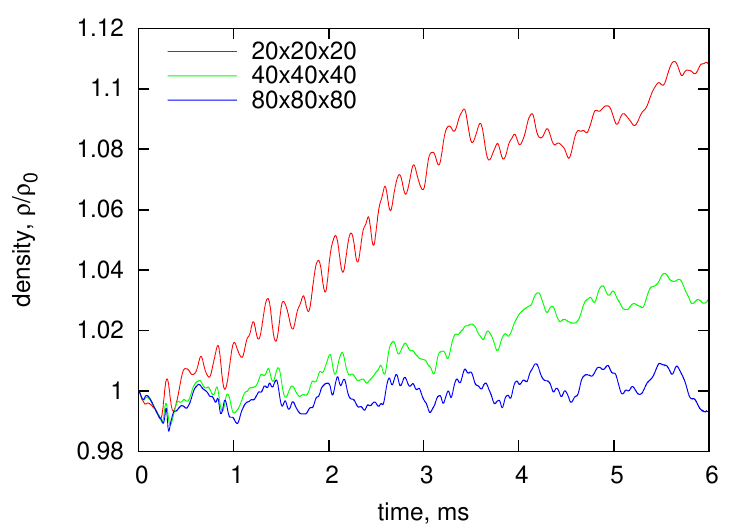} 
  \end{tabular}
  \caption{
    Time evolution of the density $\rho_c(t)$ at the center of a TOV star,
    normalized by its initial value $\rho_0$, for three different
    resolutions. Left panel: Cowling approximation. Right panel: fully
    dynamical GR case.
  }
  \label{F:TOVCentralDensity}
  \end{center}
  \end{figure*}

  \begin{figure}[!htp]
  \begin{center}
  \begin{tabular}{c}
  \includegraphics[width=0.45\textwidth]{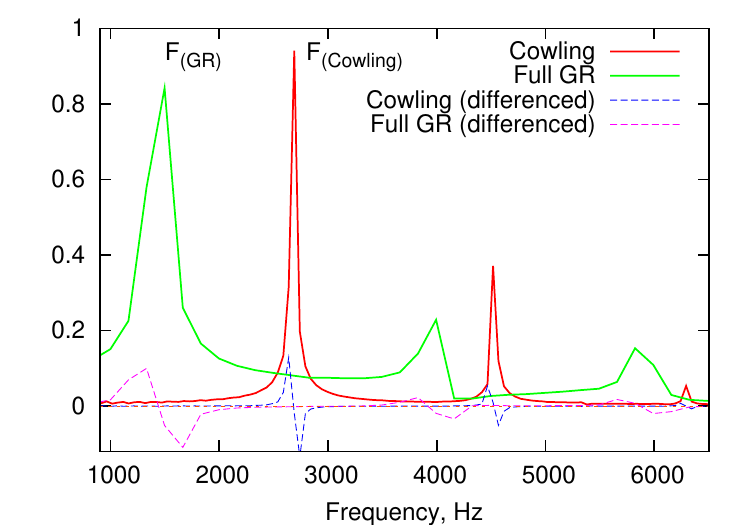}    \\
  \end{tabular}
  \caption{
    Power spectrum of the density oscillations $\rho_c(t)$ at the center 
    of a TOV star for the highest resolution simulations in Cowling
    approximation (red solid line) and in full GR (green solid line).
    Also shown are derivatives of the power spectrum with respect to the 
    frequency, obtained using the central finite-differencing scheme.
    The derivatives allow to localize peaks in the power spectrum more
    accurately.
    Vertical axis has arbitrary units.
  }
  \label{F:TOVFourier}
  \end{center}
  \end{figure}

As another test of the coupling between the GR and hydro parts of the code, we
evolved a Tolman-Oppenheimer-Volkoff (TOV) solution on a seven-block system, and
measured the frequencies of its fundamental oscillations both in the Cowling
approximation and in full GR.
In these tests, we use geometrized units in which $G=c=\MSun=1$.
We choose a star with $\Gamma=2$, $K=100$ and the value of rest-mass
density in the center $\rho_c=1.28\cdot10^{-3}$. 
These parameters produce a TOV star with gravitational mass $M=1.4$ and
circumferential radius $R_e=9.8$.
This system has already been extensively studied in the literature and used
for the assessment of relativistic hydrodynamical codes
(e.g.~\cite{Font99,Font01b}).
The seven-block cubed sphere system that we used (see
Fig.~\ref{F:MyPatchsystems}a) has the outer radius $R=12$, which makes the
star occupy $82\%$ of the domain in radial direction and leaves extra room for
small oscillations.
The size of the cubical block in the center is $a=4.8$, placing it completely
inside the star. 
The bulk of the star is therefore threaded by interpolation boundaries between
the blocks.
The cubical block contains $N^3$ volume cells, and the outer
blocks have $N^2\times(2N)$ cells.
For the tests, we used resolutions $N=20$, $40$ and $80$, which roughly
correspond to $40$, $80$ and $160$ points across the star.

To observe and measure the fundamental mode, we artificially add a small
initial perturbation, roughly corresponding to the shape of the mode:
$$
\frac{\delta\rho}{\rho} = A\cos{\frac{\pi r}{2 R_e}},
$$
The amplitude was chosen to be $A=0.005$. 
Fig.~\ref{F:TOVCentralDensity} displays the resulting oscillatory behavior of
the rest-mass density in the center of the star for three different resolutions. 
Left and right panels correspond to fixed (Cowling approximation) and
dynamical spacetime geometries, respectively. 
Oscillations of the density are accompanied by a secular drift, which reflects
accumulation of truncation errors and converges away with resolution at
approximately second-order convergence rate.
Conserved quantities such as the total rest mass and the total angular
momentum (not shown) also exhibit the second-order convergence, as
expected.
In particular, for the Cowling approximation case, the simulation continued up
to $20\ \ms$, and the final rest mass is conserved up to $8.3$, $3.7$ and
$1.1$ percent for resolutions with $N=20$, $40$ and $80$.
For the full GR case, the simulation continued for $6\ \ms$ and the rest mass
is conserved up to $3.2$, $1.3$, $0.4$ percent for the same three resolutions.
This shows that the rate of the mass loss in Cowling and full GR simulations
is approximately the same, as expected.
Because the bulk of the star is threaded by interpolation boundaries between
the blocks, the mass loss is significantly higher in this setup than in case
of a regular Cartesian grid, where we normally observe that the mass is
conserved up to 7-8 significant digits for a similar resolution.

A Fourier transform of $\rho_c(t)$ allows to measure the frequencies
of the dominant oscillation modes.
Fig.~\ref{F:TOVFourier} shows the Fourier power spectrum of $\rho_c(t)$ in
Cowling and full GR cases for simulations with the highest resolution $N=80$.
Both spectra contain three easily identifiable peaks corresponding to the
fundamental radial modes $F$, $H_1$ and $H_2$.
The same plot also shows derivatives of the spectral power with respect to the
frequency, computed using the central finite differencing scheme.
Zeroes of these numerical derivatives provide accurate estimates of the
location of frequency peaks.\ 
The frequency of the $F$-mode in Cowling approximation is 
$\nu(F) = 2.684(40)\ \kHz$, which is in agreement with the value of $2.706\
\kHz$, found in~\cite{Font01b}.\ 
In the fully general relativistic case, we obtain the frequency 
$\nu(F) = 1.440(50)\ \kHz$, which also agrees with the value $1.458\ \kHz$,
found in~\cite{Dimmelmeier06a}.
Note that the error in the values of fundamental frequencies above is
estimated as the distance between the root of the power spectrum derivative
and the nearest point with a non-zero value.

\newpage
\bibliography{}

\end{document}